\documentclass[manuscript]{aastex}

\usepackage{natbib}
\usepackage{color}
\usepackage{rotating}
\usepackage{lscape}
\usepackage{longtable}

\newcommand{\gps}{\ensuremath{g_{\rm P1}}}
\newcommand{\rps}{\ensuremath{r_{\rm P1}}}
\newcommand{\ips}{\ensuremath{i_{\rm P1}}}
\newcommand{\zps}{\ensuremath{z_{\rm P1}}}
\newcommand{\yps}{\ensuremath{y_{\rm P1}}}

\begin{document}
\title{Properties of M31. III: Candidate Beat Cepheids from PS1 PAndromeda Data and Their Implication on Metallicity Gradient}

\author{C.-H. Lee\altaffilmark{1,2}, M. Kodric\altaffilmark{1,2}, S. Seitz\altaffilmark{1,2}, A. Riffeser\altaffilmark{1,2}, J. Koppenhoefer\altaffilmark{2,1}, R. Bender\altaffilmark{1,2}, U. Hopp\altaffilmark{1,2}, C. G\"ossl\altaffilmark{1,2}, J. Snigula\altaffilmark{2,1}, W. S. Burgett\altaffilmark{3}, K. C. Chambers\altaffilmark{3}, H. Flewelling\altaffilmark{3}, K. W. Hodapp\altaffilmark{3}, N. Kaiser\altaffilmark{3}, R.-P. Kudritzki\altaffilmark{3}, P. A. Price\altaffilmark{4}, J. L. Tonry\altaffilmark{3}, R. J. Wainscoat\altaffilmark{3}}
\altaffiltext{1}{University Observatory Munich, Scheinerstrasse 1, 81679 Munich, Germany}
\altaffiltext{2}{Max Planck Institute for Extraterrestrial Physics, Giessenbachstrasse, 85748 Garching, Germany}
\altaffiltext{3}{Institute for Astronomy, University of Hawaii at Manoa, Honolulu, HI 96822, USA}
\altaffiltext{4}{Department of Astrophysical Sciences, Princeton University, Princeton, NJ 08544, USA}

\received{2013 June}
\accepted{2013 August}
\begin{abstract}
We present a sample of M31 beat Cepheids from the Pan-STARRS 1 PAndromeda campaign. 
By analyzing three years of PAndromeda data, we identify seventeen beat Cepheids, 
spreading from a galactocentric distance of 10 to 16 kpc. 
Since the relation between fundamental mode period and the ratio of 
fundamental to the first overtone period puts a tight constraint on metallicity 
we are able to derive the metallicity
at the position of the beat Cepheids using the 
relations from the model of \cite{2008ApJ...680.1412B}. 
Our metallicity estimates show sub-solar values within 15 kpc, 
similar to the metallicities from HII regions \citep{2012MNRAS.427.1463Z}.
We then use the metallicity estimates
to calculate the metallicity gradient of the M31 disk, which we find to be 
closer to the metallicity gradient derived from planetary nebula 
\citep{2012ApJ...753...12K} than the metallicity gradient 
from HII regions \citep{2012MNRAS.427.1463Z}. 
\end{abstract}


\section{Introduction}
Beat Cepheids are pulsating simultaneously in two radial modes.
Studies of beat Cepheids can be dated back to \cite{1957BAN....13..317O,1957BAN....13..320O}, 
where he introduced a beat period to explain the large scattered photometric measurements of 
U TrA and TU Cas in the Milky Way. Several attempts to search for Galactic beat Cepheids have been
conducted \citep[see e.g.][]{1979MNRAS.187..261P,1979MNRAS.189..149H,1980MNRAS.192..621H}, however, only 20 Galactic beat Cepheids 
are documented to-date (see e.g. McMaster Cepheid Data Archive\footnote{http://crocus.physics.mcmaster.ca/Cepheid/}
, where 651 Type I Cepheids and 209 Type II Cepheids are listed as well).
The first larger samples of beat Cepheids have been identified in microlensing survey data. 
For example, the MACHO project has discovered 
45 beat Cepheids in the Large Magellanic Cloud, where 30 are pulsating in the fundamental
mode and first overtone, while 15 are pulsating in the first and second overtone \citep{1995AJ....109.1653A}. 
The OGLE team found 93 beat Cepheids in the Small Magellan Clouds \citep{1999AcA....49....1U} and 
76 beat Cepheids in the Large Magellanic Clouds \citep{2000AcA....50..451S}. 
A recent study from the EROS group has increased the number of known beat 
Cepheids in the Magellanic clouds to over 200 \citep{2009A&A...495..249M}. 
The OGLE-III survey has found the largest number
of beat Cepheids so far: 271 objects in the LMC \citep{2008AcA....58..163S}
 and 277 in the SMC \citep{2010AcA....60...17S}. These
numbers will be even larger during the currently conducted
OGLE-IV phase.
 
Beat Cepheids pulsating in the fundamental mode and first overtone 
can be used as a tracer of the metallicity content within a galaxy. 
This is because from modelling, there exists only a sub-region in the parameter
spaces of mass, luminosity, temperature, and metallicity where both 
the fundamental mode and first overtone are 
linearly unstable \citep[see e.g.][]{2002A&A...385..932K}.
\cite{2006ApJ...653L.101B} have thus made use of the beat Cepheids found in the CFHT M33 survey \citep{2006MNRAS.371.1405H}
and derived the metallicity gradient of M33 to be -0.16 dex/kpc. Their metallicity gradient supports 
the HII region result from \cite{1997ApJ...489...63G} but disagrees with the much shallower 
gradient from \cite{2006ApJ...637..741C}, who also used HII region to derive the metallicity. It is important to note that both results from \cite{1997ApJ...489...63G} and \cite{2006ApJ...637..741C} are derived from HII regions, yet are inconsistent with each other. 

In this study we present a sample of beat Cepheids identified from the PS1 PAndromeda 
project. We derive the metallicity gradient of M31 and compare our results
with previous studies of HII regions and planetary nebulae. Our paper is composed as follows. In section \ref{sec.bc_sel} we demonstrate 
our method to search for beat Cepehids. We elucidate the approach to derive metallicity 
in section \ref{sec.est_z}. The metallicity gradient of M31 from 
our sample, as well as a comparison with previous HII region and planetary nebulae method is presented in 
section \ref{sec.z_grad}, followed by a conclusion and outlook in section \ref{sec.outlook}. 

\section{Beat Cepheid Identification}
\label{sec.bc_sel}
We use the optical data taken by the PAndromeda project to search for beat Cepheids. 
PAndromeda monitors the Andromeda galaxy with the 1.8m PS1 telescope with a $\sim$ 7 deg$^2$ 
field-of-view \citep[see][for a detailed description of the PS1 system, optical design, and the imager]{2010SPIE.7733E..12K,2004SPIE.5489..667H,2009amos.confE..40T}. Observations are taken in $\rps$ and $\ips$ on daily basis during July to December 
in order to search for microlensing events and variables. Several exposures in $\gps$, $\zps$, and $\yps$ 
are also taken as complementary information for studies on the stellar content.

The data reduction is based on the MDia tool \citep{2013ExA....35..329K} and is explained in \cite{2012AJ....143...89L} in detail. 
We outline our data reduction 
steps as follows. The raw data are detrended by the image processing pipeline \citep[IPP, ][]{2006amos.confE..50M} and warped to a sky-based image plane (so-called skycells). The images at the skycell stage 
are further analyzed by our sophisticated imaging subtraction pipeline \textit{mupipe} \citep{2002A&A...381.1095G}
based on the idea of image differencing analysis advocated by \cite{1998ApJ...503..325A}. This includes the creation of deep
reference images from best seeing data, stacking of observations within one visit to have better signal
to noise ratio (hereafter ``visit stacks''), subtraction of visit stacks from the reference images
to search for variability, and creating light-curves from the subtracted images.

\begin{figure*}[!h]
  \centering
  \includegraphics[scale=0.8]{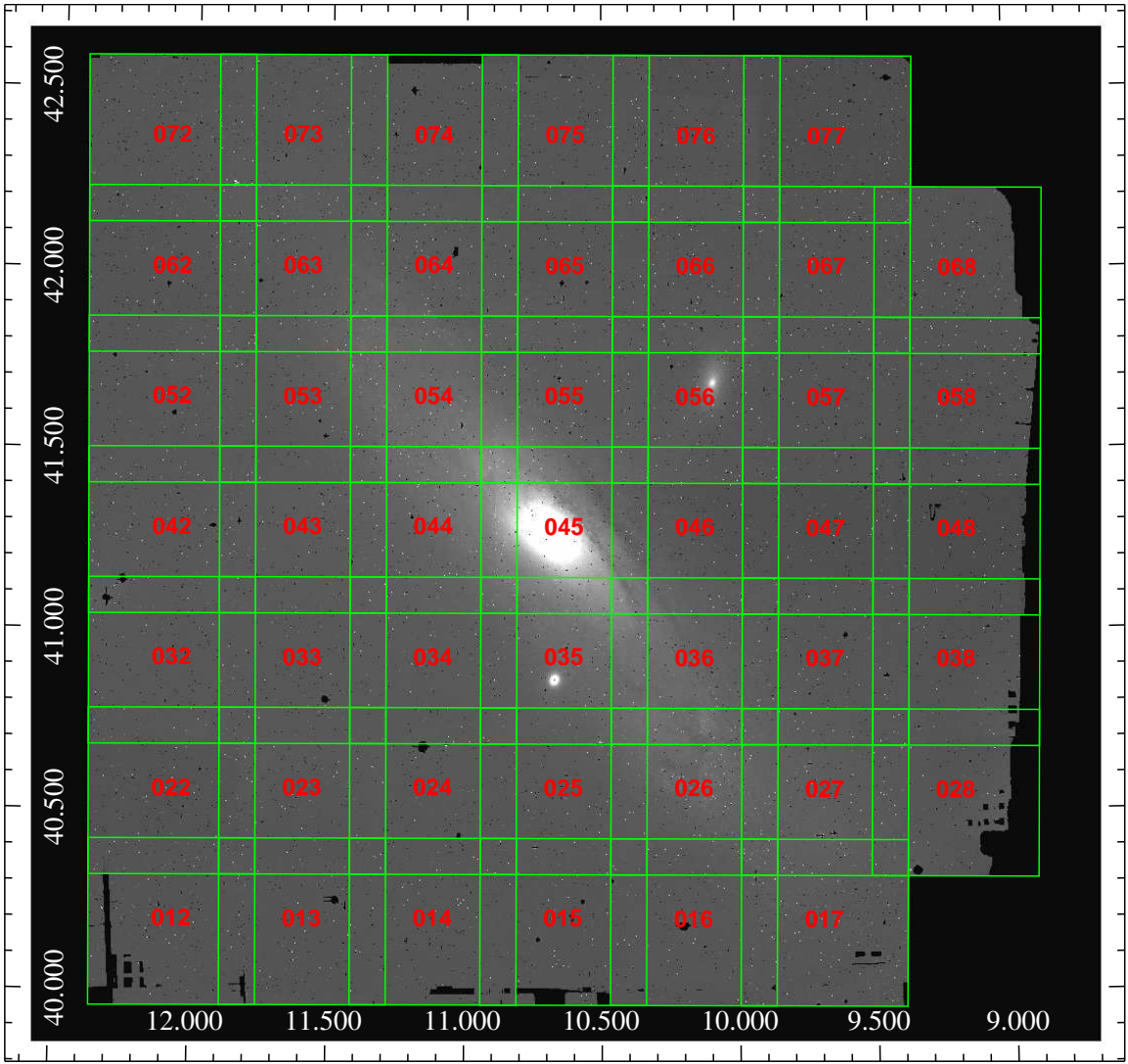}
  \caption{Illustration of our new tessellation. The background image is a mosaic of reference images from all skycells analyzed in this work.}
  \label{fig.tes}
\end{figure*}

We have shown in \cite{2013AJ....145..106K} how to obtain Cepheid light-curves in the PAndromeda data. 
The major difference is that the data-set
used in this work contains three years of PAndromeda, instead of one year and a few days from the 
second year data used in \cite{2013AJ....145..106K}.
The sky tessellation is also different, in order to have the central region of M31 in the center of a skycell 
(skycell 045), instead of at the corner of adjacent skycells (skycell number 065, 066, 077, and 078) as in \cite{2013AJ....145..106K};
the skycells are larger and overlap in the new tessellation. The new tessellation is drawn in Fig. \ref{fig.tes}. 
We have extended the analysis to 47 skycells, twice as many as the number of skycells used in \cite{2013AJ....145..106K}. 
The skycells we used are 012-017, 022-028, 032-038, 042-048, 052-058, 062-068, 072-077, which cover the whole of M31. 
The search of Cepheids is conducted in both $\rps$ and $\ips$, where we start from the resolved sources in 
the $\rps$ reference images, and require variability in both $\rps$ and $\ips$ filters. In addition 
one could search for variables also in the pixel-based light-curves. This approach would add light-curves for fainter variable
sources (among them potentially lower period Cepheids) which we do not aim to study in this work.

We use the SigSpec package \citep{2007A&A...467.1353R} to determine the period of all 
variables. For a given light-curve, we iterate the 
period search five times both in $\rps$ and $\ips$ to search for multiple periods. 
In each iteration, SigSpec computes the significance spectrum and determines the most significant period.
It then fits a multi-sine function based on this period, subtracts the best-fitted multi-sine curve to the 
input light-curve, and performs another iteration of period search based on this pre-whitened light-curve.

\begin{figure*}[!h]
  \centering
  \includegraphics[scale=0.8]{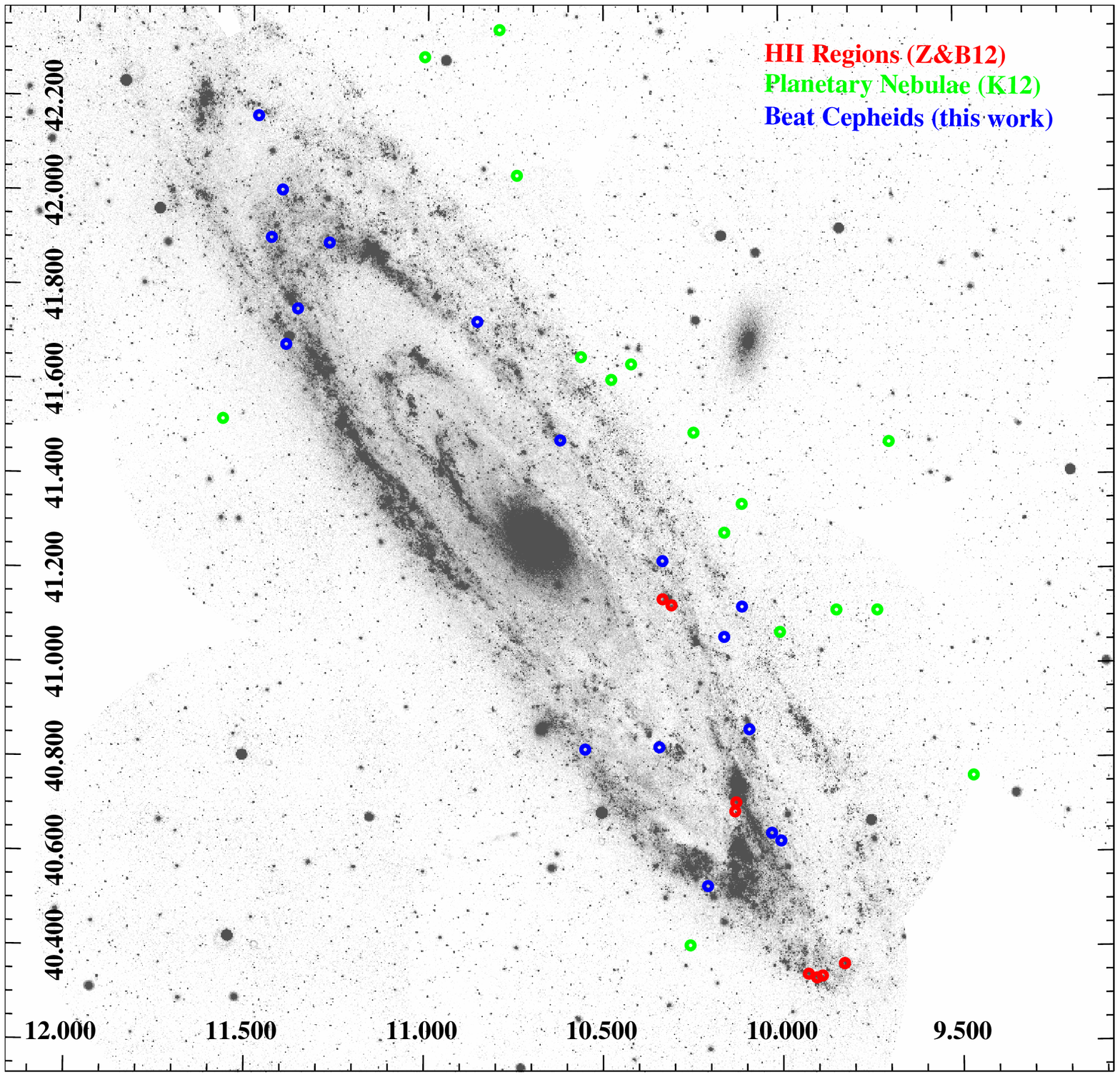}
  \caption{Spatial distribution of our sample (blue circle), HII regions in \cite{2012MNRAS.427.1463Z}, and planetary nebulae in \cite{2012ApJ...753...12K}, over-plotted with \textit{GALEX} NUV image \citep{2007ApJS..173..185G}.}
  \label{fig.spat}
\end{figure*}

For the beat Cepheids, we look for sources that are showing only two significant periods 
(i.e. where SigSpec does not find a period after the 
second iteration). We also require that both periods are found in $\rps$ and $\ips$ light-curves and are consistent 
within ten percent. 
We adopt the period derived from $\rps$ as final period, due to the better sampling and the higher amplitude than the $\ips$-band light-curves.
This leads to a sample of seventeen beat Cepheids. Their locations, periods in fundamental mode (P$_0$) and first overtone (P$_1$)
, and light-curves are shown in 
Fig. \ref{fig.spat}, Fig. 3, and Table \ref{tab.bc}. In the next section, we present 
their metallicities derived from the period and the period ratio. Given the periods and 
metallicities, we are also able to obtain an estimate of their ages.

\begin{table*}[!h]
\centering
\begin{sideways}
\begin{minipage}{255mm}
\begin{tabular}{lllllllllll}
\hline 
Name              & RA      & Dec      & \bf{P$_0 ^{\rps}$} & \bf{P$_1 ^{\rps}$} & \bf{P$_1 ^{\rps}$/P$_0 ^{\rps}$} & P$_0 ^{\ips}$ & P$_1 ^{\ips}$ &  P$_1 ^{\ips}$/P$_0 ^{\ips}$ \\
                  & (J2000) & (J2000)  & [days]      & [days]       &                   & [days]       & [days]      &  \\
\hline
PSO J010.0031+40.6271 & 10.00313 & 40.62716 & \bf{5.08121$\pm$0.00131} & \bf{3.57641$\pm$0.00049} & \bf{0.703850} & 5.08163$\pm$0.00790 & 3.57673$\pm$0.00161 & 0.703855 \\ 
PSO J010.0289+40.6434 & 10.02891 & 40.64346 & \bf{4.42890$\pm$0.00084} & \bf{3.11390$\pm$0.00071} & \bf{0.703087} & 4.42311$\pm$0.00269 & 3.11339$\pm$0.00099 & 0.703892 \\ 
PSO J010.0908+40.8632 & 10.09082 & 40.86323 & \bf{3.82737$\pm$0.00050} & \bf{2.69071$\pm$0.00028} & \bf{0.703018} & 3.82750$\pm$0.00103 & 2.69091$\pm$0.01056 & 0.703046 \\ 
PSO J010.1097+41.1233 & 10.10973 & 41.12339 & \bf{3.88787$\pm$0.00057} & \bf{2.76180$\pm$0.00049} & \bf{0.710363} & 3.88655$\pm$0.00129 & 2.76147$\pm$0.00076 & 0.710520 \\ 
PSO J010.1601+41.0591 & 10.16016 & 41.05914 & \bf{4.81211$\pm$0.00068} & \bf{3.37280$\pm$0.00057} & \bf{0.700898} & 4.81242$\pm$0.00120 & 3.37283$\pm$0.00140 & 0.700859 \\ 
PSO J010.2081+40.5311 & 10.20820 & 40.53114 & \bf{3.73484$\pm$0.00059} & \bf{2.66164$\pm$0.00085} & \bf{0.712652} & 3.73517$\pm$0.00191 & 2.66181$\pm$0.01328 & 0.712634 \\ 
PSO J010.3333+41.2202 & 10.33331 & 41.22027 & \bf{3.96209$\pm$0.00066} & \bf{2.82765$\pm$0.00021} & \bf{0.713676} & 3.96449$\pm$0.00257 & 2.82771$\pm$0.00043 & 0.713259 \\ 
PSO J010.3431+40.8255 & 10.34310 & 40.82556 & \bf{8.68802$\pm$0.00299} & \bf{6.02351$\pm$0.00240} & \bf{0.693312} & 8.69612$\pm$0.00515 & 6.02638$\pm$0.00361 & 0.692996 \\ 
PSO J010.5507+40.8208 & 10.55071 & 40.82087 & \bf{4.68226$\pm$0.00140} & \bf{3.28483$\pm$0.00052} & \bf{0.701548} & 4.67941$\pm$0.00300 & 3.28542$\pm$0.00074 & 0.702101 \\ 
PSO J010.6214+41.4763 & 10.62146 & 41.47634 & \bf{5.86549$\pm$0.00125} & \bf{4.08473$\pm$0.00099} & \bf{0.696400} & 5.86891$\pm$0.00338 & 4.08356$\pm$0.02249 & 0.695795 \\ 
PSO J010.8571+41.7272 & 10.85714 & 41.72723 & \bf{4.12749$\pm$0.00071} & \bf{2.93815$\pm$0.00049} & \bf{0.711849} & 4.12611$\pm$0.00164 & 2.93745$\pm$0.00260 & 0.711918 \\ 
PSO J011.2784+41.8935 & 11.27840 & 41.89359 & \bf{4.77328$\pm$0.00088} & \bf{3.37092$\pm$0.00056} & \bf{0.706206} & 4.77551$\pm$0.02909 & 3.34082$\pm$0.01690 & 0.699573 \\ 
PSO J011.3670+41.7533 & 11.36709 & 41.75335 & \bf{8.26314$\pm$0.00167} & \bf{5.78110$\pm$0.00110} & \bf{0.699625} & 8.26723$\pm$0.00300 & 5.78289$\pm$0.00169 & 0.699495 \\ 
PSO J011.3993+41.6778 & 11.39932 & 41.67789 & \bf{4.81283$\pm$0.00192} & \bf{3.39273$\pm$0.00053} & \bf{0.704935} & 4.81013$\pm$0.00356 & 3.39258$\pm$0.00090 & 0.705299 \\ 
PSO J011.4131+42.0052 & 11.41317 & 42.00529 & \bf{3.66634$\pm$0.00081} & \bf{2.60609$\pm$0.00043} & \bf{0.710815} & 3.66630$\pm$0.00272 & 2.60564$\pm$0.00130 & 0.710700 \\ 
PSO J011.4436+41.9044 & 11.44369 & 41.90446 & \bf{2.37187$\pm$0.00066} & \bf{1.69231$\pm$0.00032} & \bf{0.713492} & 2.37164$\pm$0.00096 & 1.69201$\pm$0.00750 & 0.713435 \\ 
PSO J011.4835+42.1621 & 11.48356 & 42.16218 & \bf{6.09759$\pm$0.00176} & \bf{4.25145$\pm$0.00131} & \bf{0.697234} & 6.09709$\pm$0.00277 & 4.25231$\pm$0.00364 & 0.697433 \\
\hline
\multicolumn{11}{l}{Table 1: Location and periods of our beat Cepheid sample. We high-lighted the $\rps$-band columns in because these are the ones we adopt for the final analysis.}
\end{tabular}
\end{minipage}
\end{sideways}
\label{tab.bc}
\end{table*}

\clearpage
\begin{figure*}[!h]
  \centering
  \includegraphics[scale=0.3]{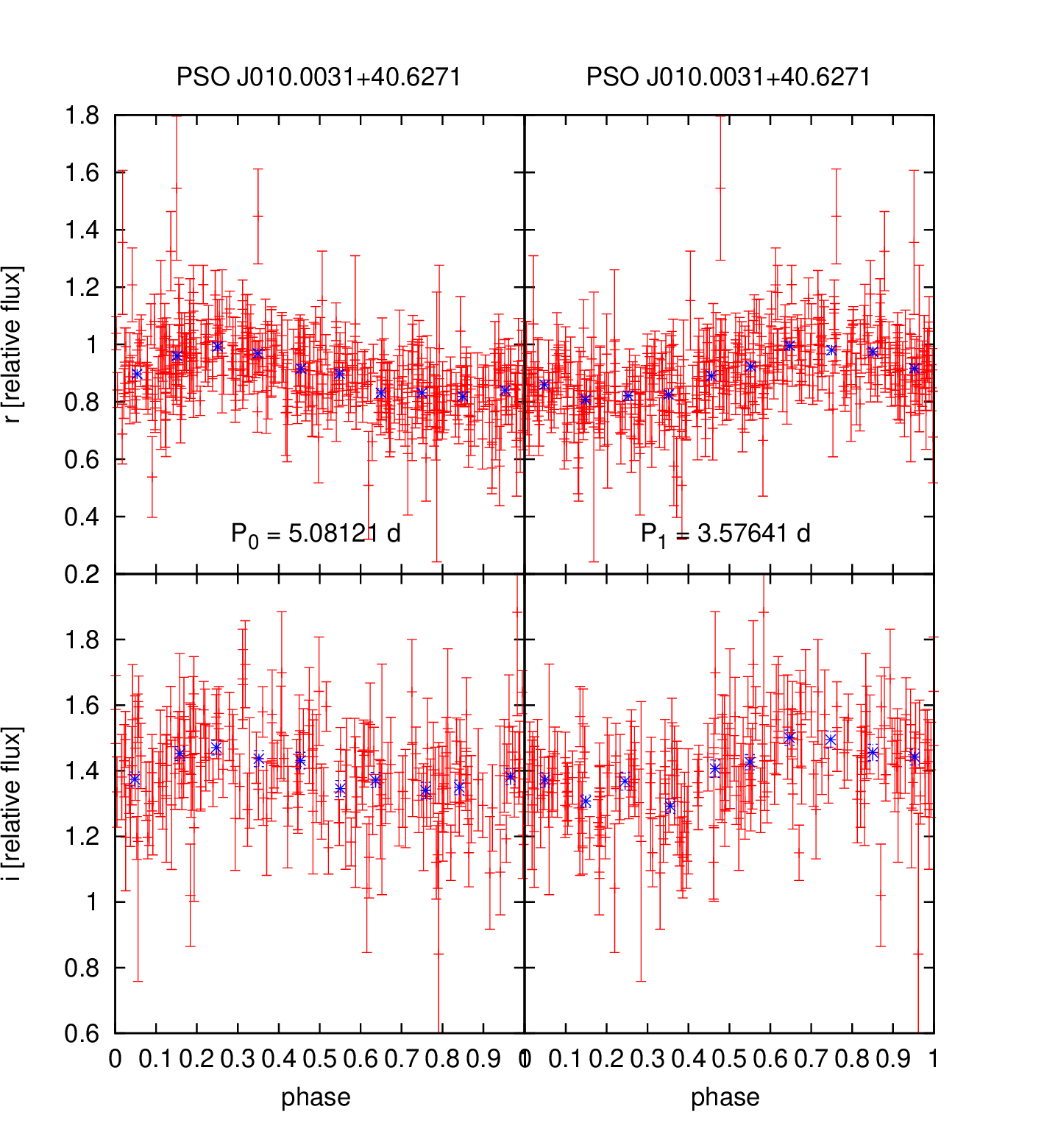}
  \includegraphics[scale=0.3]{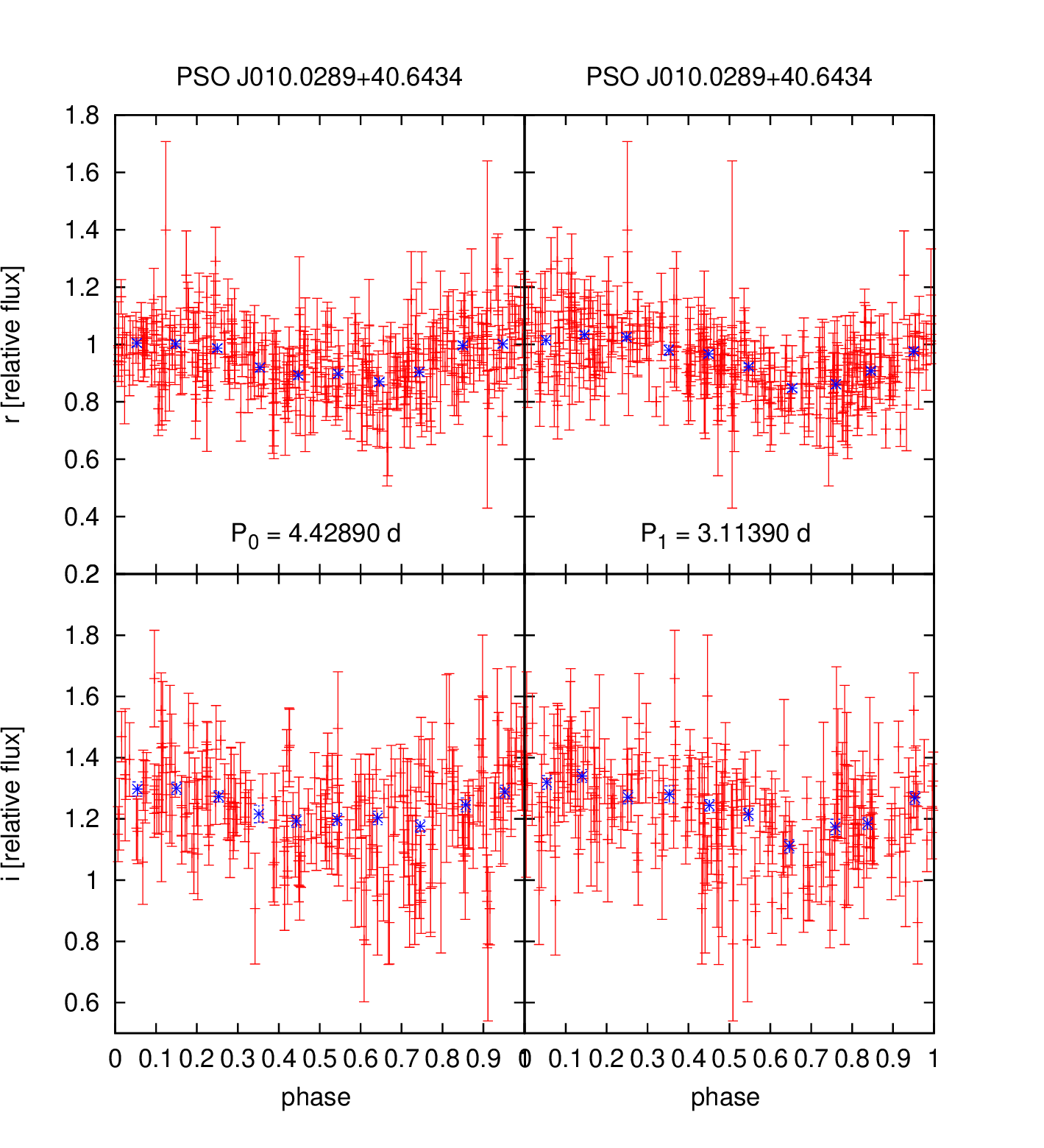}
  \includegraphics[scale=0.3]{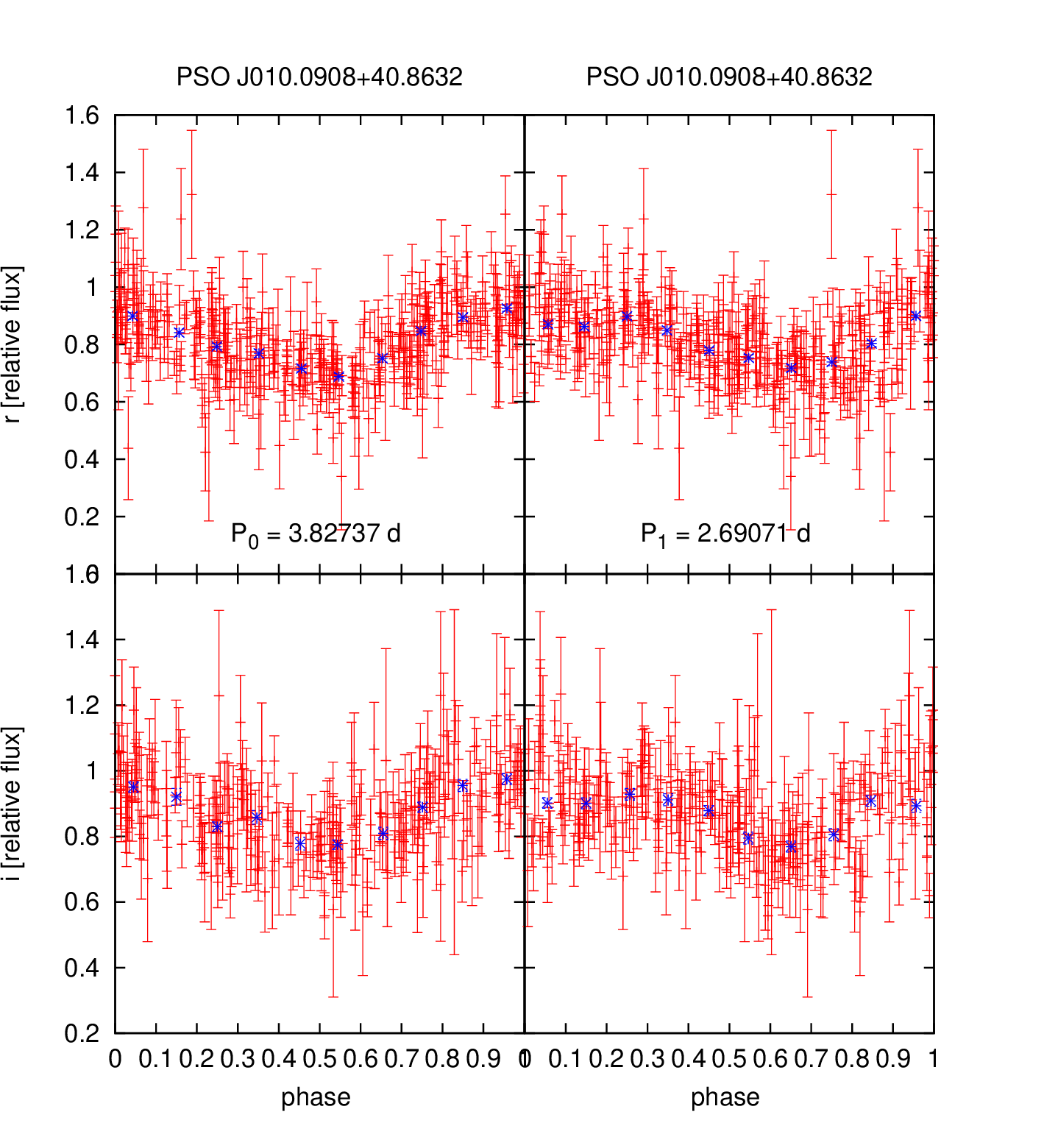}
  \includegraphics[scale=0.3]{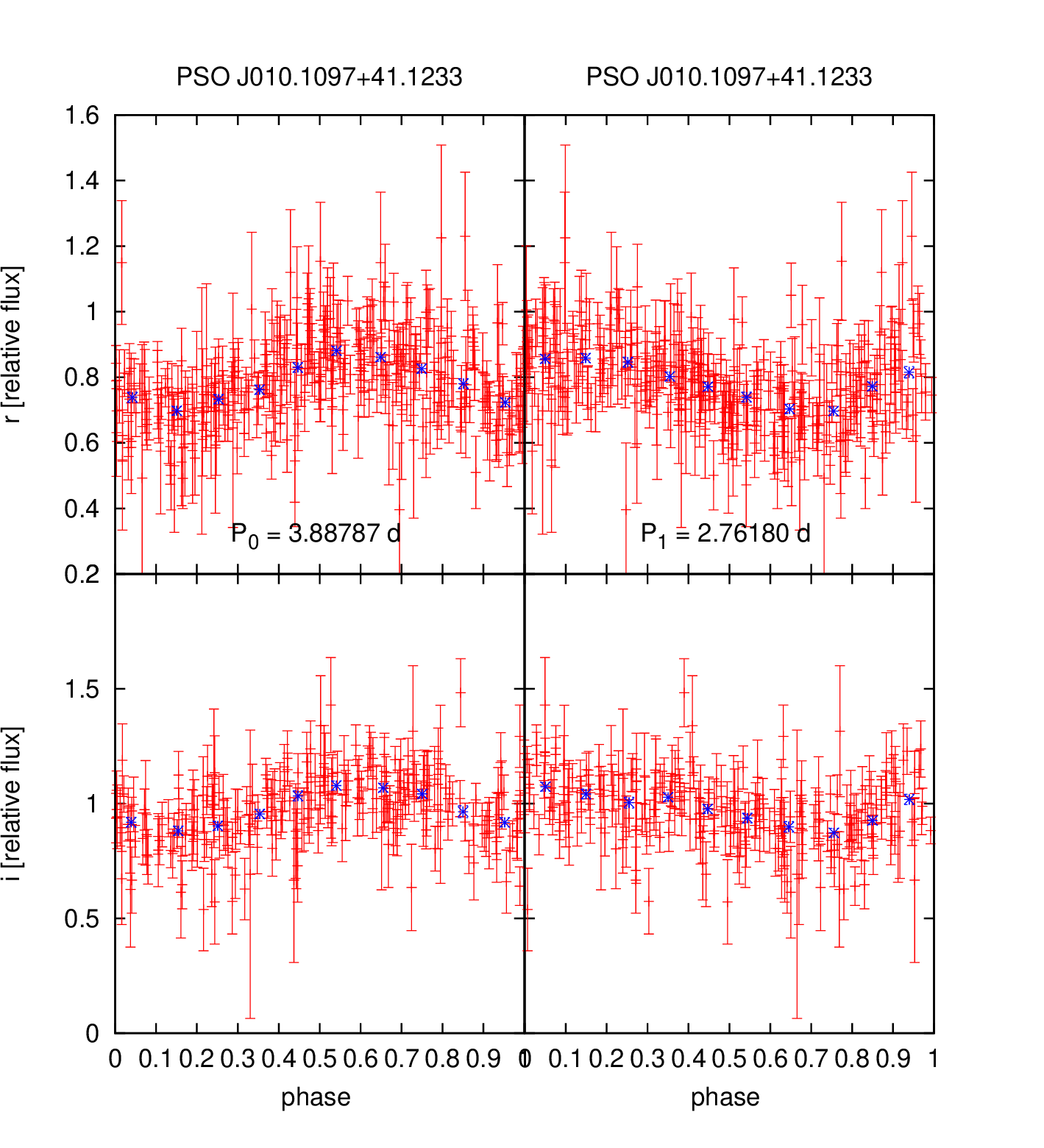}
  \caption{Light-curves of our sample. We fold the light-curves with the $\rps$-band period of fundamental mode P$_0$ (left-hand side) and first overtone P$_1$ (right-hand side). The red points are un-binned data, while the blue points are data binned with 0.1 phase interval.}
  \label{fig.lc1}
\end{figure*}

\addtocounter{figure}{-1}
\begin{figure*}[!h]
  \centering
  \includegraphics[scale=0.3]{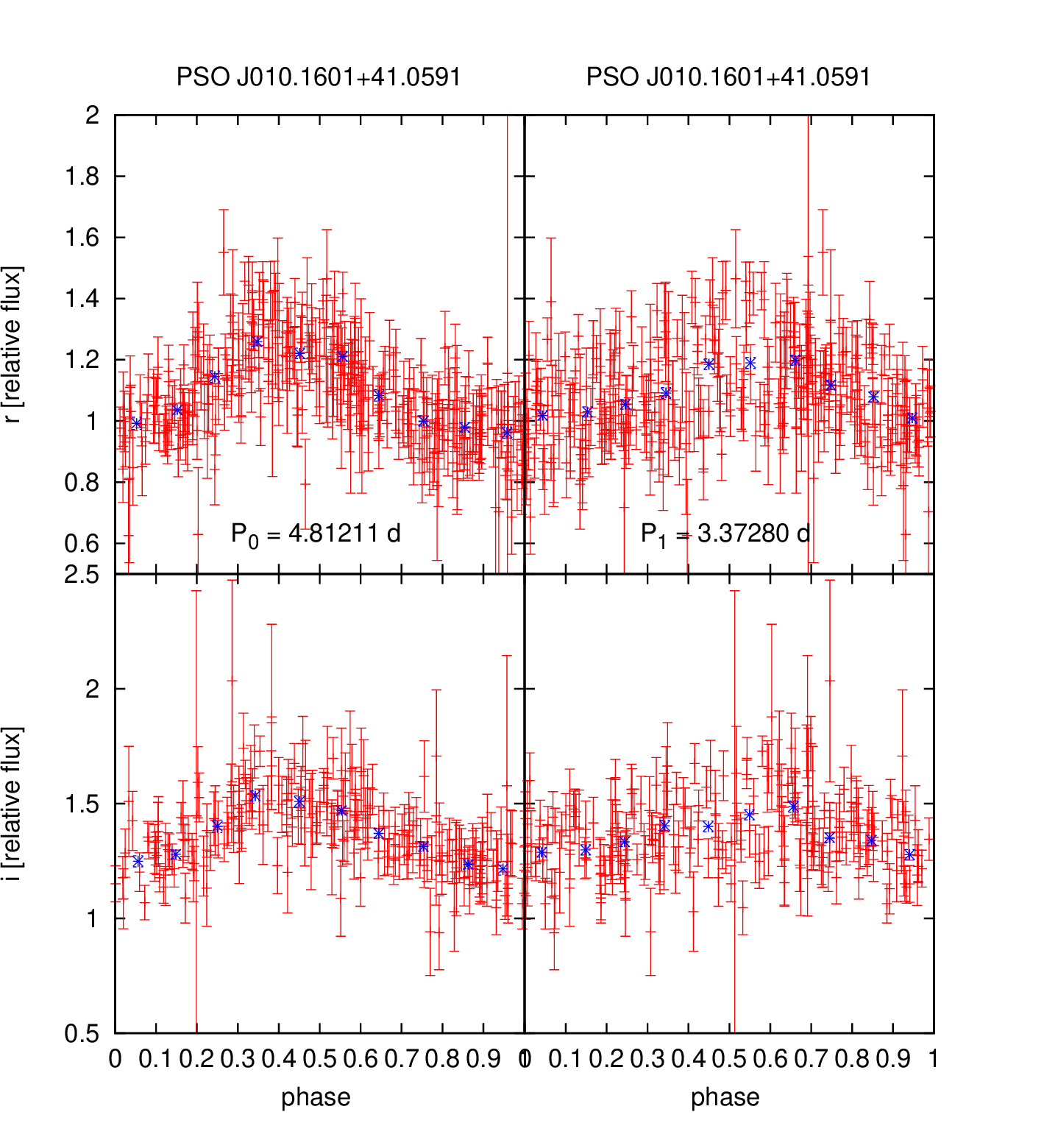}
  \includegraphics[scale=0.3]{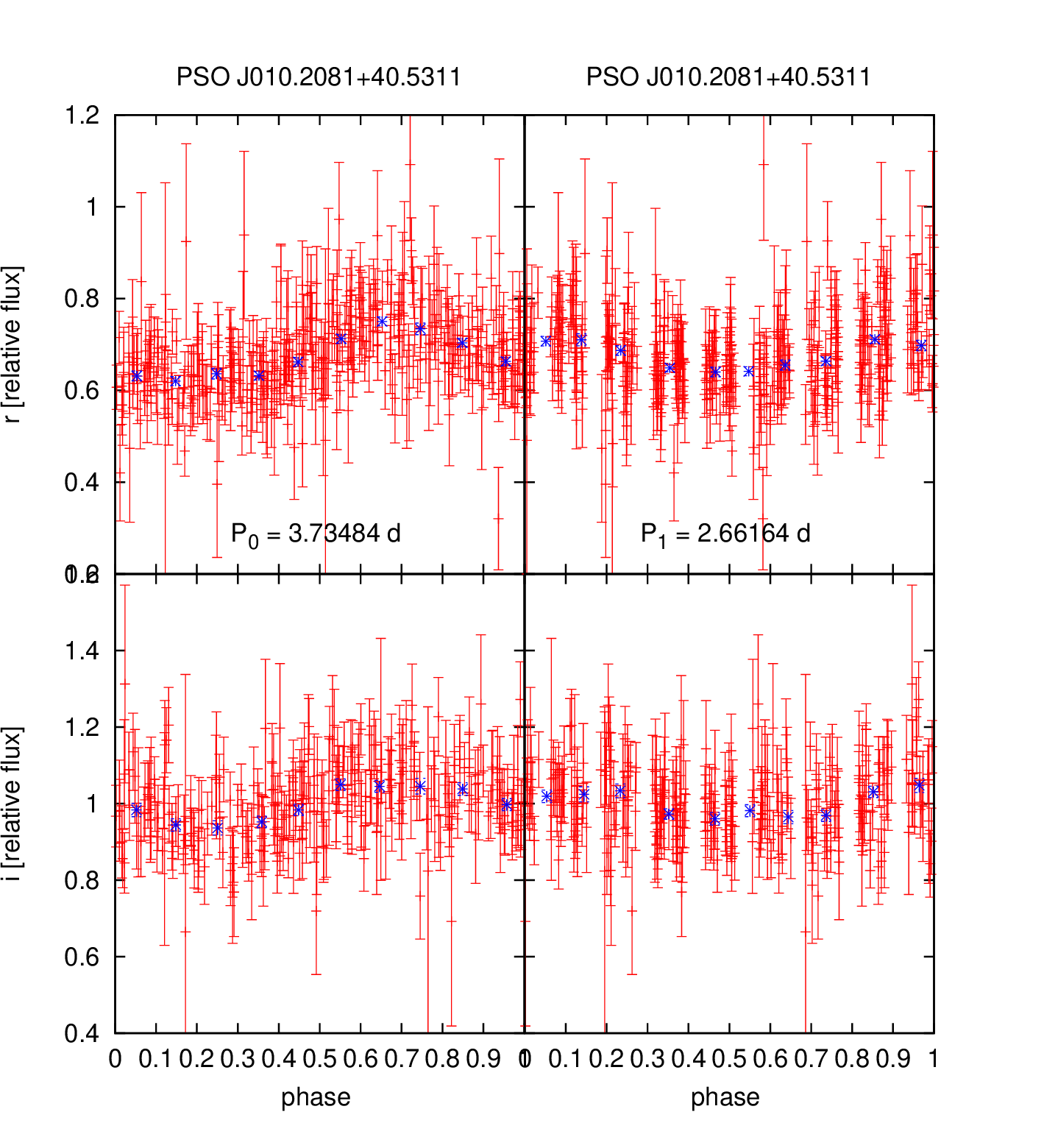}
  \includegraphics[scale=0.3]{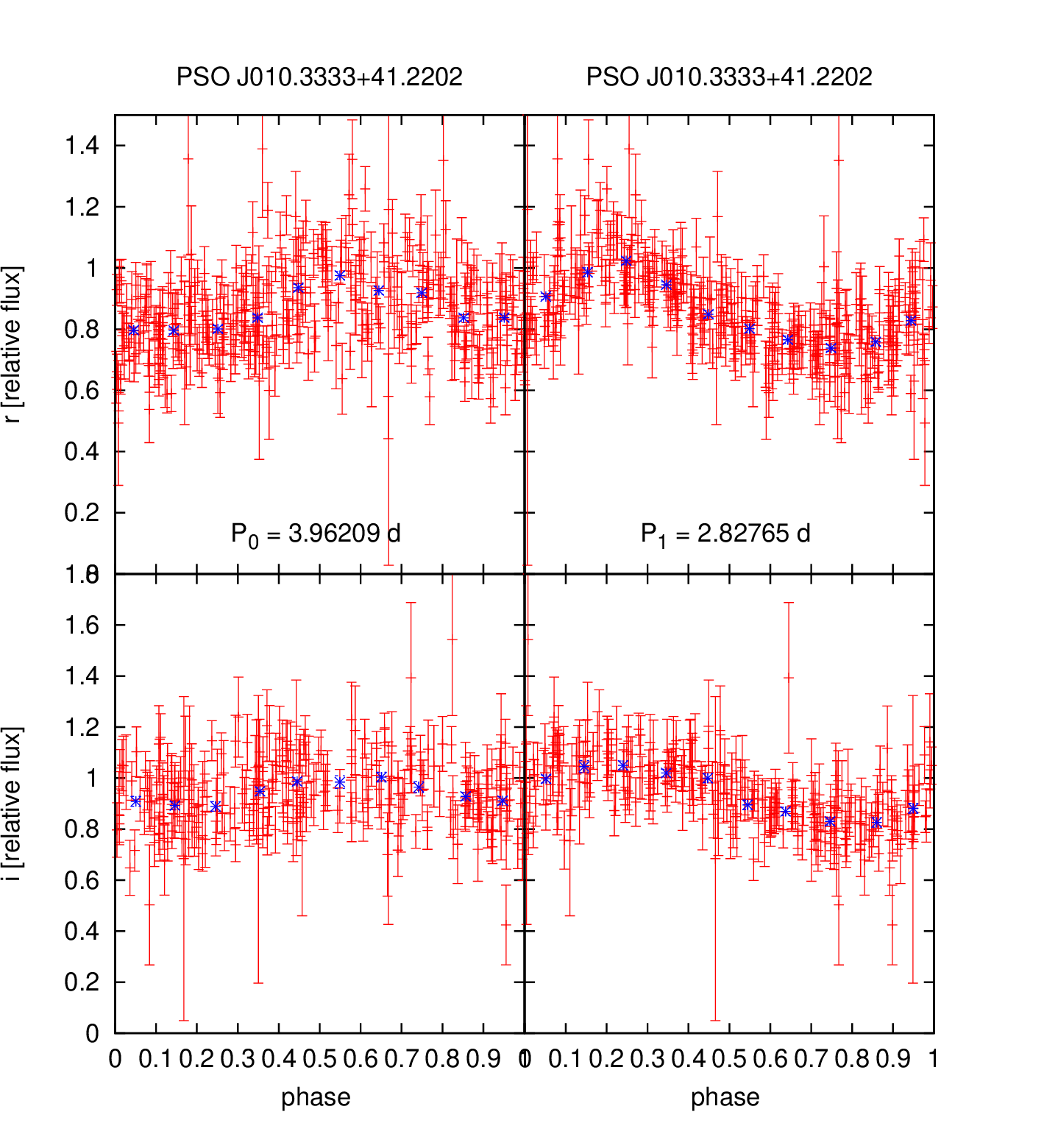}
  \includegraphics[scale=0.3]{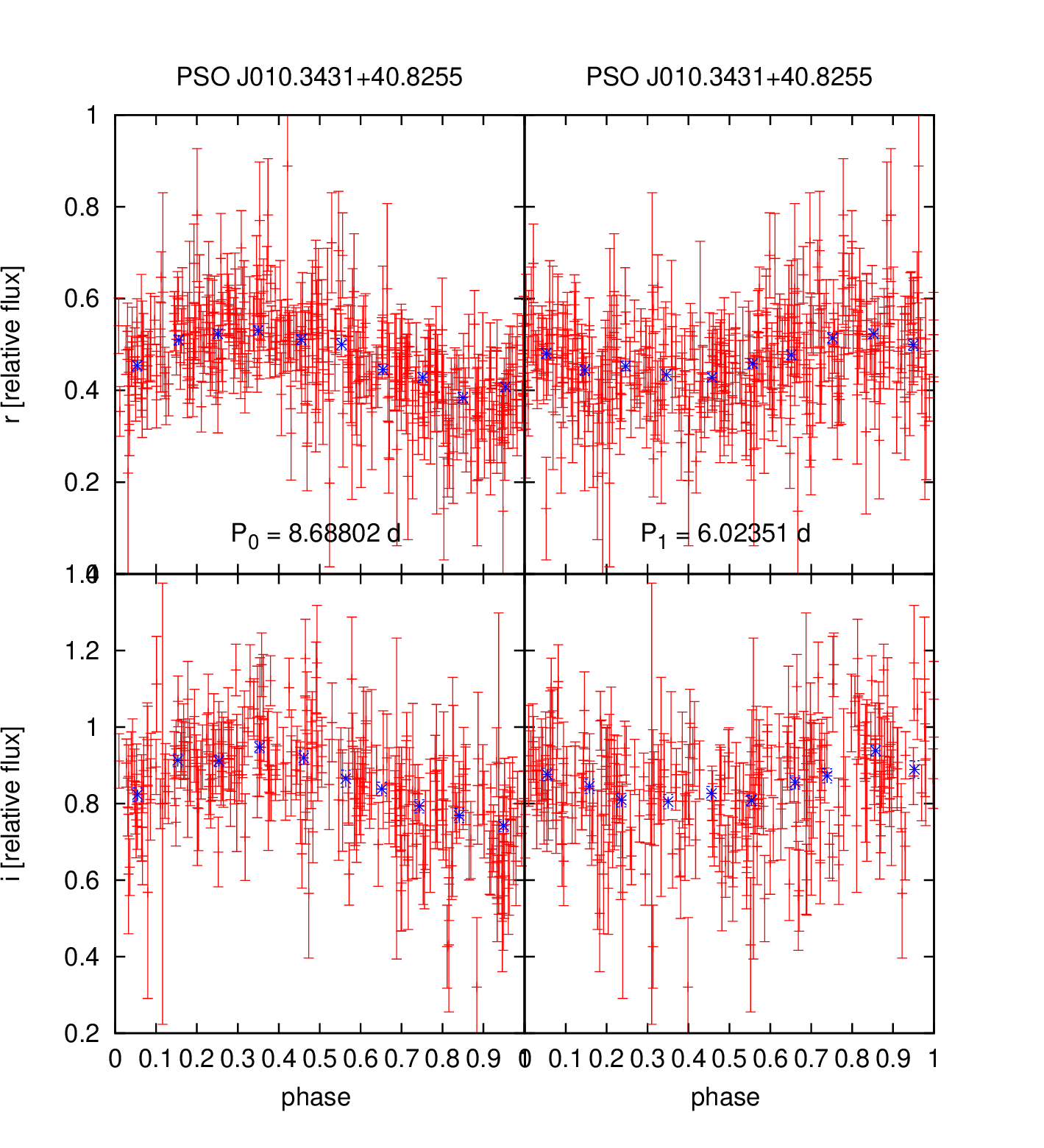}
  \caption{Light-curves of our sample - \textit{continued}.}
  \label{fig.lc}
\end{figure*}

\addtocounter{figure}{-1}
\begin{figure*}[!h]
  \centering
  \includegraphics[scale=0.3]{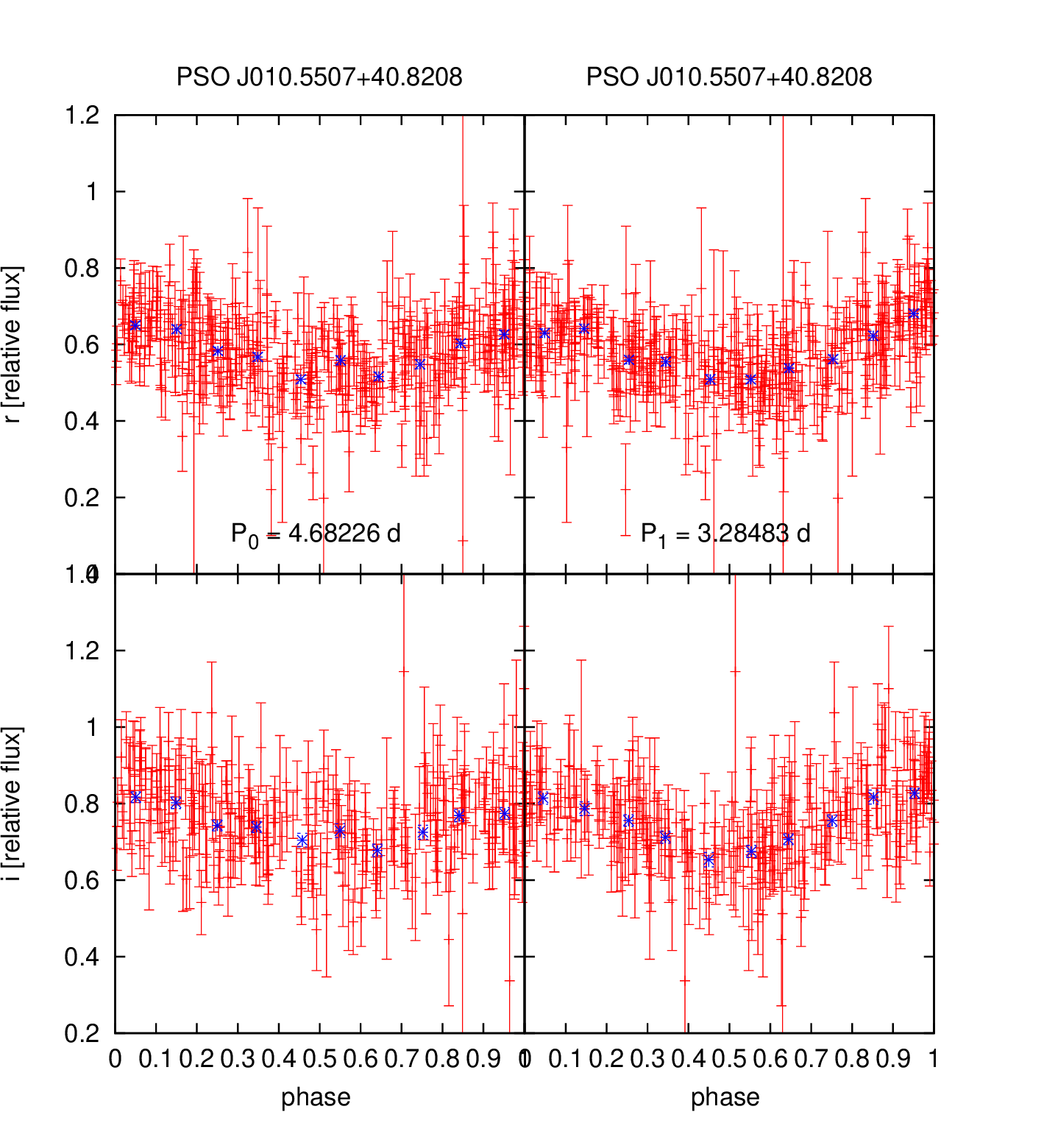}
  \includegraphics[scale=0.3]{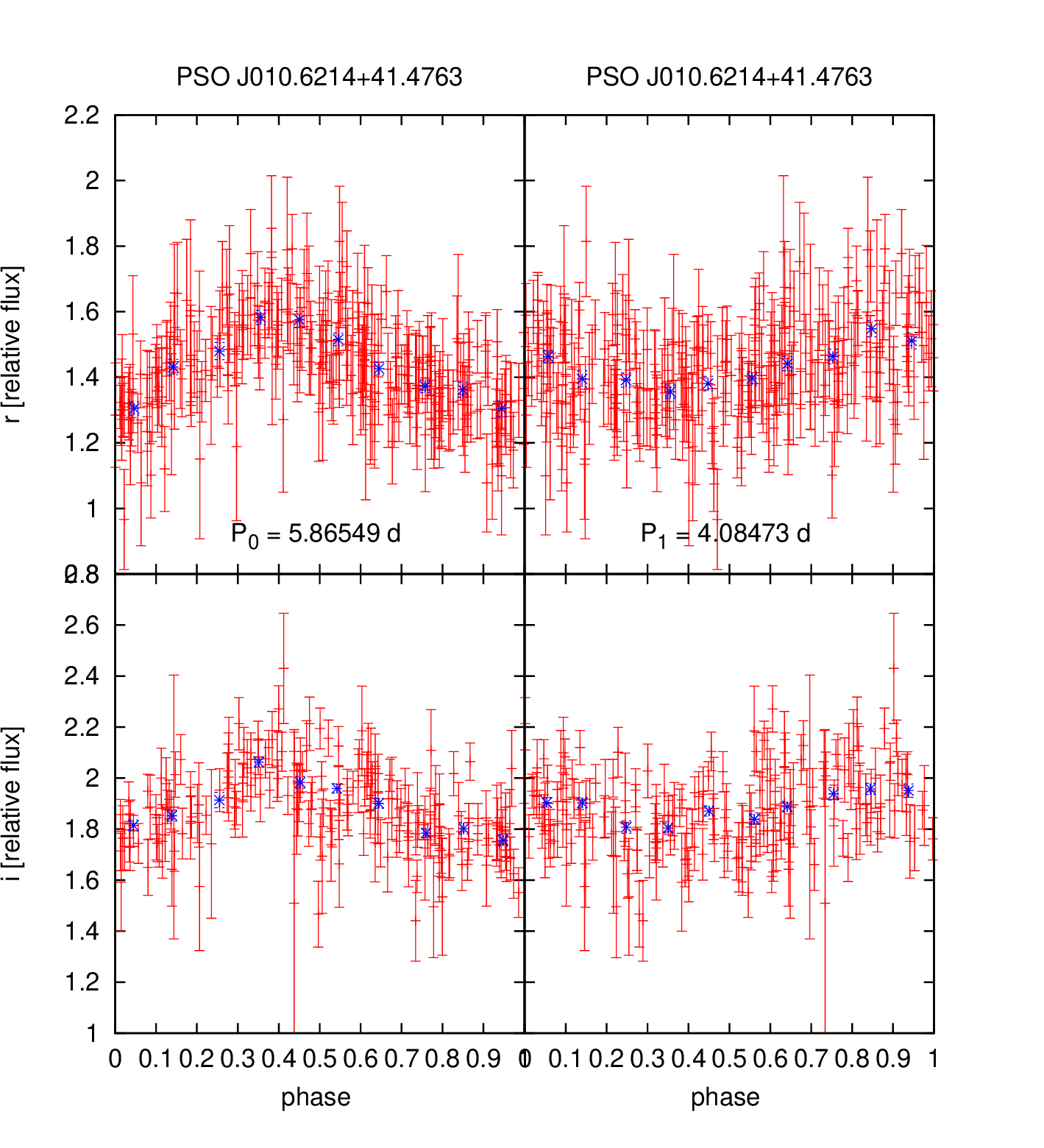}
  \includegraphics[scale=0.3]{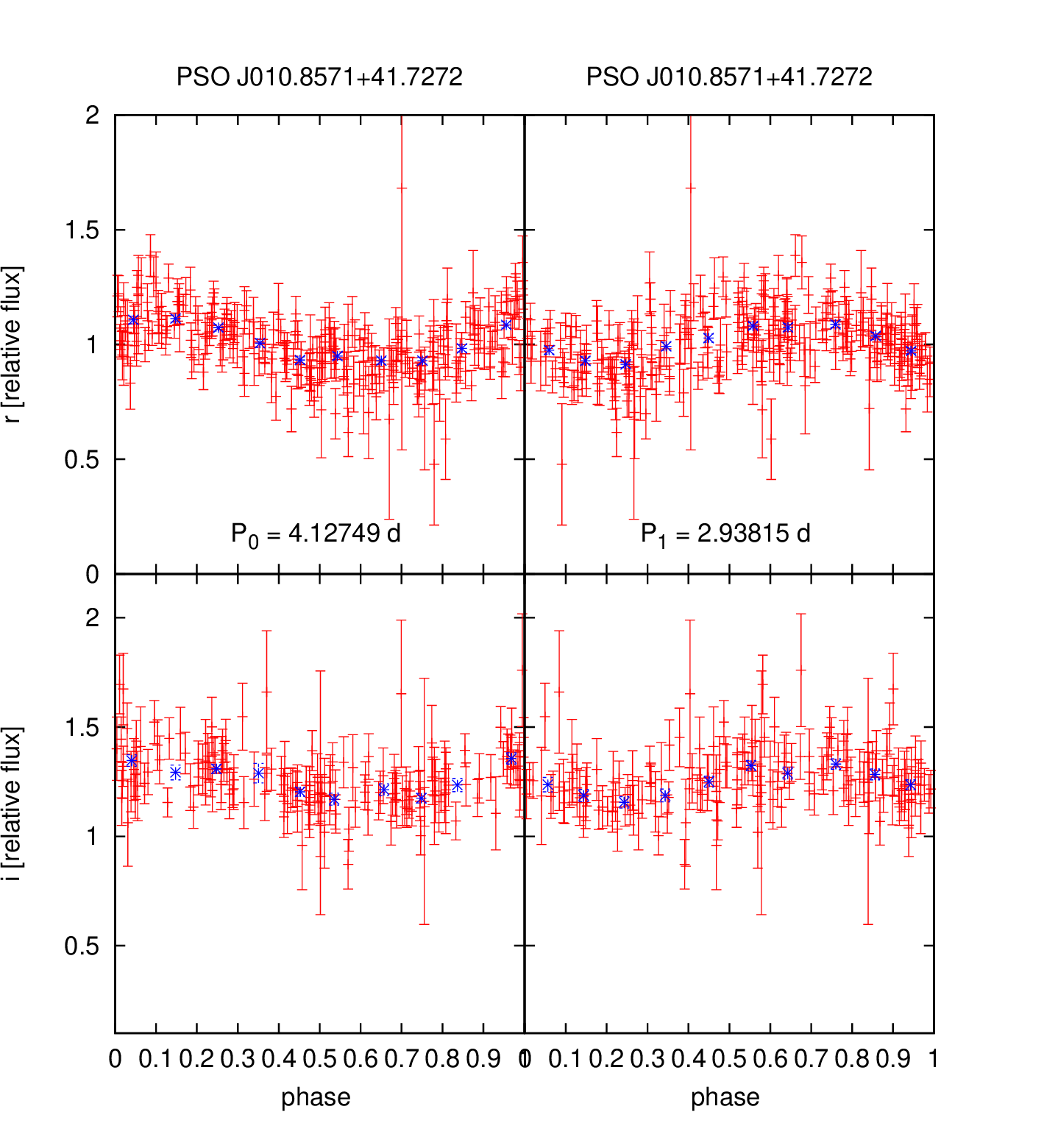}
  \includegraphics[scale=0.3]{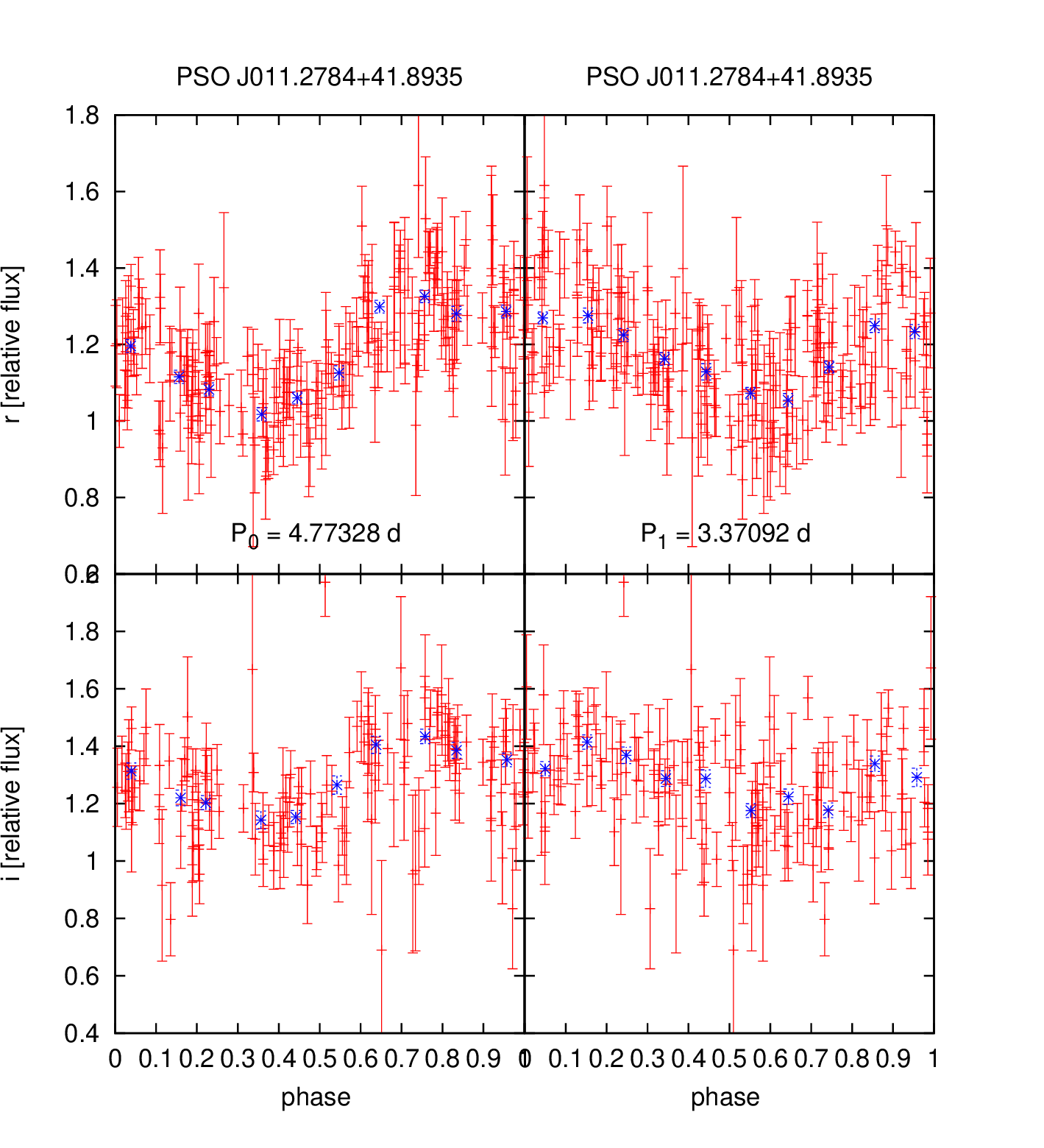}
  \caption{Light-curves of our sample - \textit{continued}.}
  \label{fig.lc}
\end{figure*}
\addtocounter{figure}{-1}
\begin{figure*}[!h]
  \centering
  \includegraphics[scale=0.3]{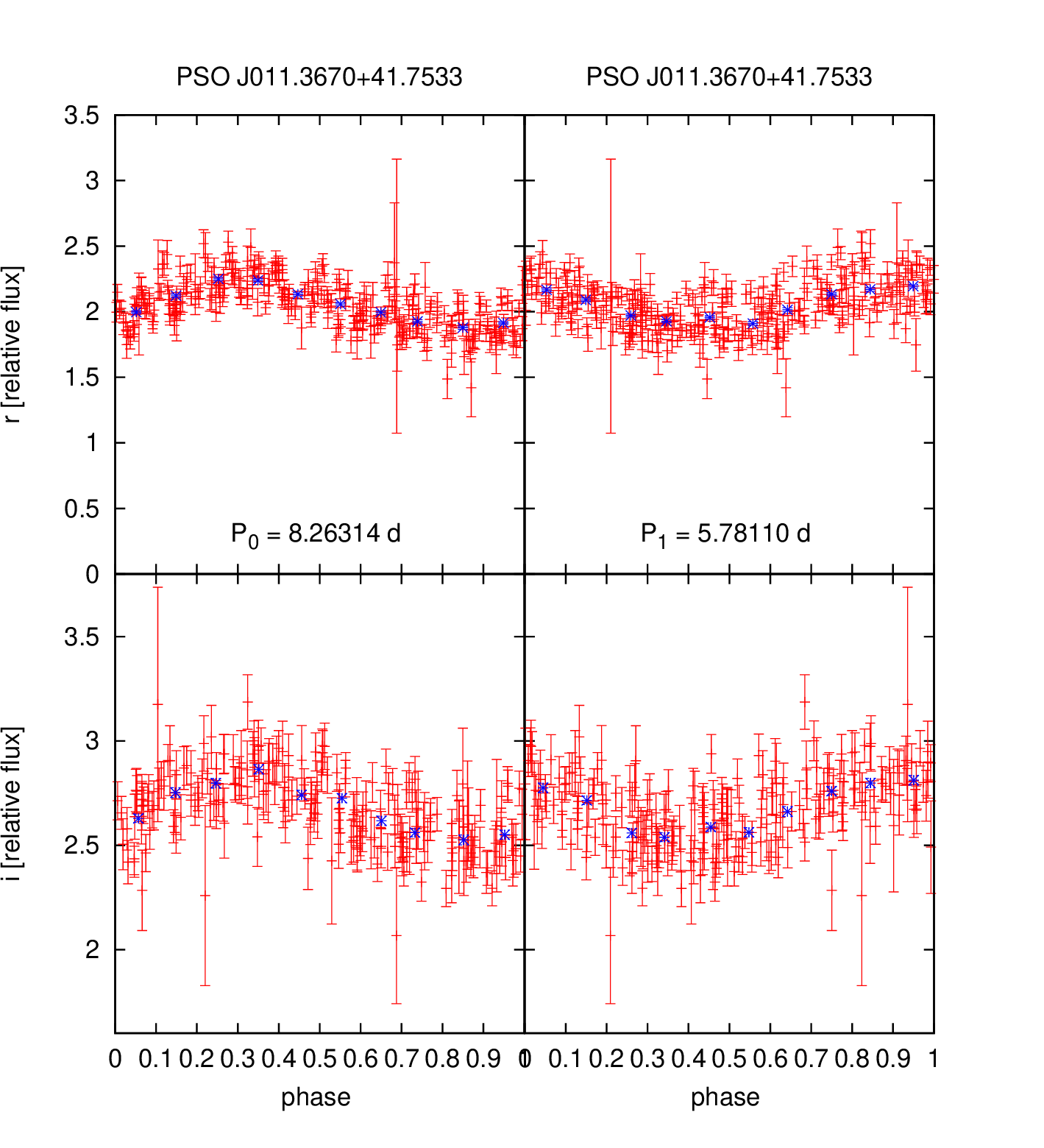}
  \includegraphics[scale=0.3]{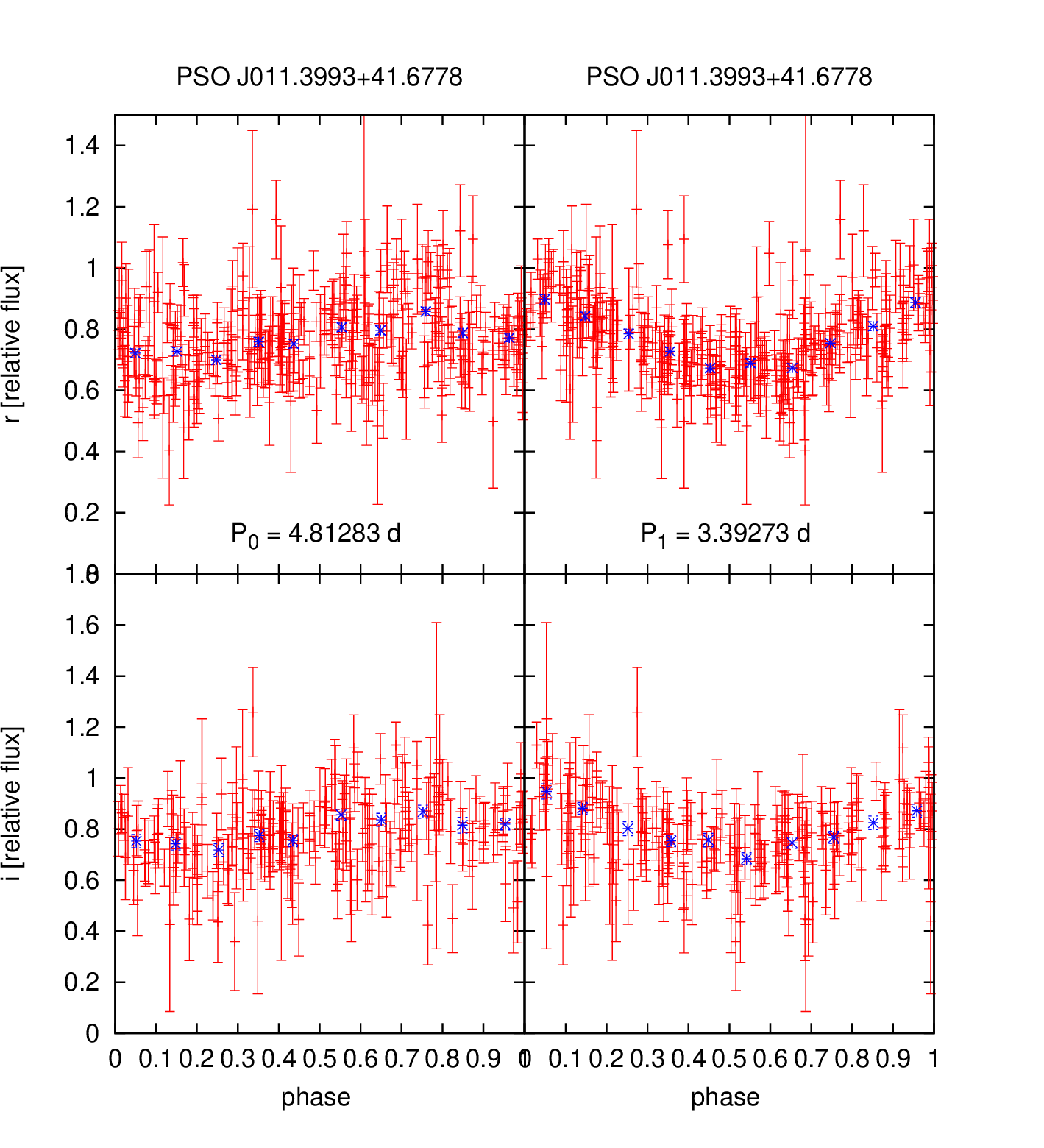}
  \includegraphics[scale=0.3]{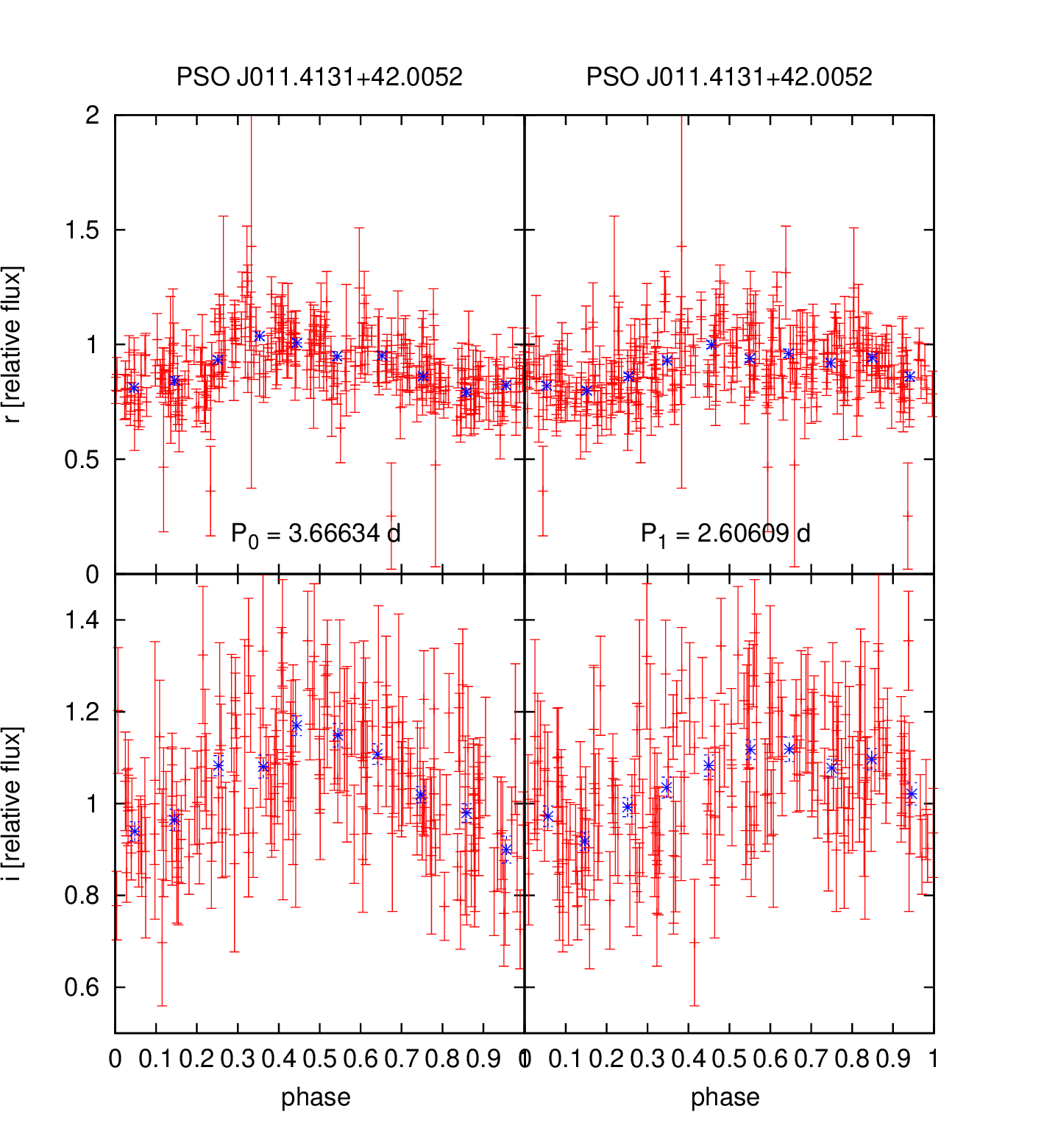}
  \includegraphics[scale=0.3]{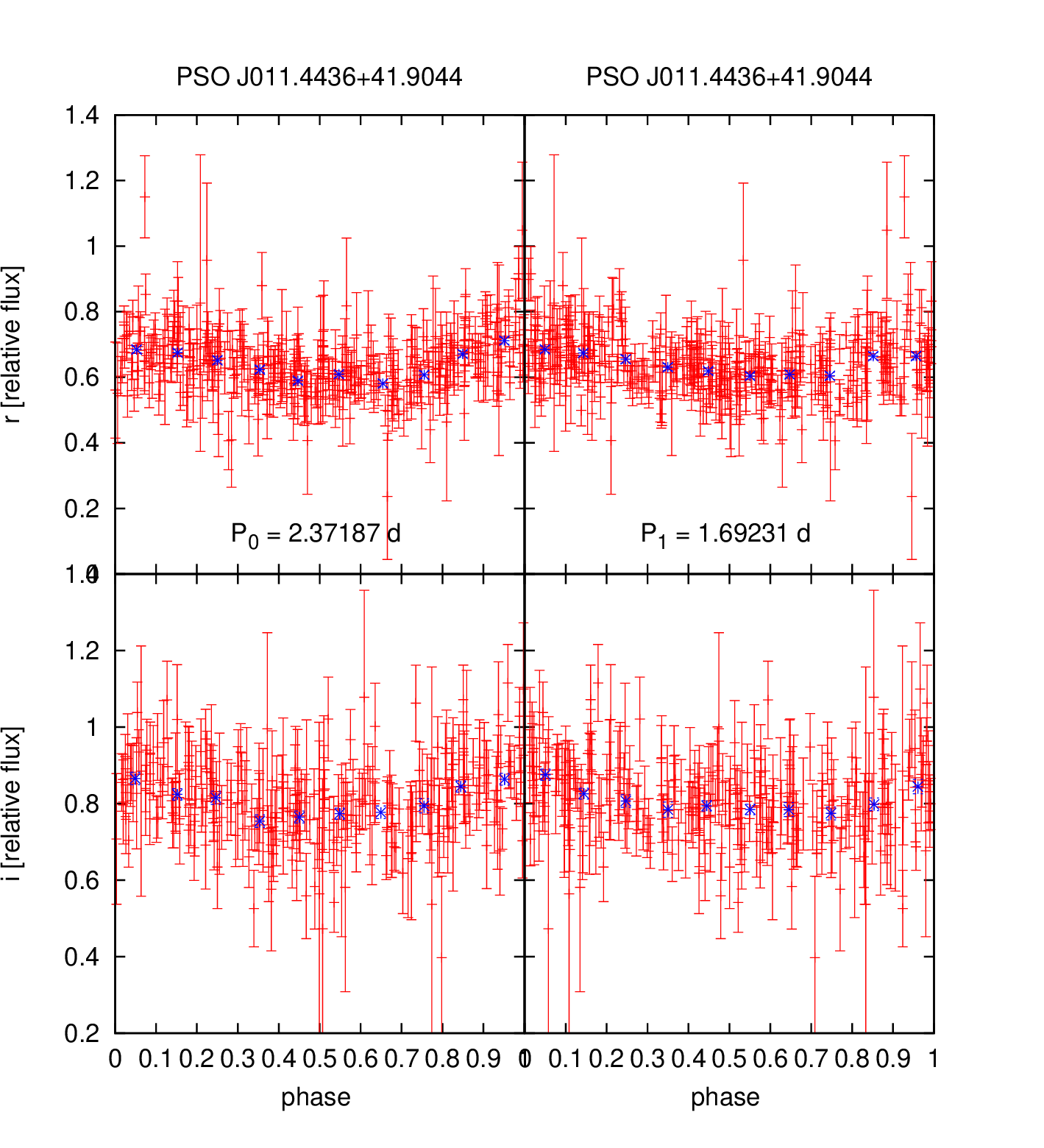}
  \caption{Light-curves of our sample - \textit{continued}.}
  \label{fig.lc}
\end{figure*}

\addtocounter{figure}{-1}
\begin{figure*}[!h]
  \centering
  \includegraphics[scale=0.3]{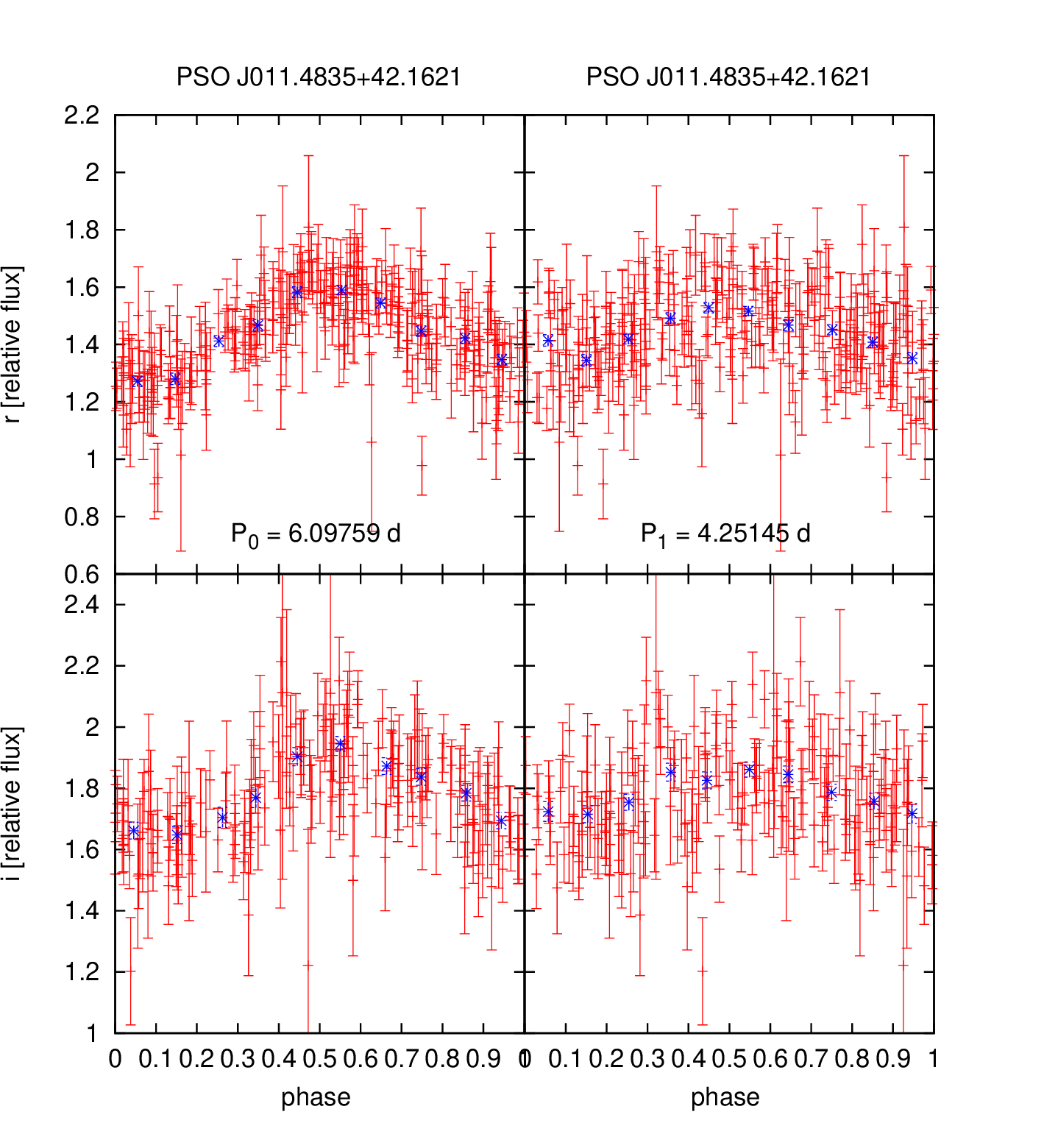}
  \caption{Light-curves of our sample - \textit{continued}.}
  \label{fig.lc}
\end{figure*}

\clearpage
\section{Metallicity Estimate}
\label{sec.est_z}
Given a uniquely measured period (P$_0$) and period ratio (P$_1$/P$_0$) of a beat Cepheid, the pulsation 
models only allow a sub-region in the parameter spaces of mass, luminosity, temperature, and metallicity for stable double mode pulsations.
This enables us to narrow down the metallicity of the beat Cepheids \citep{2006ApJ...653L.101B}. 
As has been shown by \cite{2008ApJ...680.1412B}, one can derive the upper and lower
metallicity limits simply by the location of a beat Cepheid on the log($P_0$) v.s. 
P$_{1}$/P$_{0}$ diagram (the so-called Petersen diagram, \citealt{1973A&A....27...89P}). 
In the paper of \citep{2007ApJ...660..723B,2008ApJ...680.1412B}, they have shown that the metallicity estimates from this method fall in the generally accepted ballpark for Magellanic Clouds and M33.
Fig. \ref{fig.BC} shows our M31 sample on the Petersen 
diagram, as well as beat Cepheids from Milky Way, Large and Small Magellanic Clouds, and M33 
with tracks of different metallicities taken from \cite{2008ApJ...680.1412B}.
Our sample - similar to the beat Cepheids in the Milky Way - is on the metal rich side.
On the other hand, the beat Cepheids in the Magellanic Clouds appear to be metal poor.

We interpolate the theoretical tracks by \cite{2008ApJ...680.1412B} to derive 
the limits on the metallicity for our beat Cepheids. 
In \cite{2008ApJ...680.1412B}, two different solar mixtures are compared, one from
\cite{1993oee..conf...15G}, and the other from \cite{2005ASPC..336...25A}. 
In this work we use the metallicity tracks based on the solar mixture
of \cite{1993oee..conf...15G}, which agree better with the commonly used values.
{The metallicity is derived as follows. From the theoretical tracks by \cite{2008ApJ...680.1412B}, one can delimit the 
lower (Z$_{min}$, Fig. 4, left-hand side) and upper (Z$_{max}$, Fig. 4, right-hand side) boundaries of a given position in the Petersen diagram
by interpolating between isometallictiy lines. We adopt the average of Z$_{min}$ and Z$_{max}$ as the metallicity estimate \textit{Z}. The uncertainty is taken
as $\frac{Z_{max}-Z_{min}}{2}$.
For example, PSO J011.4436+41.9044 has log P$_0$ $\sim$ 0.38 and P$_1$/P$_0$ $\sim$ 0.7135; 
On Fig. 4, its lower boundary \textit{Z$_{min}$} is bracketed by isometallicity lines 
\textit{Z} = 0.009 and 0.010, while its upper boundary \textit{Z$_{max}$} is between \textit{Z} = 
0.011 and 0.012, as also shown in the zoom-in for Petersen diagram in Fig. 5. By interpolation, we thus
derive \textit{Z$_{min}$} = 0.0098 and \textit{Z$_{max}$} = 0.01152. The metallicity estimate is thus 
\textit{Z} = $\frac{Z_{max}+Z_{min}}{2}$ = 0.01066 and the uncertainty is $\frac{Z_{max}-Z_{min}}{2}$ = 0.00086. The derived metallicity and its error for our sample are shown in Table 2. 
We also explore the impact on the metallicity estimates from errors in P$_1$/P$_0$ and present the results in the appendix. In Fig. 7, when calculating lower boundary \textit{Z}$_{min}$, we use P$_1$/P$_0$ + (error of P$_1$/P$_0$) instead of P$_1$/P$_0$; and for upper boundary \textit{Z}$_{max}$, we use P$_1$/P$_0$ - (error of P$_1$/P$_0$). The results are shown in Table 3 in the appendix, where the metallicity estimates \textit{Z} remain the same, with or without taking into account of error of P$_1$/P$_0$. Only the uncertainty of the metallicity estimates changes very slightly.

The fact that the uncertainties in the metallicity become 
large when log($P_0$) $\sim$ 0.84 only allows us to determine the value of \textit{Z} 
for fifteen out of seventeen beat Cepheids in our sample.

Once we have the period and metallicity, we can 
use the period-age relation from Table 4 of \cite{2005ApJ...621..966B}:\\
\begin{equation}
\mathrm{log}(t) = \alpha + \beta \mathrm{log}(P)
\end{equation}
to derive the age of our sample. 
Here we use P$_0 ^{\rps}$ to calculate the age. However, one should bear in mind that this period-age relation is for fundamental mode, but not specially for beat Cepheids.
We adopt ($\alpha$, $\beta$) = (8.49, -0.79) for \textit{Z} $<$ 0.007
, (8.41, -0.78) for \textit{Z} between 0.007 and 0.015, and (8.31, -0.67) for \textit{Z} between 0.015 and 0.025. The ages of our beat Cepheids are all in the order of 
$\sim$ 100 Myr, showing that they are tracing a rather young stellar population. The age estimates can be 
found in Table. 2. 

\begin{figure*}[!h]
  \centering
  \includegraphics[scale=0.45]{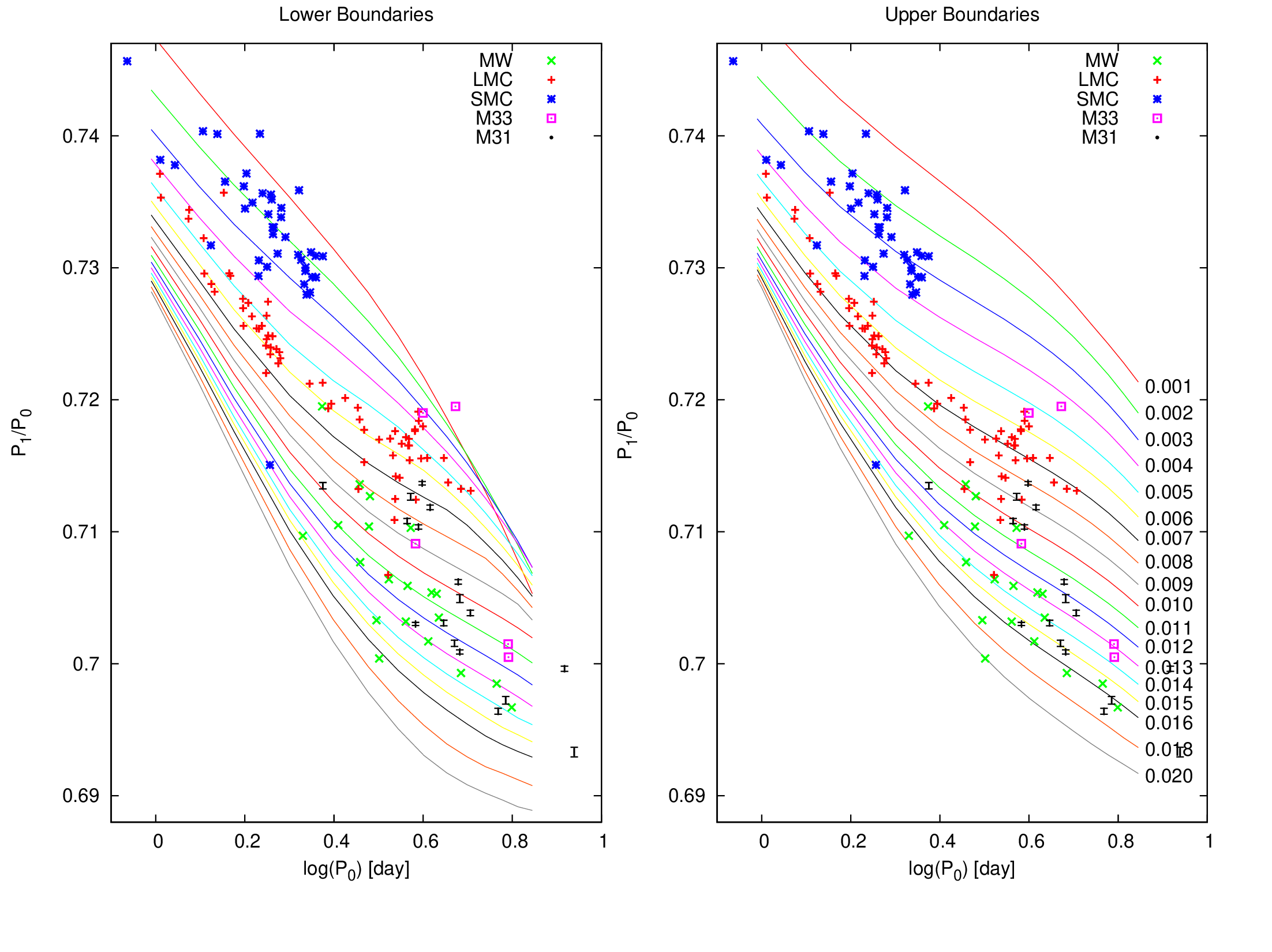}
  \caption{Petersen diagram of our sample (black symbols) and beat Cepheids of Milky Way (green crosses, from McMaster Cepheid Data Archive), Large and Small Magellanic Clouds \citep[red and blue crosses,][]{2009A&A...495..249M}, and M33 \citep[violet squares,][]{2006ApJ...653L.101B}. The period errors of our sample in terms of log(P$_0$) are too small to be seen in this figure. Track of different metallicities \citep{2008ApJ...680.1412B} are shown as solid lines in different colors, where the corresponding metallicities are given in the right panel.} 
  \label{fig.BC}
\end{figure*}

\clearpage
\begin{figure*}[!h]
  \centering
  \includegraphics[scale=0.2]{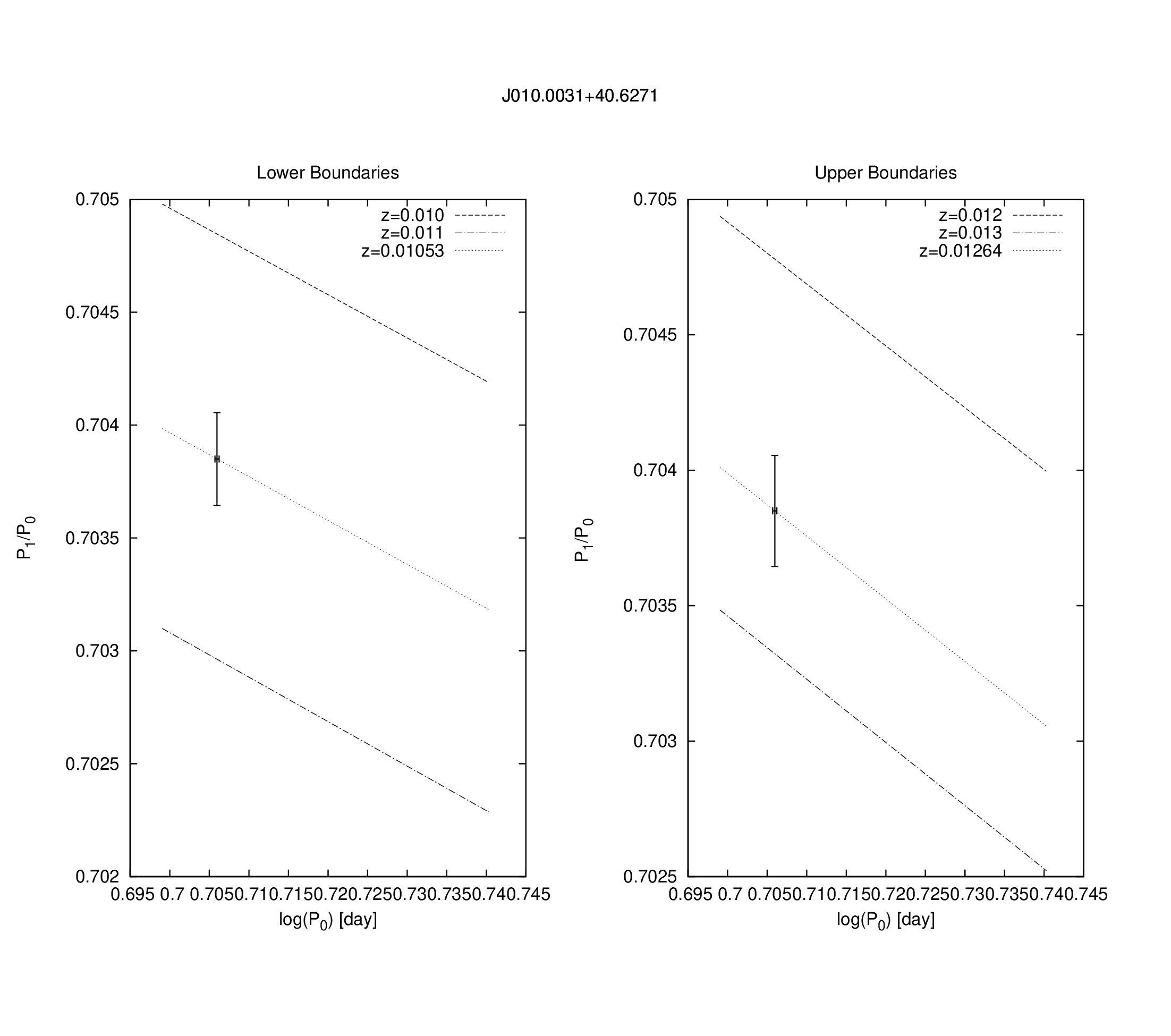}
  \includegraphics[scale=0.2]{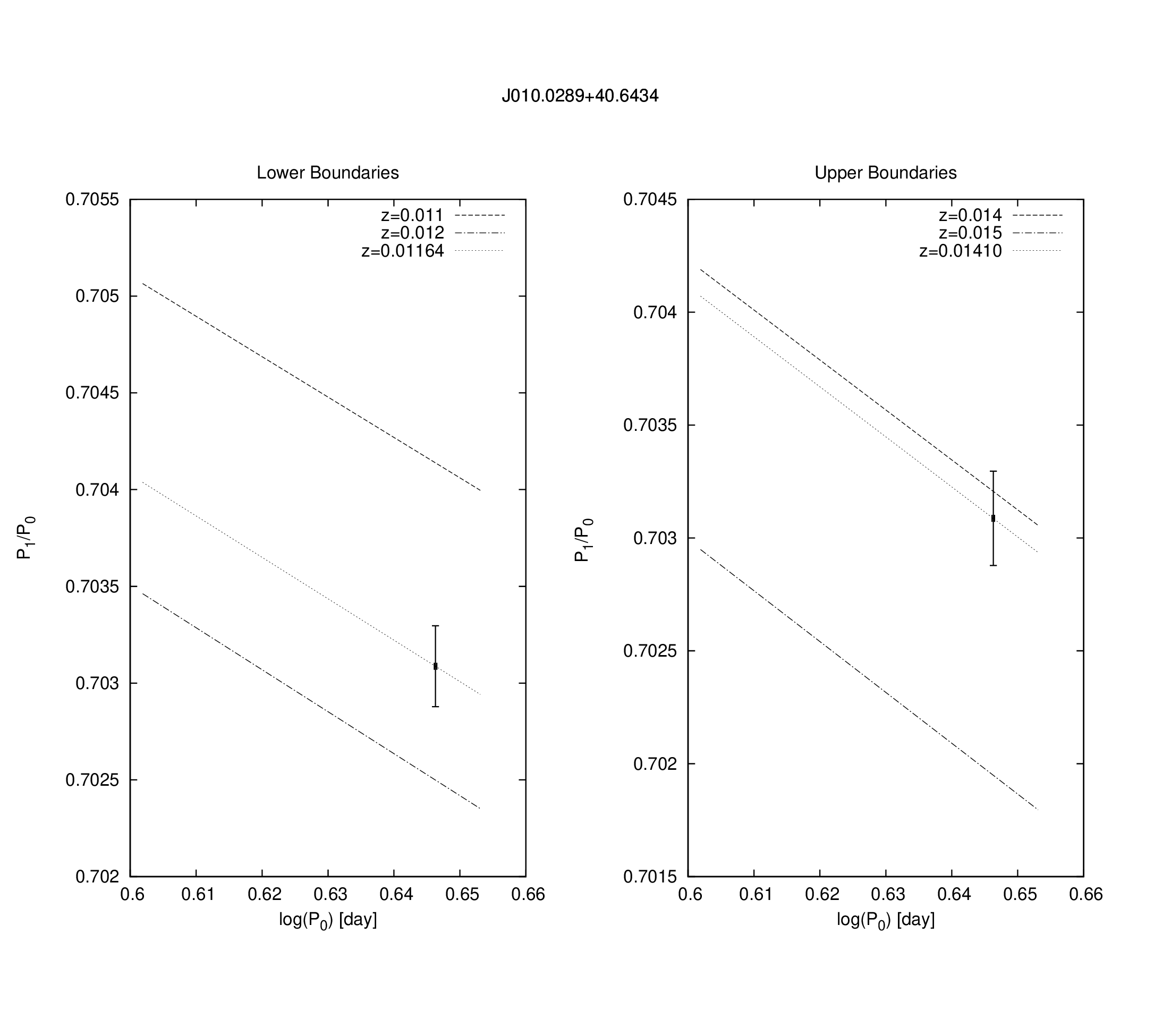}
  \includegraphics[scale=0.2]{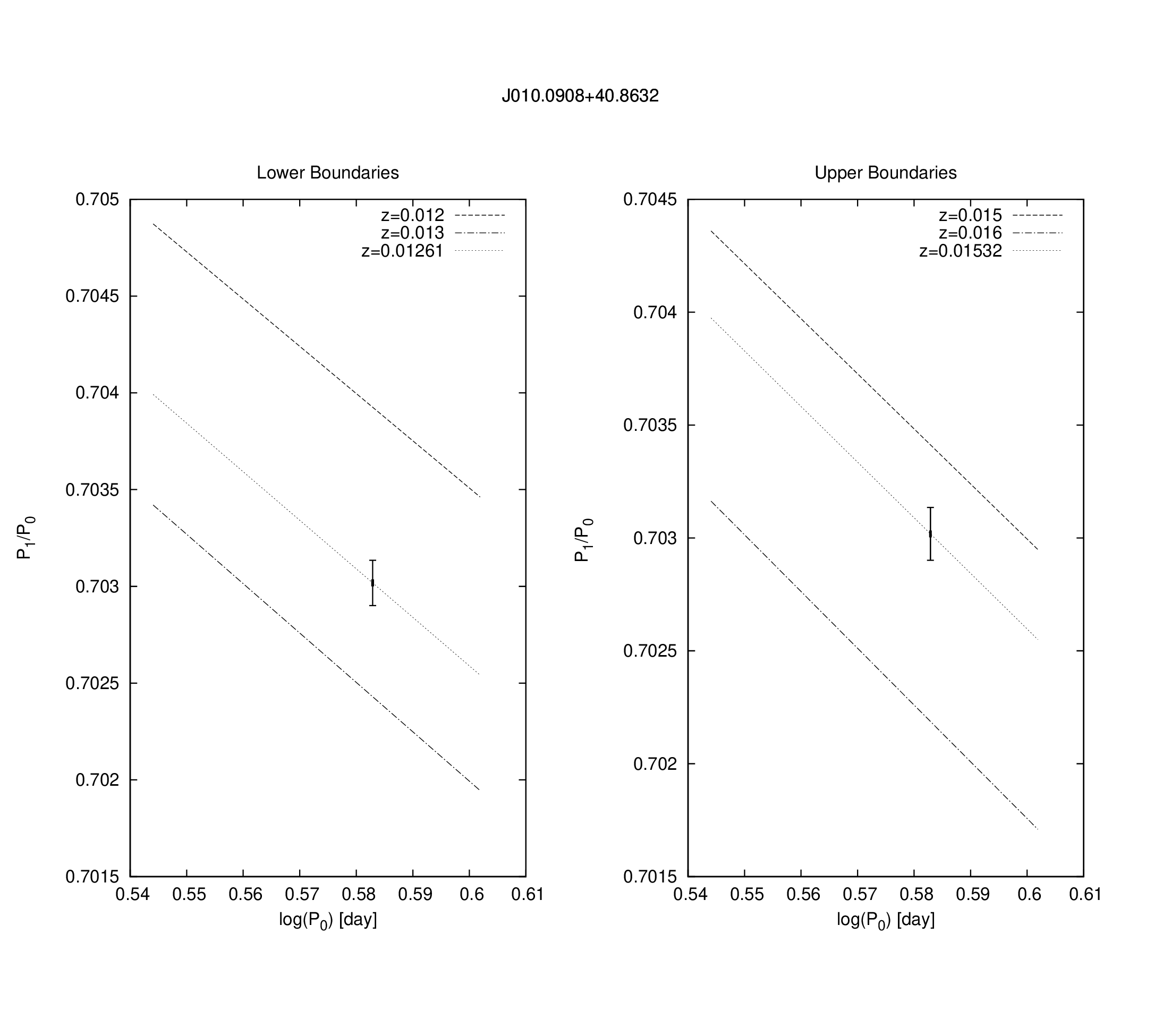}
  \includegraphics[scale=0.2]{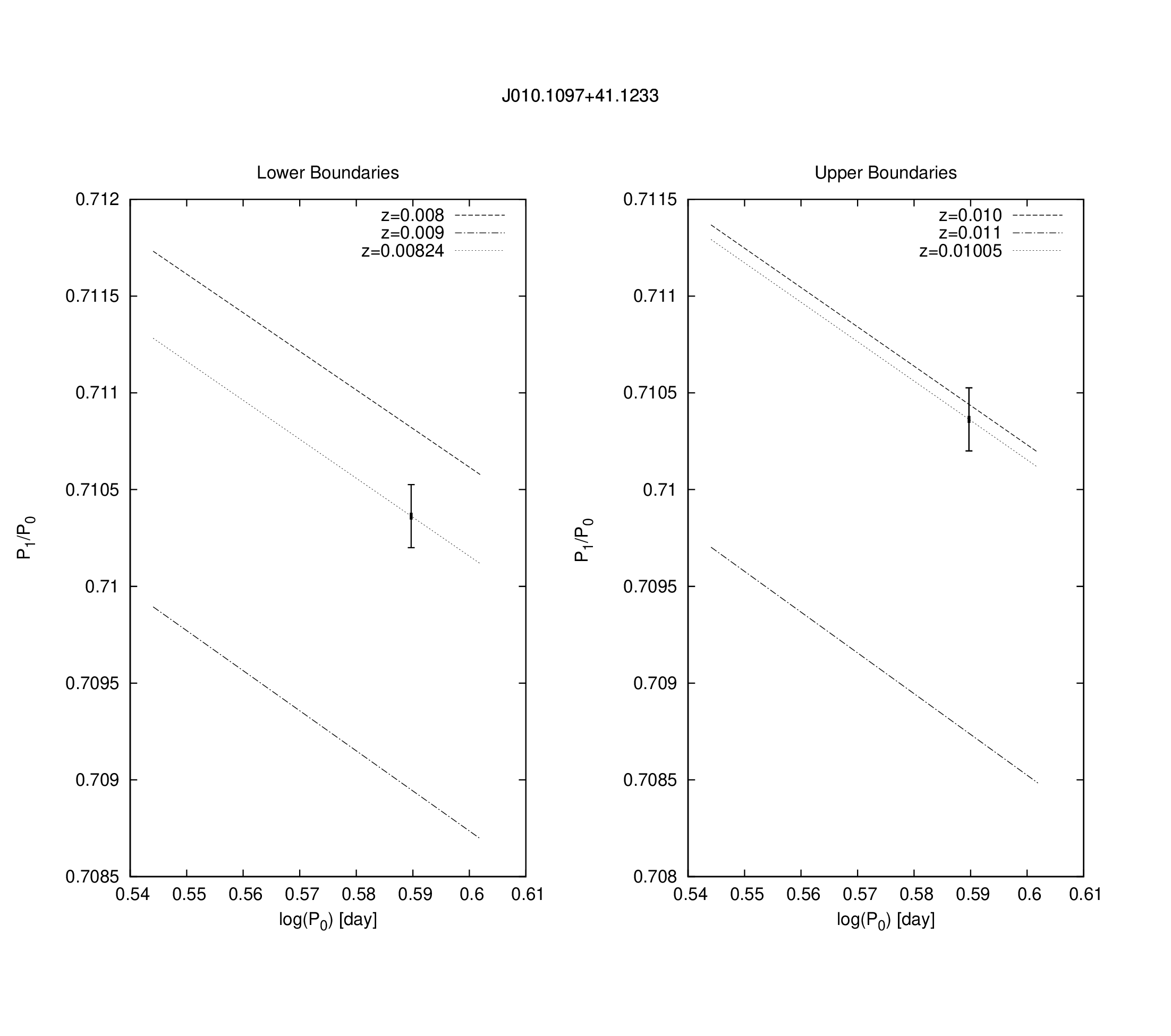}
  \caption{Zoom-in of the Petersen diagram on each candidate beat Cepheid. 
We show the adjacent theoretical isometallicity tracks and the interpolated \textit{Z} values at the  
position of the beat Cepheids. The dashed and dash-dotted curves are isometallicity tracks from the 
theoretical work of \cite{2008ApJ...680.1412B}, which are the higher and lower isometallicity tracks 
adjacent to our measured log P$_0$ and P$_1$/P$_0$ values shown in black. The dotted isometallicity line
is the interpolation that passes through our measured log P$_0$ and P$_1$/P$_0$ values. The 
estimated lower (\textit{Z$_{min}$}, left subfigures) and upper (\textit{Z$_{max}$}, right subfigures)
metallicity limits are obtained from these interpolated values. We adopt the average of Z$_{min}$ and Z$_{max}$ as the metallicity estimate \textit{Z}. The uncertainty is taken
as $\frac{Z_{max}-Z_{min}}{2}$. }
  \label{fig.lc1}
\end{figure*}

\clearpage
\addtocounter{figure}{-1}
\begin{figure*}[!h]
  \centering
  \includegraphics[scale=0.2]{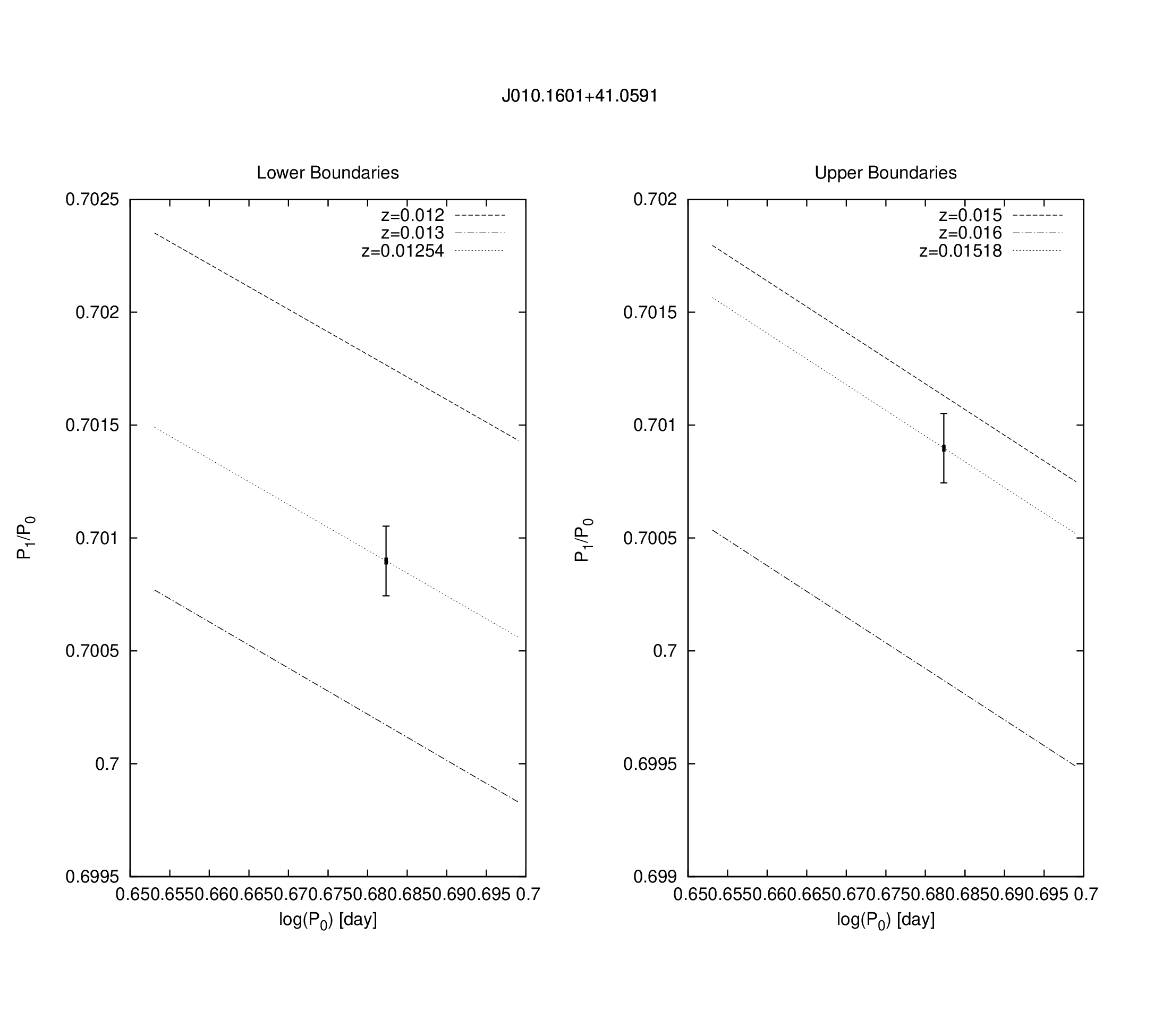}
  \includegraphics[scale=0.2]{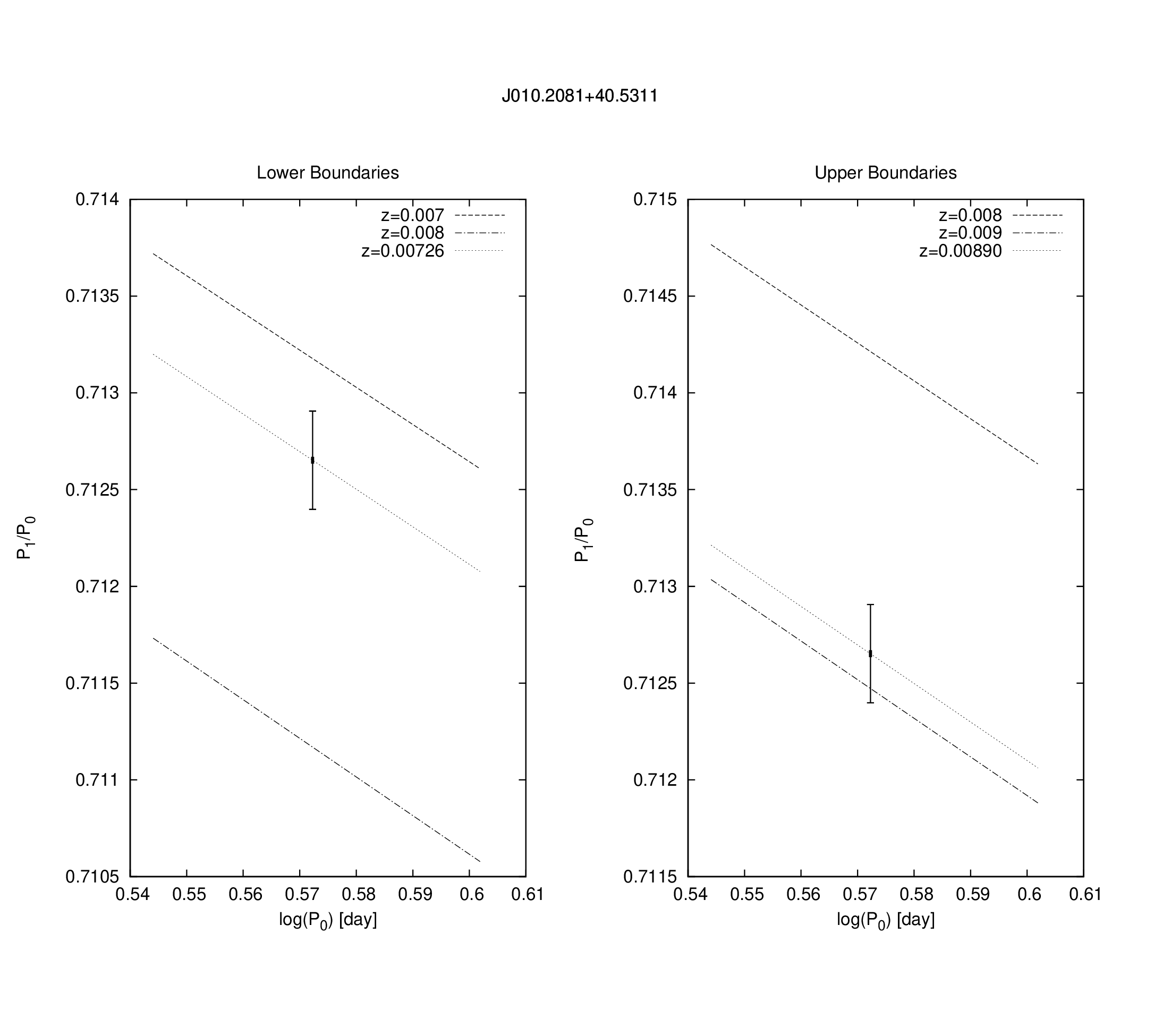}
  \includegraphics[scale=0.2]{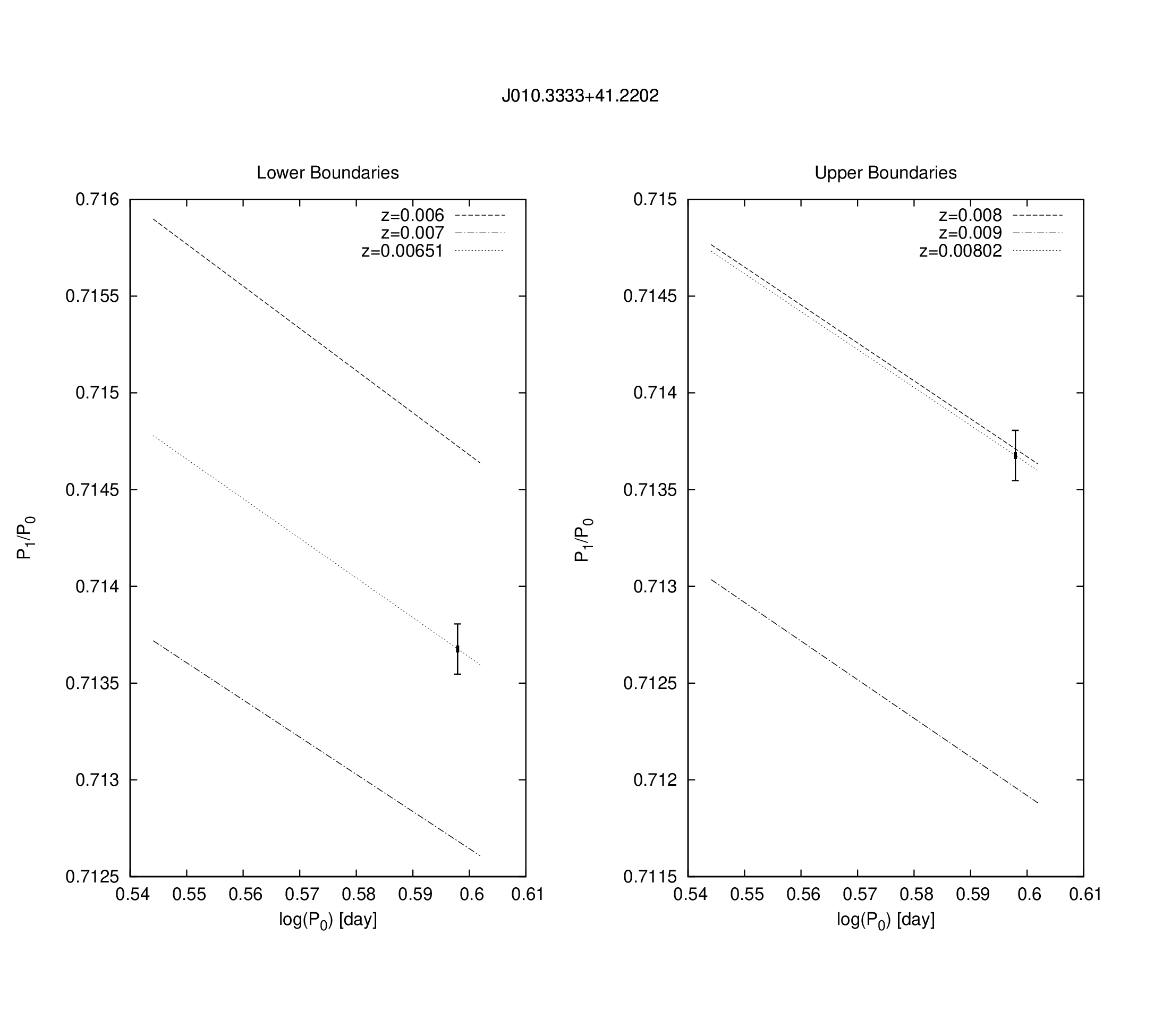}
  \includegraphics[scale=0.2]{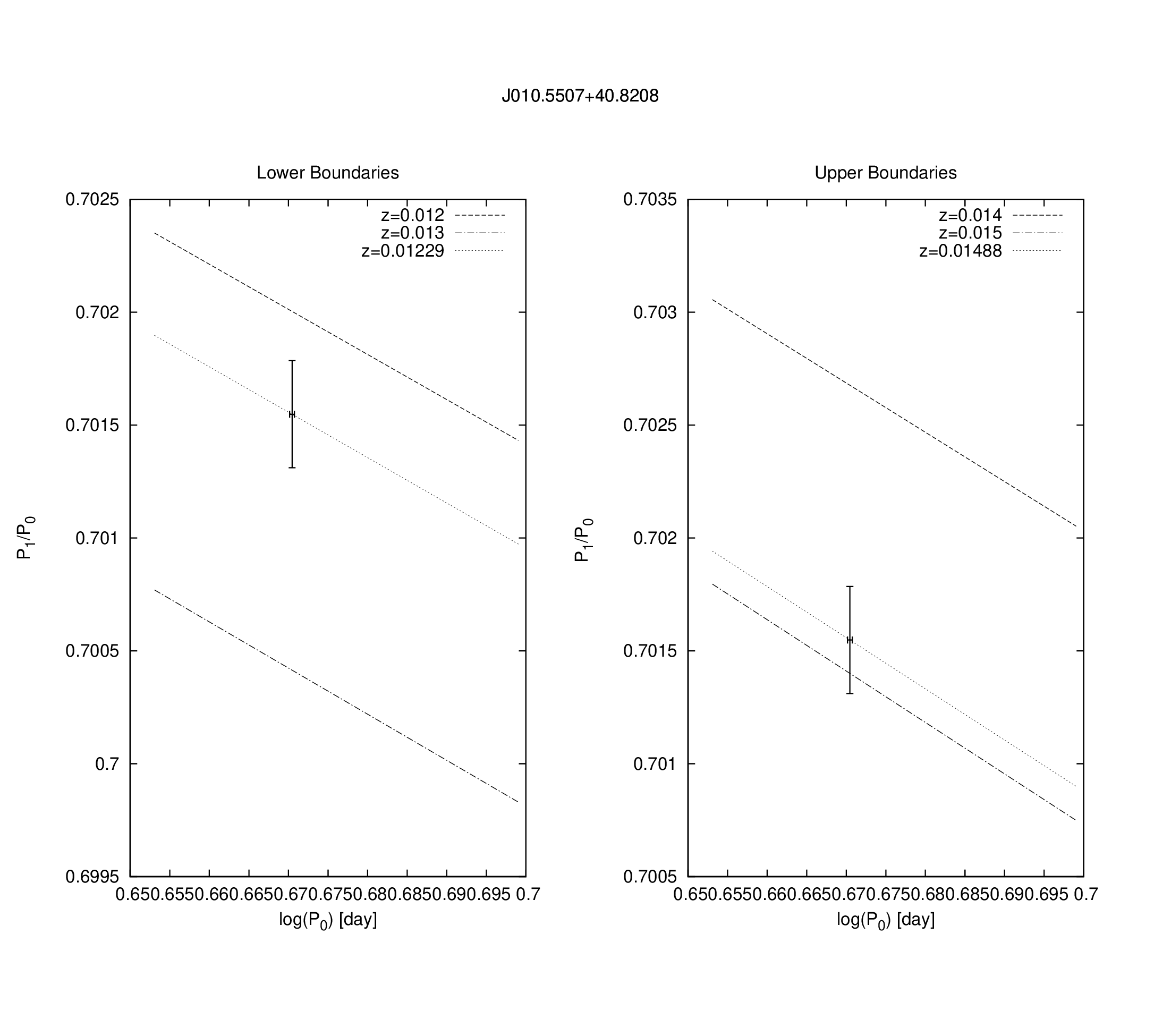} 
  \includegraphics[scale=0.2]{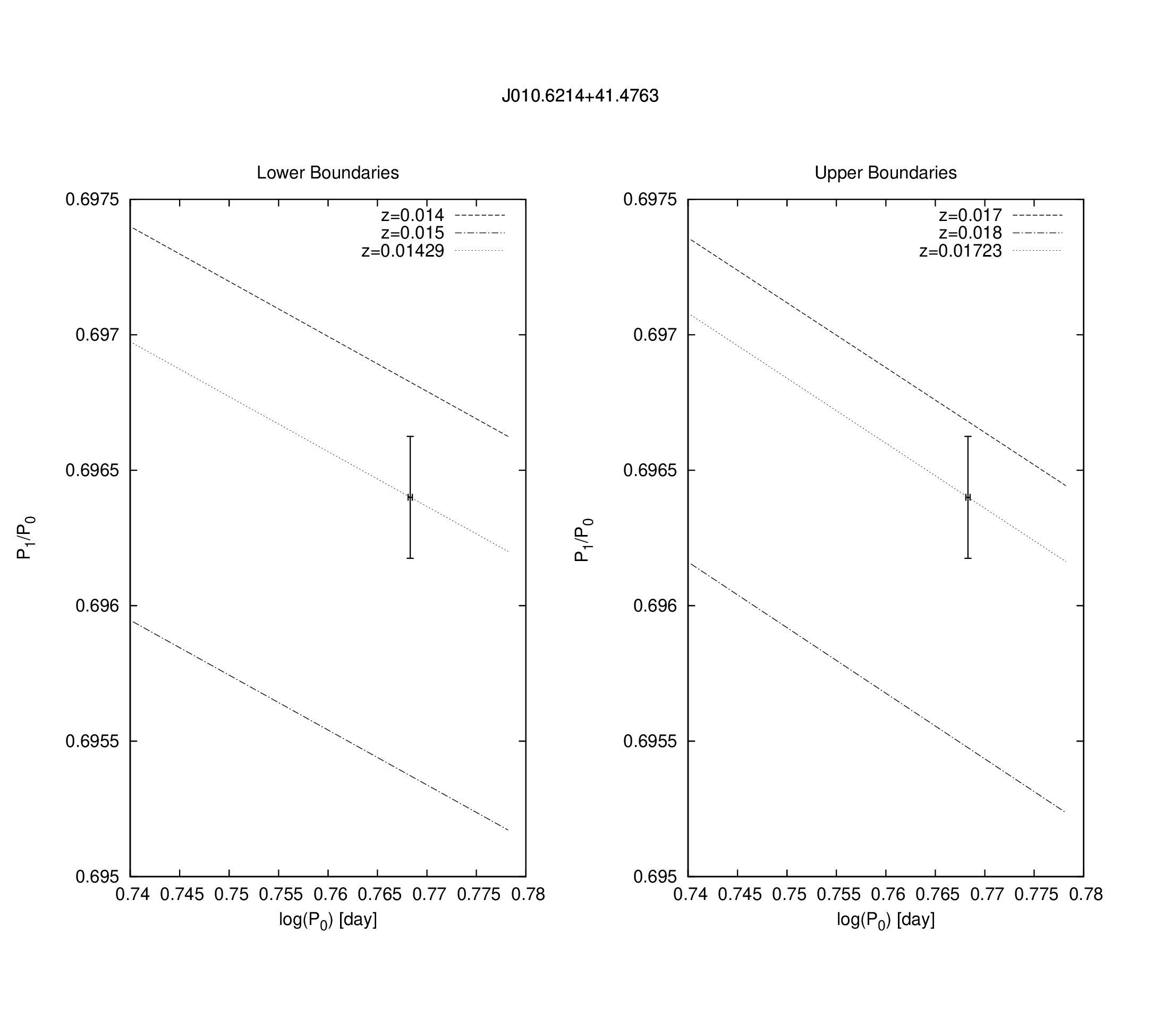}
  \includegraphics[scale=0.2]{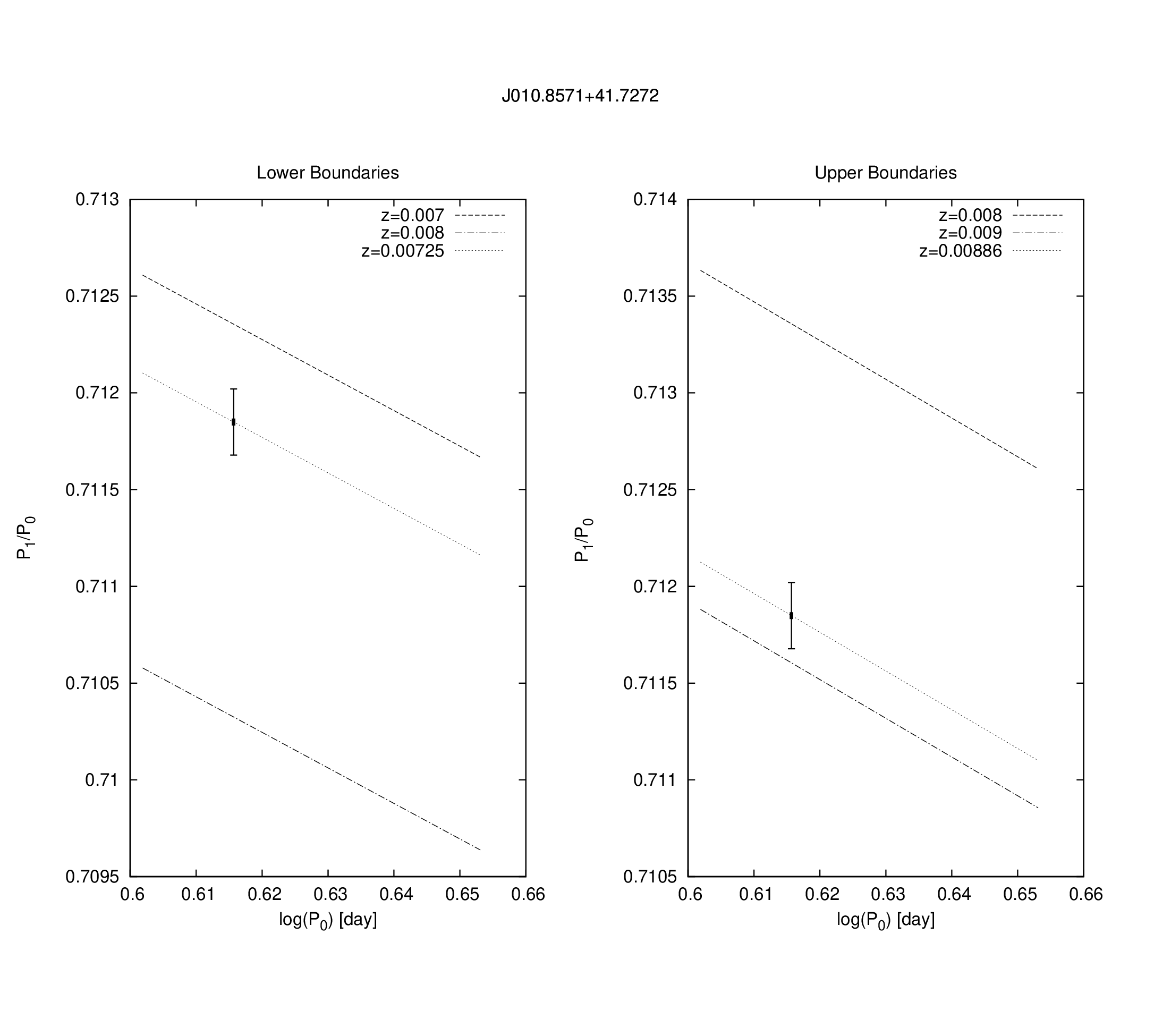}

  \caption{Zoom-in of the Petersen diagram - \textit{continued}.}
  \label{fig.lc}
\end{figure*}

\clearpage
\addtocounter{figure}{-1}
\begin{figure*}[!h]
  \centering
  \includegraphics[scale=0.2]{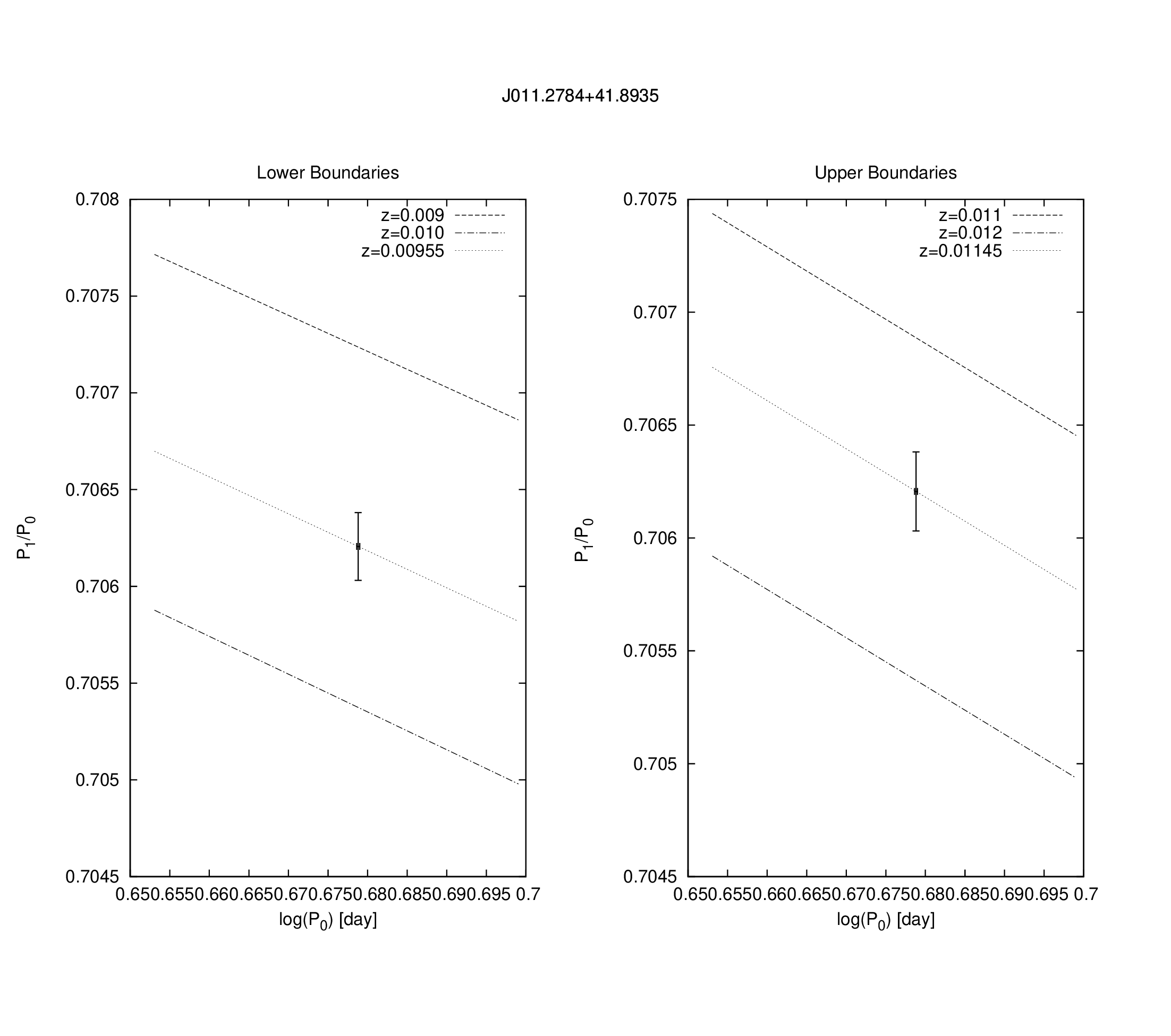}
  \includegraphics[scale=0.2]{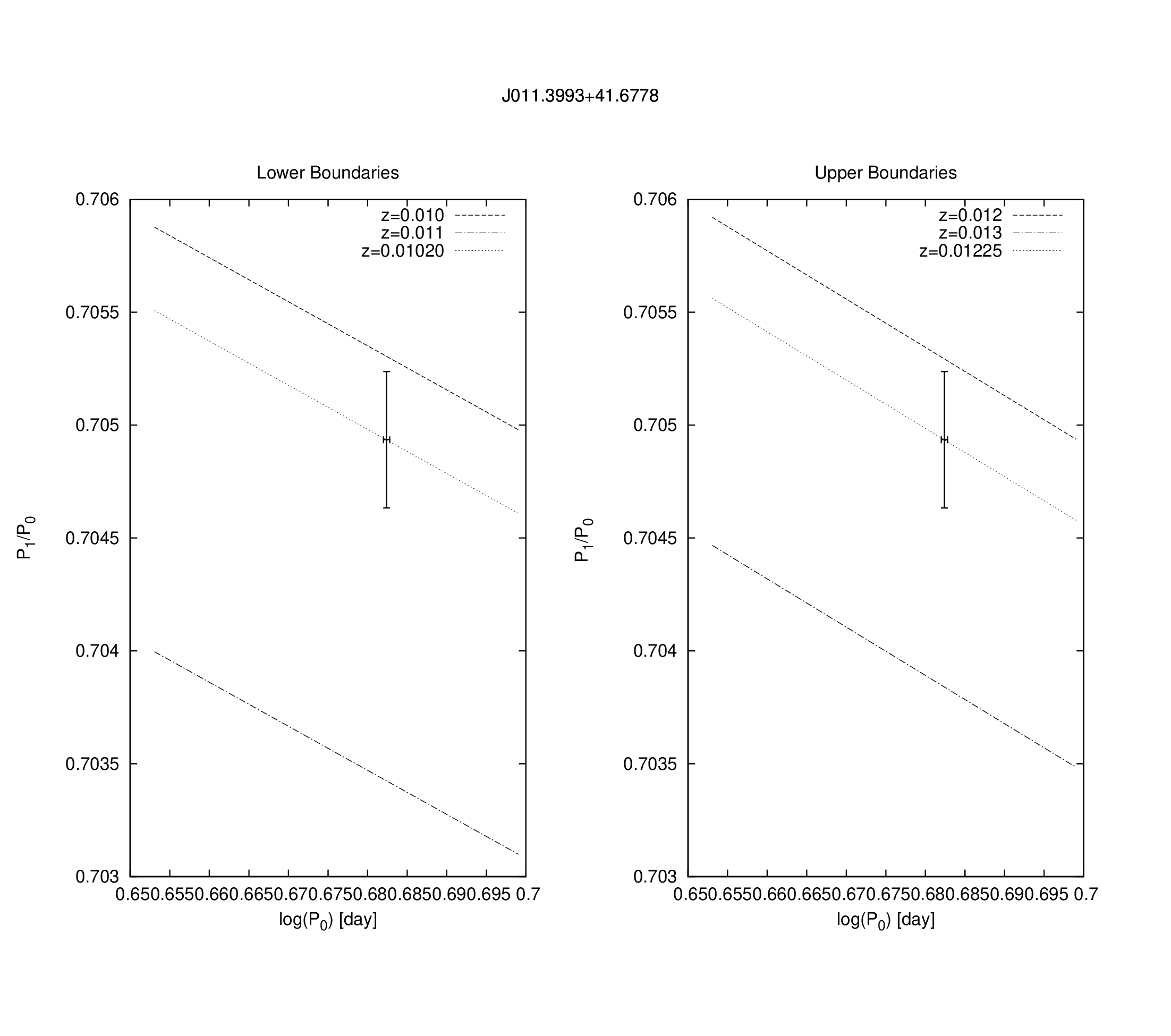}
  \includegraphics[scale=0.2]{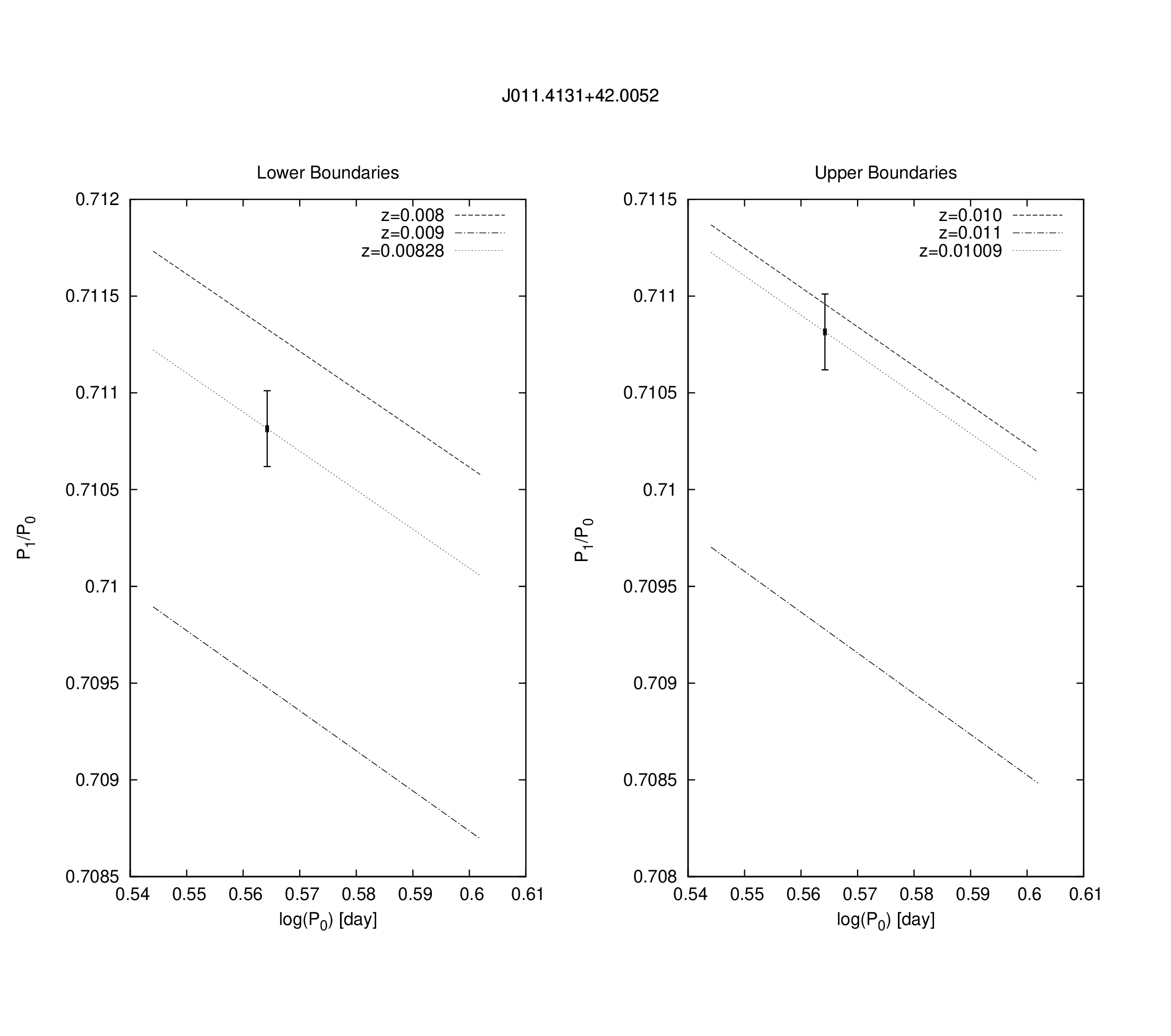}
  \includegraphics[scale=0.2]{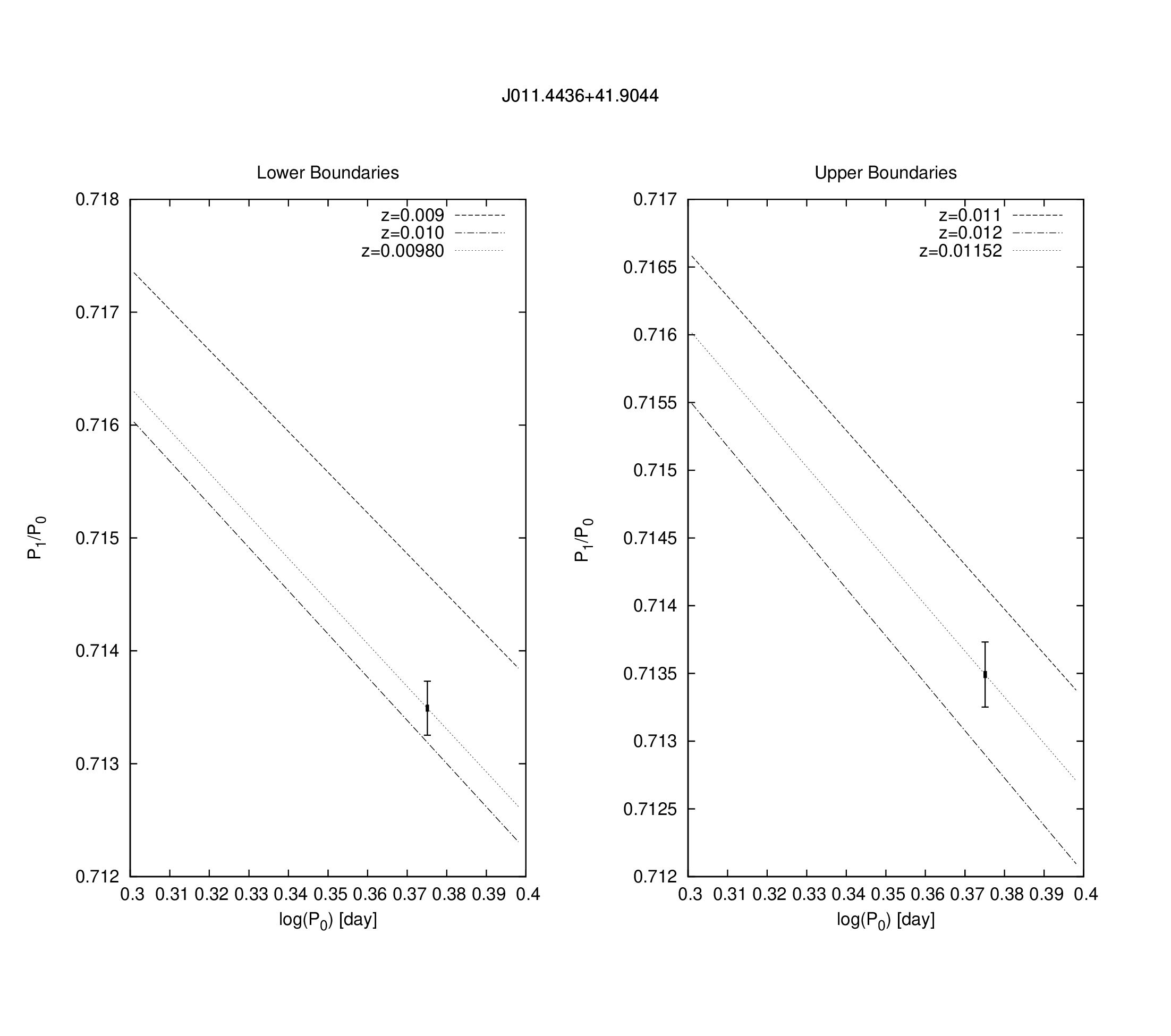}
  \includegraphics[scale=0.2]{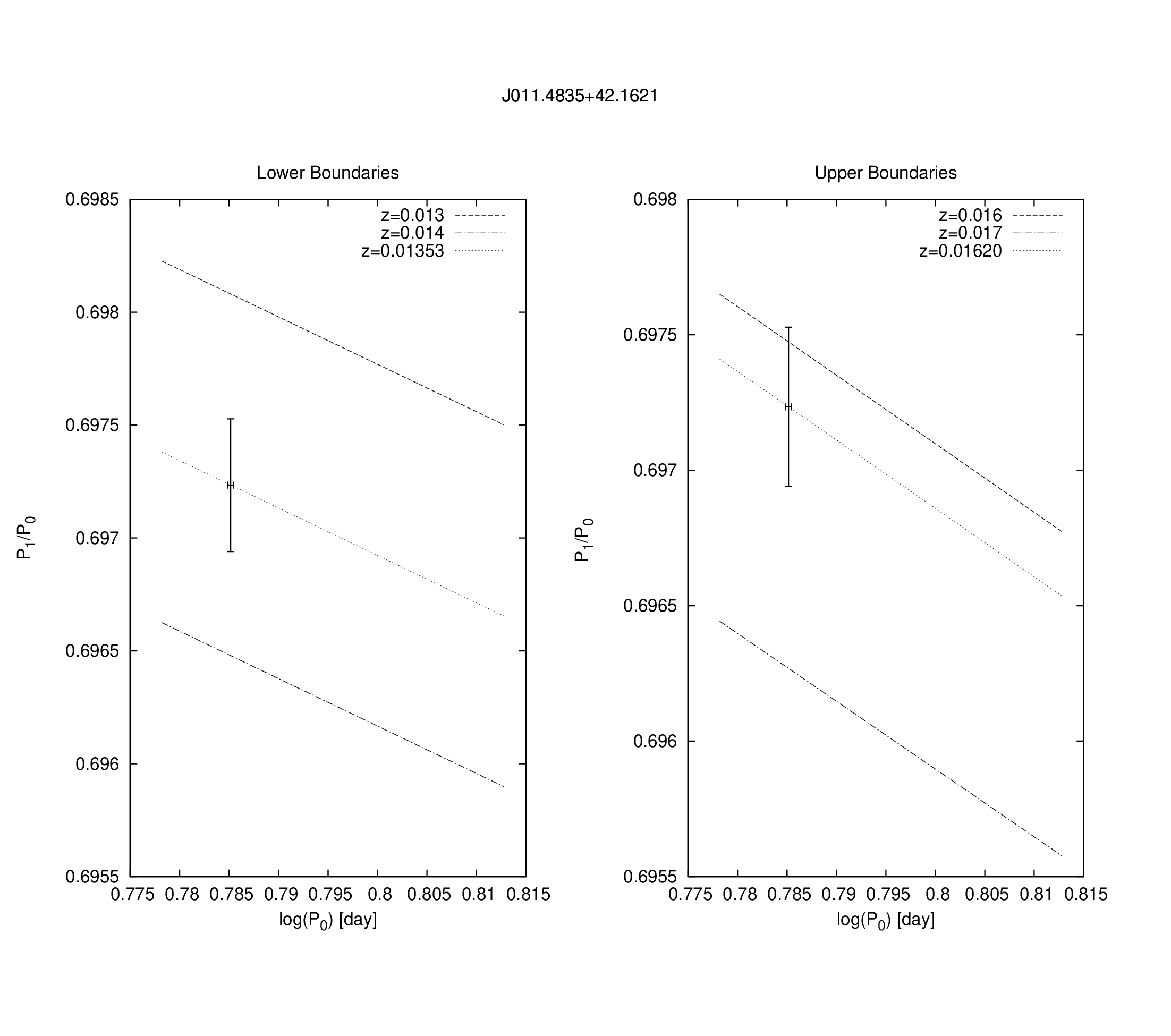} 
  \caption{Zoom-in of the Petersen diagram - \textit{continued}.}
  \label{fig.lc}
\end{figure*}

\clearpage
\addtocounter{table}{+1}
\begin{table*}[!h]
\centering
\begin{sideways}
\begin{minipage}{255mm}
\begin{tabular}[t]{llllllll}
\hline 
Name              & log P$_0$  & \textit{Z} & log($t$) & Distance & log(O/H)+12$^{\dag}$\\
                  & [day] &           & [yr]     & [kpc]    & \\
\hline
PSO J010.0031+40.6271 & 0.70597$\pm$0.00011 & 0.0116$\pm$0.0011 & 7.85935$\pm$0.00009 & 11.120 & 8.572$\pm$0.030\\ 
PSO J010.0289+40.6434 & 0.64630$\pm$0.00008 & 0.0129$\pm$0.0012 & 7.90589$\pm$0.00006 & 10.757 & 8.604$\pm$0.030\\ 
PSO J010.0908+40.8632 & 0.58290$\pm$0.00006 & 0.0140$\pm$0.0014 & 7.95534$\pm$0.00004 & 10.454 & 8.630$\pm$0.030\\ 
PSO J010.1097+41.1233 & 0.58971$\pm$0.00006 & 0.0091$\pm$0.0009 & 7.95002$\pm$0.00005 & 16.627 & 8.499$\pm$0.031\\ 
PSO J010.1601+41.0591 & 0.68234$\pm$0.00006 & 0.0139$\pm$0.0013 & 7.87778$\pm$0.00005 & 12.765 & 8.627$\pm$0.029\\ 
PSO J010.2081+40.5311 & 0.57227$\pm$0.00007 & 0.0081$\pm$0.0008 & 7.96363$\pm$0.00005 & 14.849 & 8.461$\pm$0.031\\ 
PSO J010.3333+41.2202 & 0.59792$\pm$0.00007 & 0.0073$\pm$0.0008 & 7.94362$\pm$0.00006 & 11.497 & 8.429$\pm$0.028\\ 
PSO J010.5507+40.8208 & 0.67046$\pm$0.00013 & 0.0136$\pm$0.0013 & 7.88704$\pm$0.00010 & 13.257 & 8.621$\pm$0.030\\ 
PSO J010.6214+41.4763 & 0.76830$\pm$0.00009 & 0.0158$\pm$0.0015 & 7.79524$\pm$0.00006 & 10.390 & 8.666$\pm$0.025\\ 
PSO J010.8571+41.7272 & 0.61569$\pm$0.00007 & 0.0081$\pm$0.0008 & 7.92976$\pm$0.00006 & 12.597 & 8.460$\pm$0.028\\ 
PSO J011.2784+41.8935 & 0.67882$\pm$0.00008 & 0.0105$\pm$0.0009 & 7.88052$\pm$0.00006 & 10.495 & 8.542$\pm$0.028\\ 
PSO J011.3993+41.6778 & 0.68240$\pm$0.00017 & 0.0112$\pm$0.0010 & 7.87773$\pm$0.00014 & 13.731 & 8.561$\pm$0.029\\ 
PSO J011.4131+42.0052 & 0.56423$\pm$0.00010 & 0.0092$\pm$0.0009 & 7.96990$\pm$0.00007 & 12.386 & 8.501$\pm$0.032\\ 
PSO J011.4436+41.9044 & 0.37509$\pm$0.00012 & 0.0107$\pm$0.0009 & 8.11743$\pm$0.00009 & 11.941 & 8.547$\pm$0.029\\ 
PSO J011.4835+42.1621 & 0.78516$\pm$0.00013 & 0.0149$\pm$0.0013 & 7.79758$\pm$0.00010 & 15.221 & 8.648$\pm$0.028\\ 
\hline
\end{tabular}
\caption{Beat Cepheid properties; \dag~See section 4 for a detailed explanation.}
\end{minipage}
\end{sideways}
\label{tab.z_grad}
\end{table*}

\clearpage
\section{Metallicity gradient}
\label{sec.z_grad}
To derive the metallicity gradient, we first de-project the coordinates of the beat Cepheids to 
galactocentric distances using the transformation of \cite{1981Ap&SS..76..477H}. We assume that the center of M31 is 
located at RA=00h42'44''.52 (J2000) and Dec=+41d16'08''.69 (J2000), with a position angle of 37d42'54''.
We also assume an inclination angle of 12.5 degrees \citep{1978A&A....67...73S} and a distance of 770 kpc \citep{1990ApJ...365..186F}. 

To compare with previous results from HII region studies \citep{2012MNRAS.427.1463Z, 2012ApJ...758..133S},
which are shown in log(O/H), we first convert our \textit{Z} values to [O/H] by using a 
\textit{Z$_{\odot}$} value of 0.017 and [O/H] = [Fe/H]/1.417 from \cite{2003A&A...397..667M}. We then use 
log(O/H)$_{\odot}$ + 12 = 8.69 \citep{2009ARA&A..47..481A} to calculate the values of log(O/H) + 12
for our sample, and compare them with previous results from the HII regions and planetary nebulae 
observations of M31 shown in Fig. \ref{fig.z_grad}. 

There are several ways to extract the chemical abundance from spectroscopic observations of HII regions. 
For example, \cite{2012MNRAS.427.1463Z} have determined the electron temperature of the gas from HII regions 
and derive the chemical abundance accordingly (so-called direct-$T_e$ method). On the other hand, one 
can use the flux ratio between strong lines to infer the chemical abundance of certain elements. For example, 
\cite{2012ApJ...758..133S} have used the flux ratio between [N II] and H$_{\alpha}$ proposed by \cite{2006A&A...459...85N} 
to obtain the log(O/H) values from HII regions. 
In Fig. \ref{fig.spat} and Fig. \ref{fig.z_grad} we only show the eight HII region samples from \cite{2012MNRAS.427.1463Z},
because they are the only ones who derive $T_e$ and [O/H] values from the faint [O III] line directly. 
In \cite{2012ApJ...758..133S}, they have hundreds of HII region measurements, but their [O/H] value varies depending on which
strong-lines are used.
Our beat Cepheid result is closer to the metallicities 
from the direct method of \cite{2012MNRAS.427.1463Z} than the strong-line mentioned. 
Our errors are much smaller than those for traditional metallicity measurement methods. 
As a consequence, the difference of our metallicity values to that of \cite{2012MNRAS.427.1463Z} is significant.

In addition to the HII regions, chemical abundance can be derived from 
the planetary nebulae as well. We also compare our result to the metallicities 
from \cite{2012ApJ...753...12K} in Fig. \ref{fig.z_grad}. Contrary to the metallicities from planetary nebulae,
our result shows sub-solar log(O/H) value within 15 kpc, similar to the result from HII regions. The mean log(O/H)+12 value
from our sample is 8.56, while observations from planetary nebulae give a higher value (8.64). Our sample has a gradient of 
-0.008$\pm$0.004 dex/kpc, close to the value of -0.011$\pm$0.004 dex/kpc from 
planetary nebulae \citep{2012ApJ...753...12K}. Our result shows scatter around the linear gradient, which could originate from the
intrinsic variation of \textit{in situ} metallicity.

\begin{figure*}[!h]
  \centering
  \includegraphics[scale=1.2]{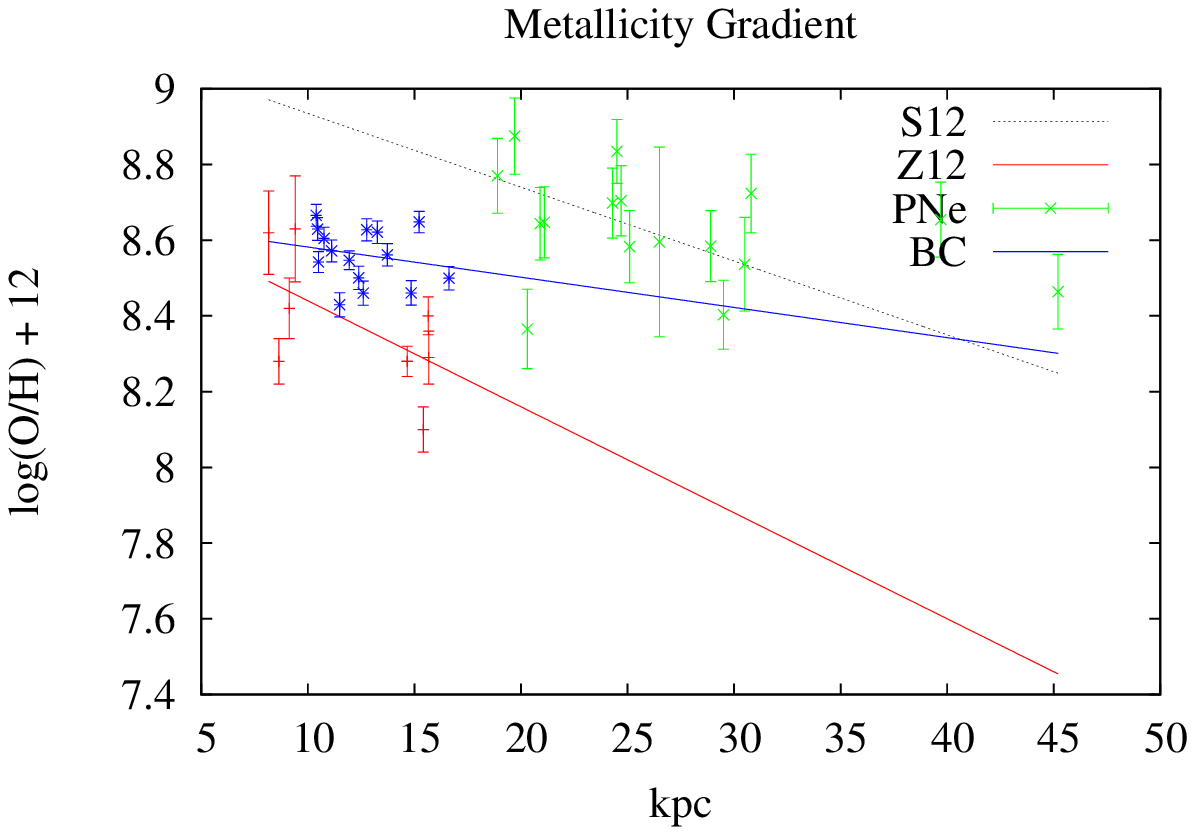}
  \caption{Metallicity as a function of de-projected distance to the center of M31. The blue crosses are 
derived from our sample (see section \ref{sec.z_grad} for a detailed description). The black line is from \cite{2012ApJ...758..133S}, 
where they use the N2 diagnosis (strong line method) to obtain the log(O/H) values from HII regions. The red 
line and red data points (see our Fig. \ref{fig.spat} for their positions)
are from \cite{2012MNRAS.427.1463Z}. They determine the electron temperature of the gas from HII regions and 
derive the chemical abundance accordingly (direct method). 
The green points mark chemical abundances derived from planetary nebulae by \cite{2012ApJ...753...12K}. Since planetary nebulae and beat Cepheids are
tracing different stellar population, we expect different metallicity estimates from these two methods. Our result is closer to the results obtained with the direct method from HII region than to results obtained with the strong line method (grey dashed line). Also, our metallicities are significantly differ from the metallicities derived for planetary nebulae in the outer M31 disk.}
  \label{fig.z_grad}
\end{figure*}

The detailed properties of our sample, including the metallicity, galactocentric distance, and age are shown in Table 2.

\section{Conclusion and Outlook}
\label{sec.outlook}

We present a sample of the beat Cepheids based on the PAndromeda data. We use the 
P$_1$/P$_0$ to P$_0$ relations from pulsation models of \cite{2008ApJ...680.1412B} to estimate the Cepheid metallicities. 
We de-project the location of beat Cepheids, and derive the metallicity gradient of M31. 
Our result is closer to the results from the planetary nebulae of \cite{2012ApJ...753...12K}.

In this work we only concentrate on searching beat Cepheids from a sample of resolved sources. 
In a future work we will also conduct searches for variables from pixel-based 
light-curves. In this case we could find fainter variables.

Because the beat Cepheids are pulsating at relative short periods, they are intrinsically 
very faint, and with a 2-m class telescope like PS1 it is difficult to find a large sample 
at the distance of M31. To increase the number of beat Cepheids in M31, it requires 
deeper surveys. Our understanding of beat Cepheid content in M31 
can be improved with the CFHT POMME survey \citep{2012Ap&SS.341...57F} 
and the up-coming \textit{LSST} project \citep{2008arXiv0805.2366I}. 

\acknowledgments
We acknowledge comments from the anonymous referee. 
We are grateful to Andrzej Udalski for his useful comments. This
work was supported by the DFG cluster of excellence ``Origin
and Structure of the Universe'' (www.universe-cluster.de).

The Pan-STARRS1 Surveys (PS1) have been made possible through contributions 
of the Institute for Astronomy, the University of Hawaii, the 
Pan-STARRS Project Office, the Max-Planck Society and its 
participating institutes, the Max Planck Institute for Astronomy, 
Heidelberg and the Max Planck Institute for Extraterrestrial 
Physics, Garching, The Johns Hopkins University, Durham University, 
the University of Edinburgh, Queen's University Belfast, the 
Harvard-Smithsonian Center for Astrophysics, the Las Cumbres 
Observatory Global Telescope Network Incorporated, the National 
Central University of Taiwan, the Space Telescope Science Institute, 
the National Aeronautics and Space Administration under Grant No. 
NNX08AR22G issued through the Planetary Science Division of the 
NASA Science Mission Directorate,  the National Science Foundation 
under Grant No. AST-1238877, and the University of Maryland.

\section{Appendix}

In this section we explore the impact on the metallicity estimates from errors in P$_1$/P$_0$ and present the results. 
In Fig. 7, when calculating lower boundary \textit{Z}$_{min}$, we use P$_1$/P$_0$ + (error of P$_1$/P$_0$) instead of P$_1$/P$_0$; and for upper 
boundary \textit{Z}$_{max}$, we use P$_1$/P$_0$ - (error of P$_1$/P$_0$). The results are shown in Table 3, 
where the metallicity estimates \textit{Z} remain the same, with or  without taking into account of error of P$_1$/P$_0$. 
Only the uncertainty of the metallicity 
estimates changes very slightly.

\begin{table*}[!h]
\center
\begin{tabular}[t]{lll}
\hline 
Name              & \textit{Z} & \textit{Z}\\
                  & without $\frac{P_1}{P_0}$-err &  with $\frac{P_1}{P_0}$-err\\
\hline
J010.0031+40.6271 & 0.0116$\pm$0.0011 & 0.0116$\pm$0.0012 \\ 
J010.0289+40.6434 & 0.0129$\pm$0.0012 & 0.0129$\pm$0.0014 \\ 
J010.0908+40.8632 & 0.0140$\pm$0.0014 & 0.0140$\pm$0.0014 \\ 
J010.1097+41.1233 & 0.0091$\pm$0.0009 & 0.0091$\pm$0.0010 \\ 
J010.1601+41.0591 & 0.0139$\pm$0.0013 & 0.0139$\pm$0.0014 \\ 
J010.2081+40.5311 & 0.0081$\pm$0.0008 & 0.0081$\pm$0.0010 \\ 
J010.3333+41.2202 & 0.0073$\pm$0.0008 & 0.0073$\pm$0.0008 \\ 
J010.5507+40.8208 & 0.0136$\pm$0.0013 & 0.0136$\pm$0.0015 \\ 
J010.6214+41.4763 & 0.0158$\pm$0.0015 & 0.0158$\pm$0.0016 \\ 
J010.8571+41.7272 & 0.0081$\pm$0.0008 & 0.0081$\pm$0.0009 \\ 
J011.2784+41.8935 & 0.0105$\pm$0.0009 & 0.0105$\pm$0.0011 \\ 
J011.3993+41.6778 & 0.0112$\pm$0.0010 & 0.0112$\pm$0.0012 \\ 
J011.4131+42.0052 & 0.0092$\pm$0.0009 & 0.0092$\pm$0.0010 \\ 
J011.4436+41.9044 & 0.0107$\pm$0.0009 & 0.0107$\pm$0.0010 \\ 
J011.4835+42.1621 & 0.0149$\pm$0.0013 & 0.0149$\pm$0.0015 \\ 
\hline
\end{tabular}
\caption{\textit{Z} of beat Cepheid properties; derived with and without taking into account errors in P$_1$/P$_0$.}
\label{tab.z_err}
\end{table*}

\clearpage
\begin{figure*}[!h]
  \centering
  \includegraphics[scale=0.2]{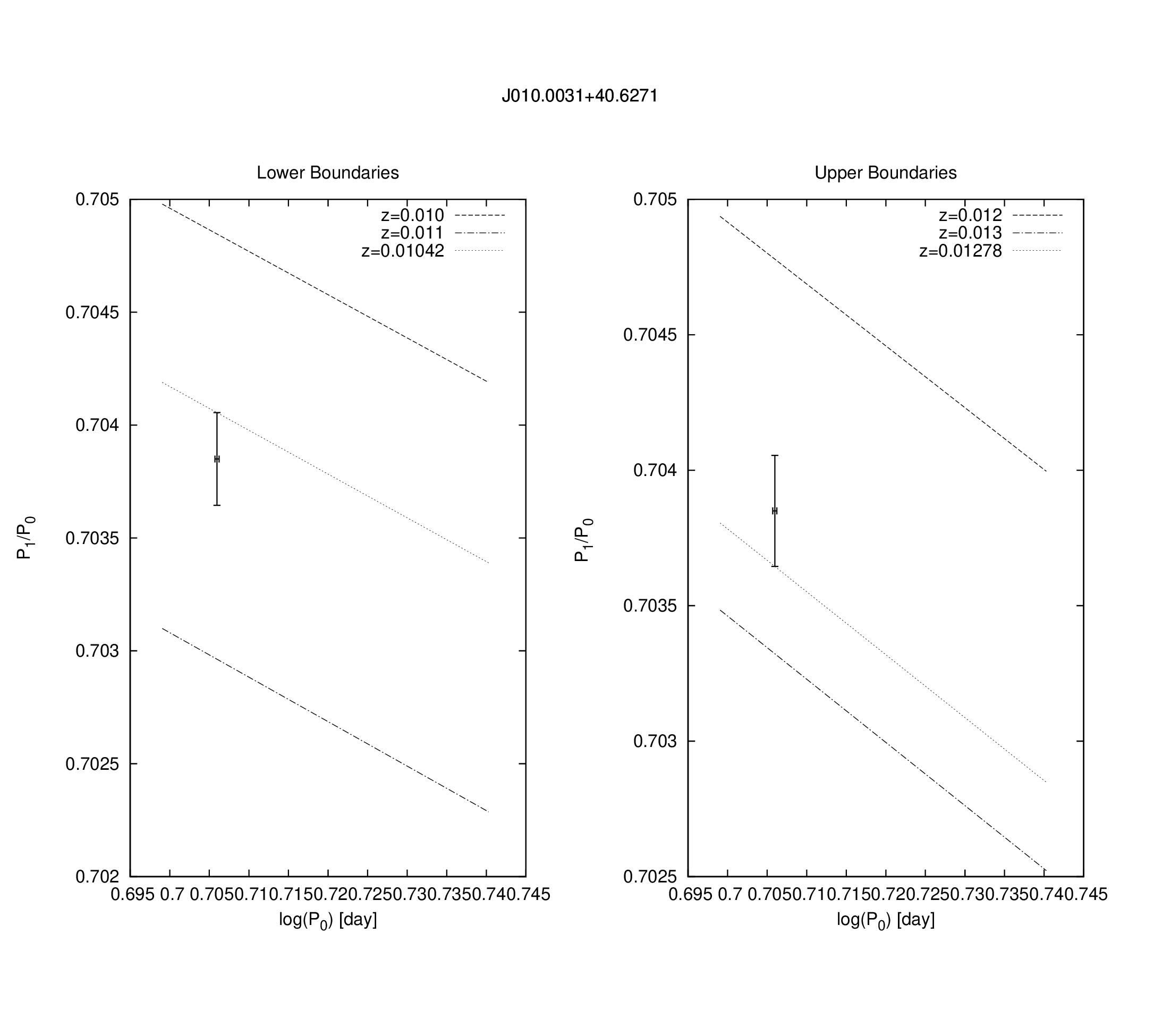}
  \includegraphics[scale=0.2]{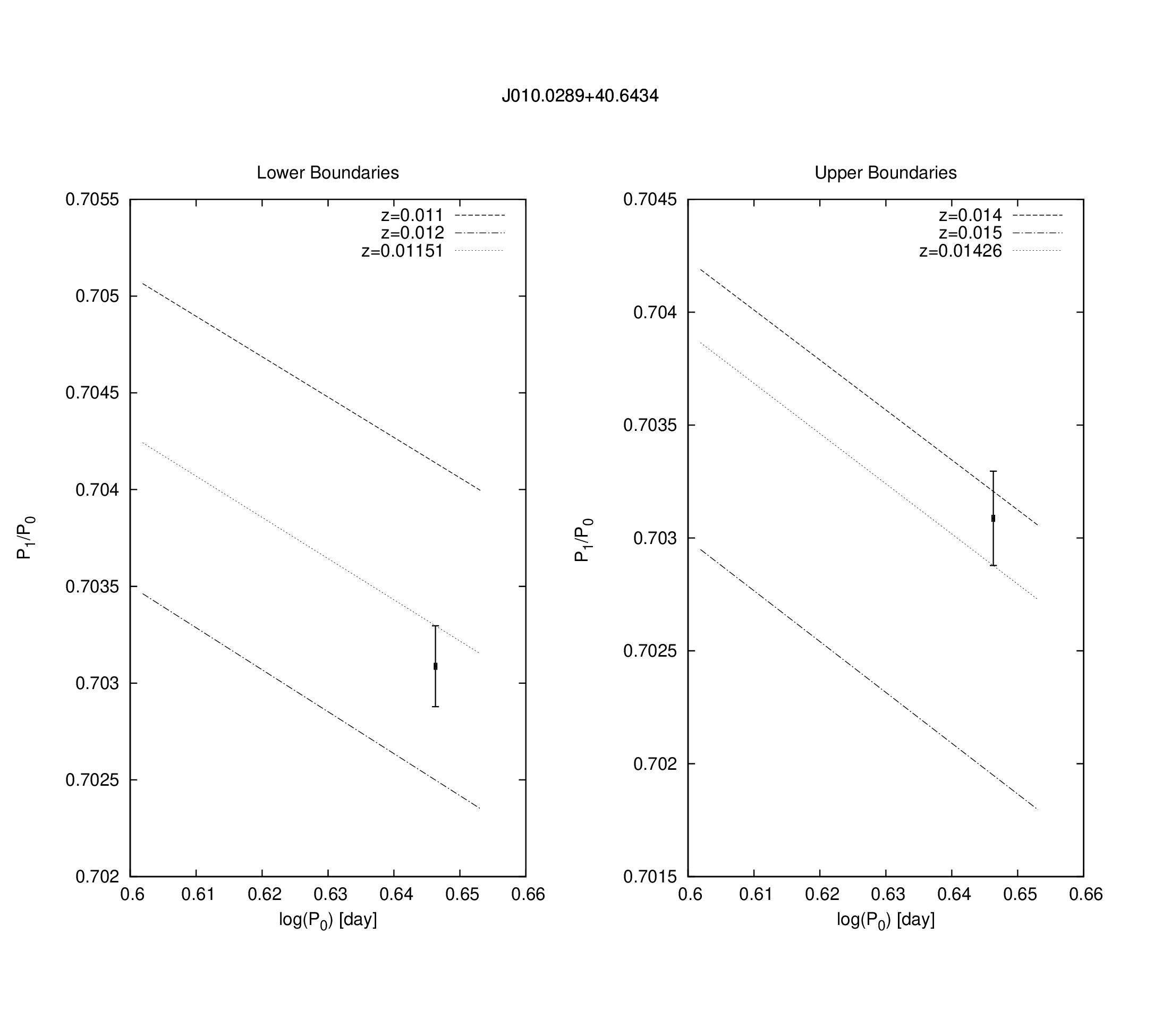}
  \includegraphics[scale=0.2]{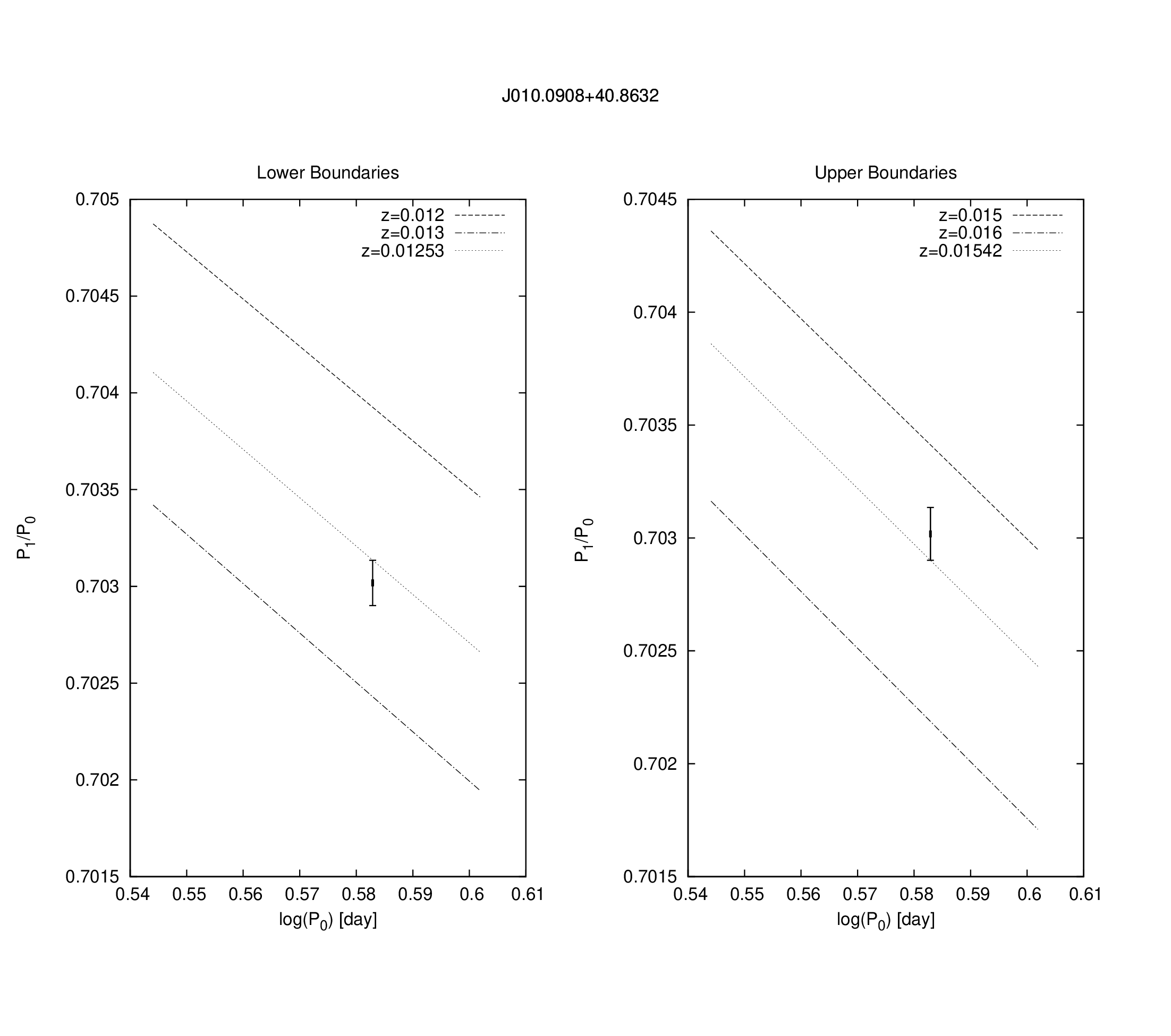}
  \includegraphics[scale=0.2]{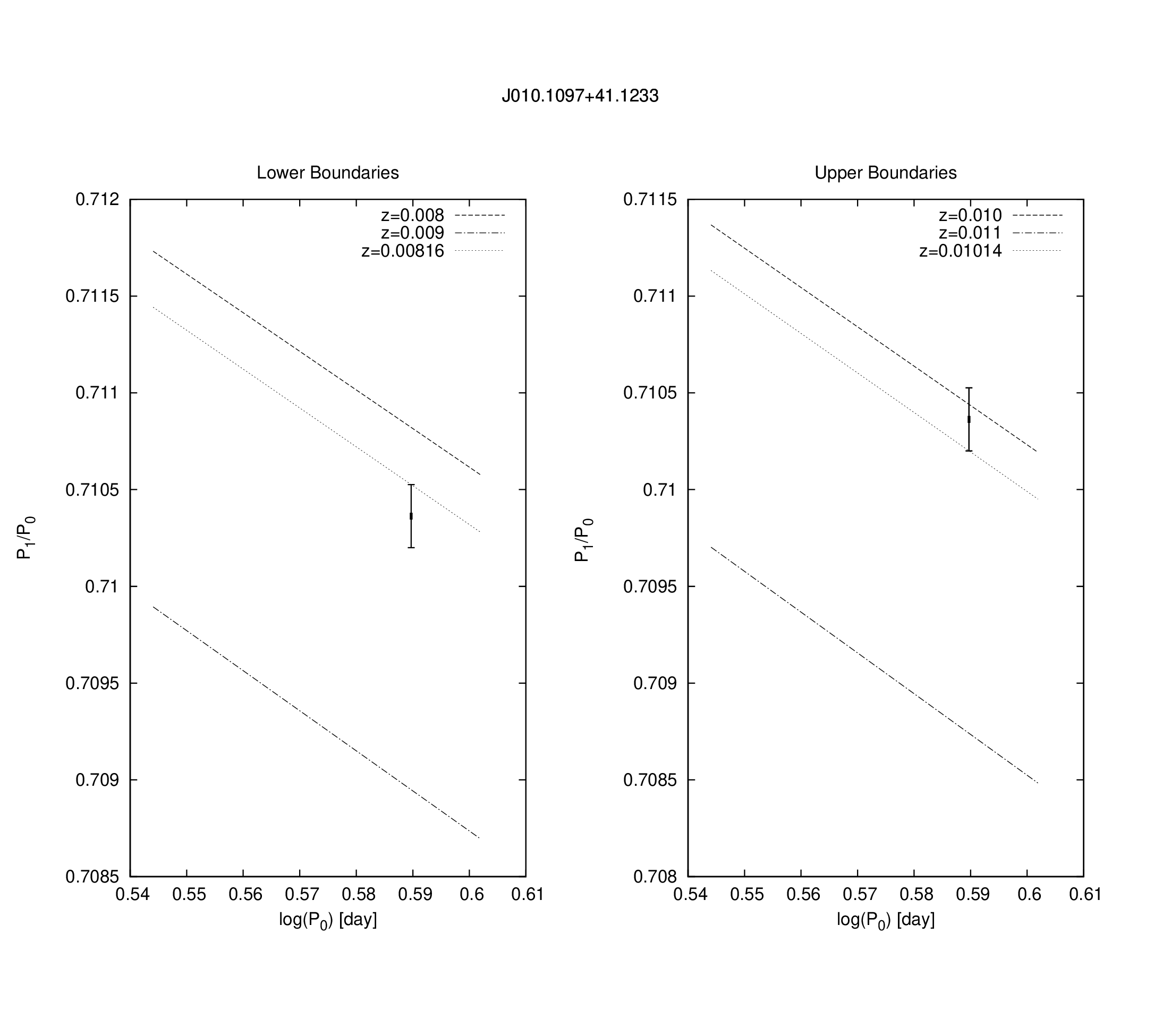}
  \caption{Zoom-in of the Petersen diagram on each candidate beat Cepheid. 
We show the adjacent theoretical isometallicity tracks and the interpolated \textit{Z} values at the  
position of the beat Cepheids. The dashed and dash-dotted curves are isometallicity tracks from the 
theoretical work of \cite{2008ApJ...680.1412B}, which are the higher and lower isometallicity tracks 
adjacent to our measured log P$_0$ and P$_1$/P$_0$ values shown in black. The dotted isometallicity line
is the interpolation that passes through our measured log P$_0$ and P$_1$/P$_0$ values. The 
estimated lower (\textit{Z$_{min}$}, left subfigures) and upper (\textit{Z$_{max}$}, right subfigures)
metallicity limits are obtained from these interpolated values.}
  \label{fig.lc1}
\end{figure*}

\clearpage
\addtocounter{figure}{-1}
\begin{figure*}[!h]
  \centering
  \includegraphics[scale=0.2]{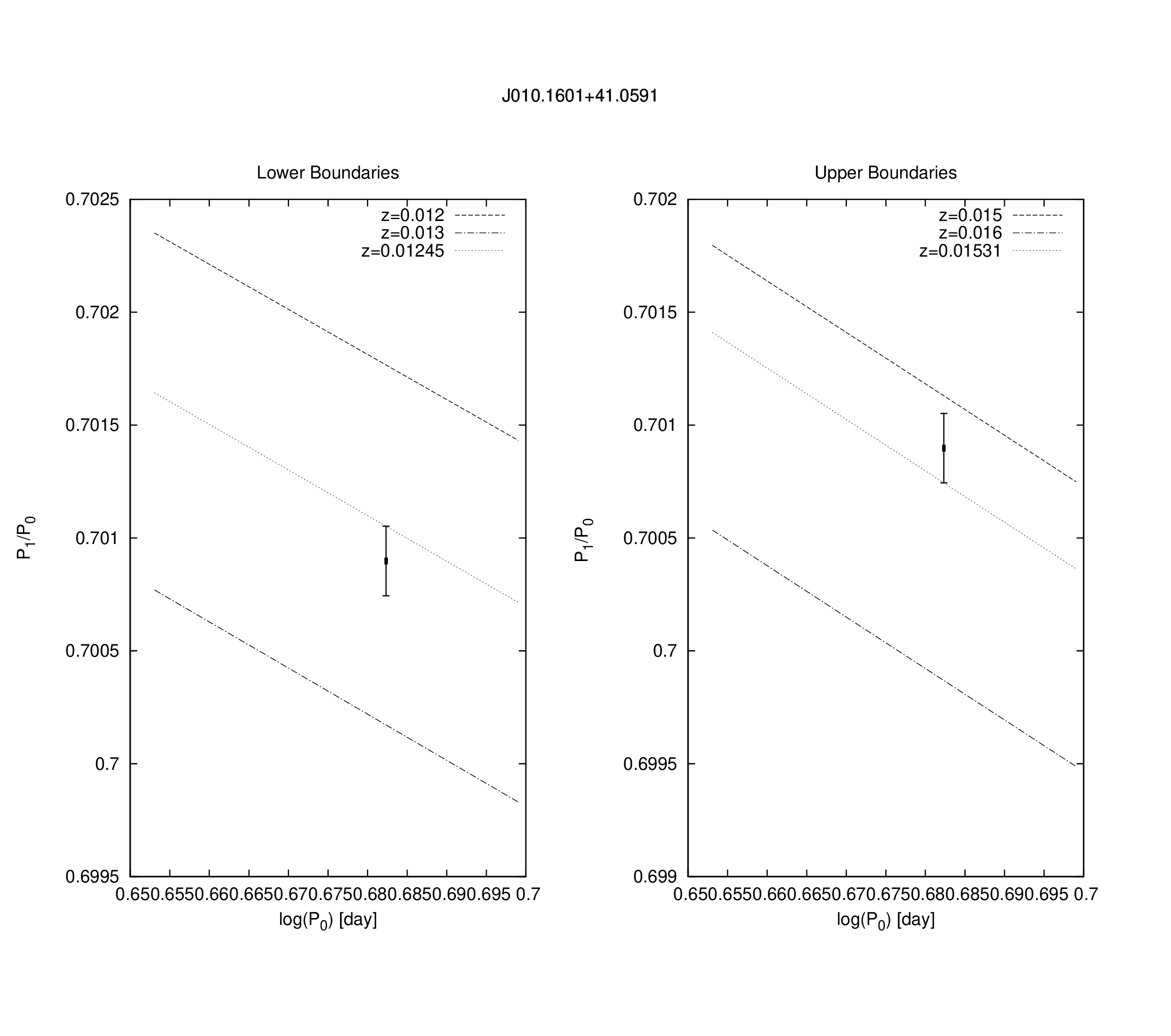}
  \includegraphics[scale=0.2]{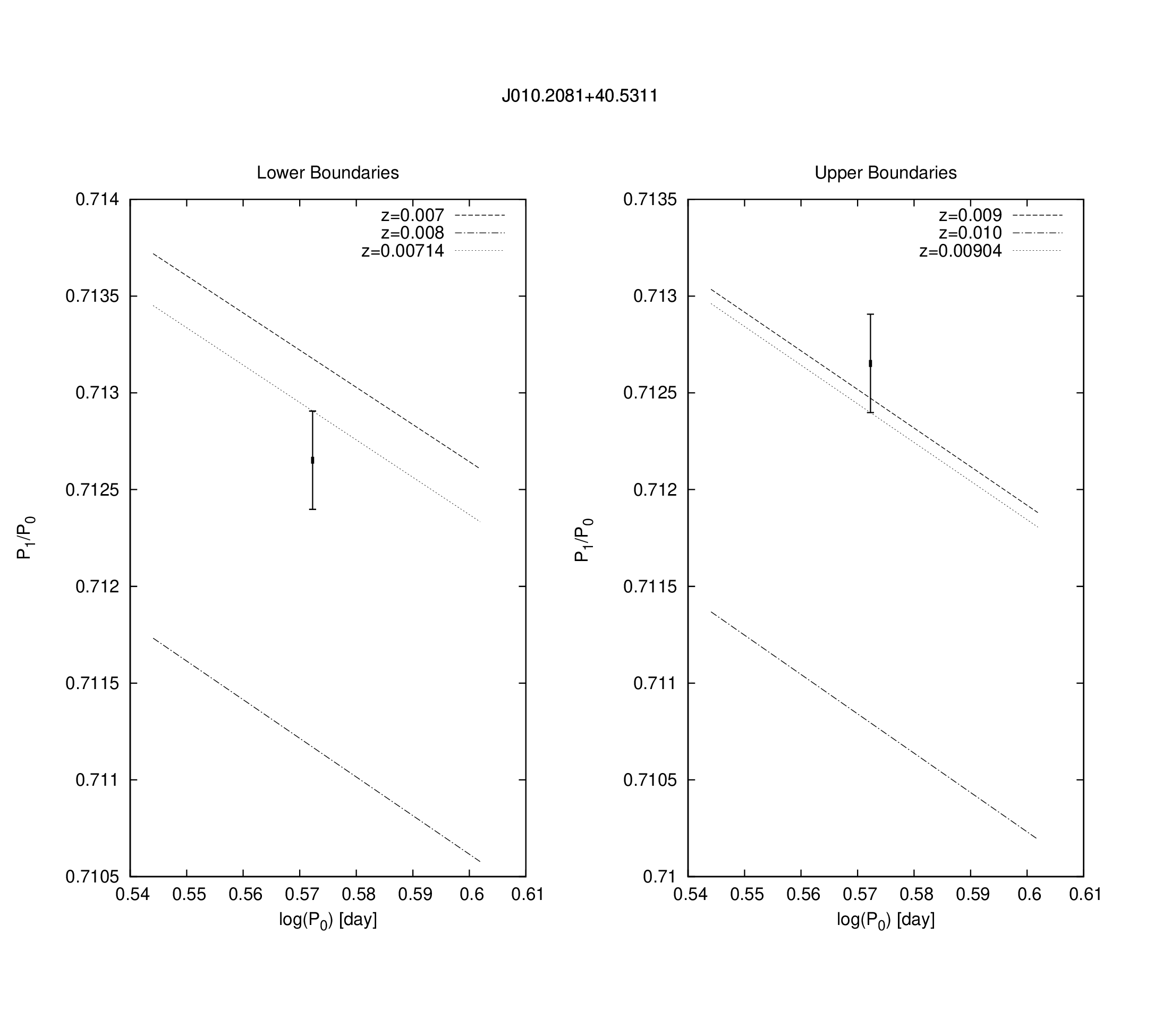}
  \includegraphics[scale=0.2]{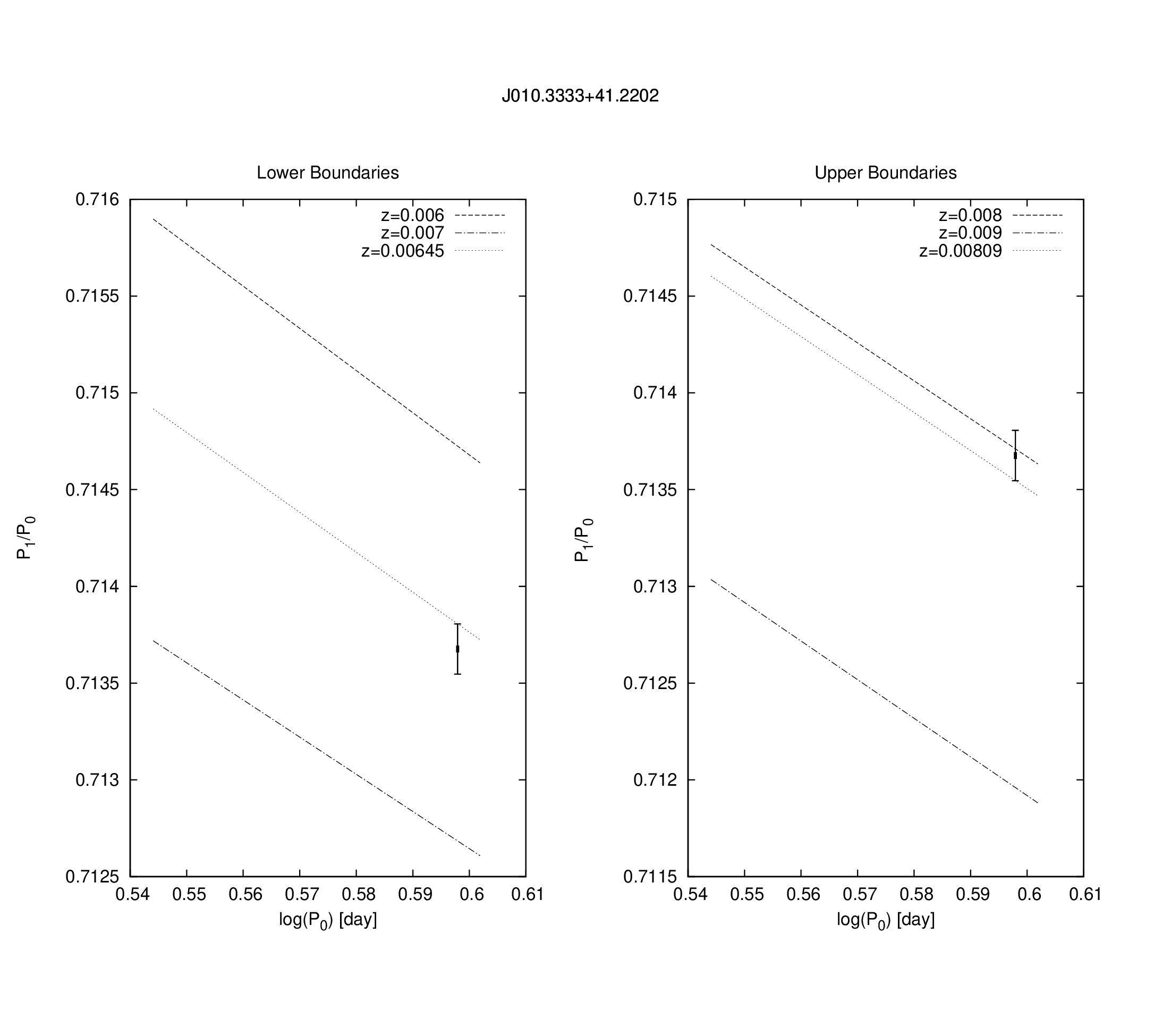}
  \includegraphics[scale=0.2]{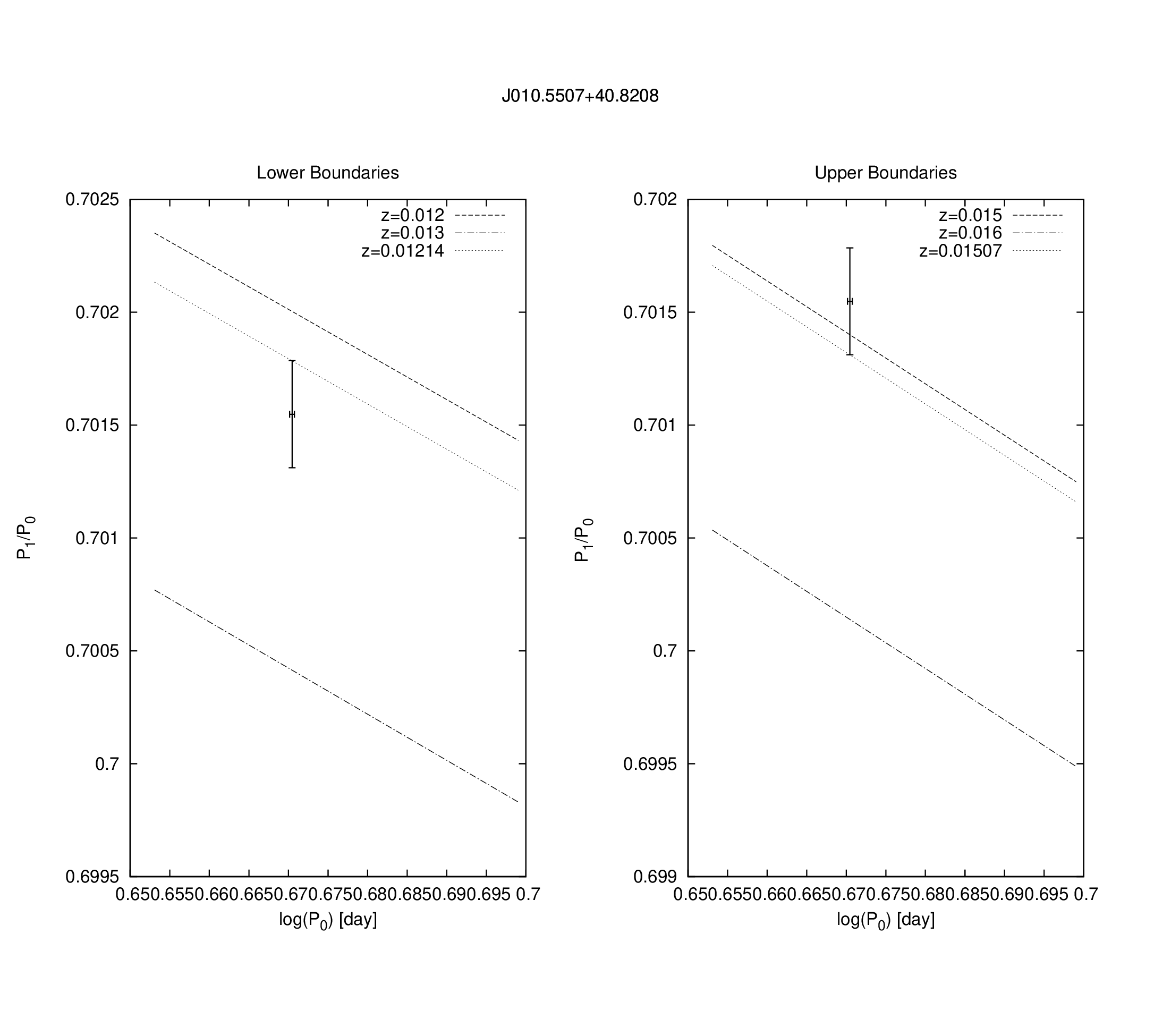} 
  \includegraphics[scale=0.2]{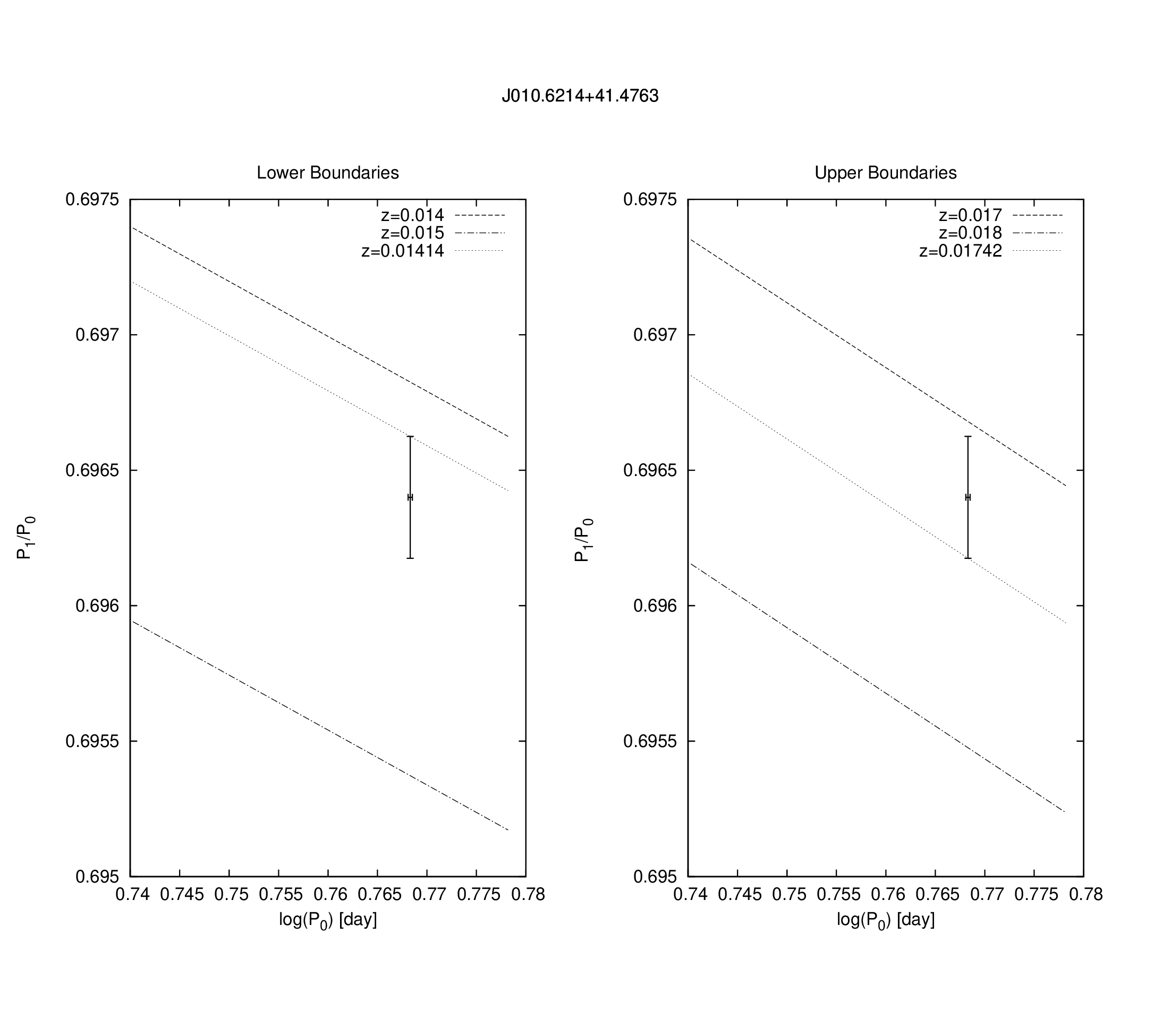}
  \includegraphics[scale=0.2]{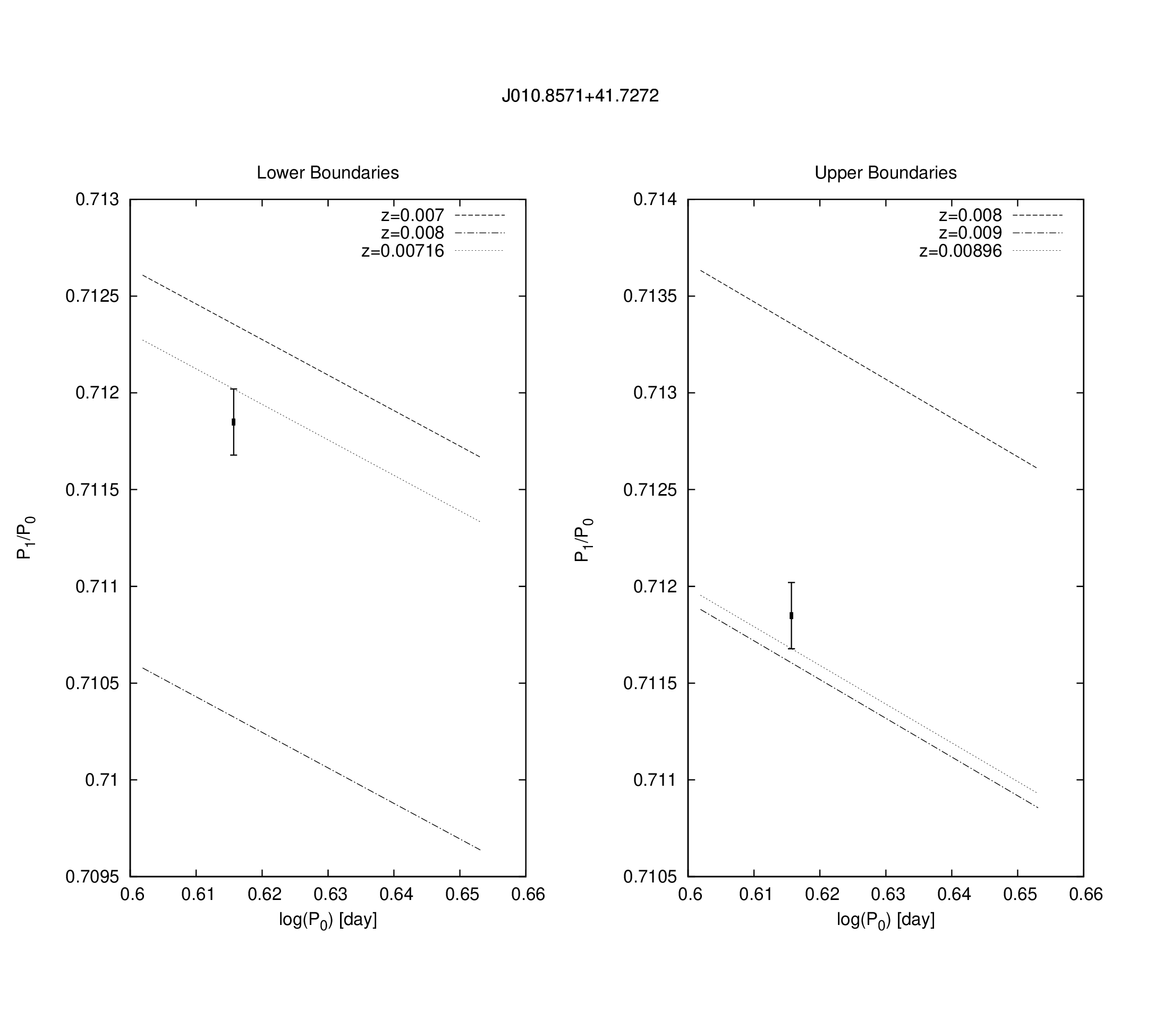}

  \caption{Zoom-in of the Petersen diagram - \textit{continued}.}
  \label{fig.lc}
\end{figure*}

\clearpage
\addtocounter{figure}{-1}
\begin{figure*}[!h]
  \centering
  \includegraphics[scale=0.2]{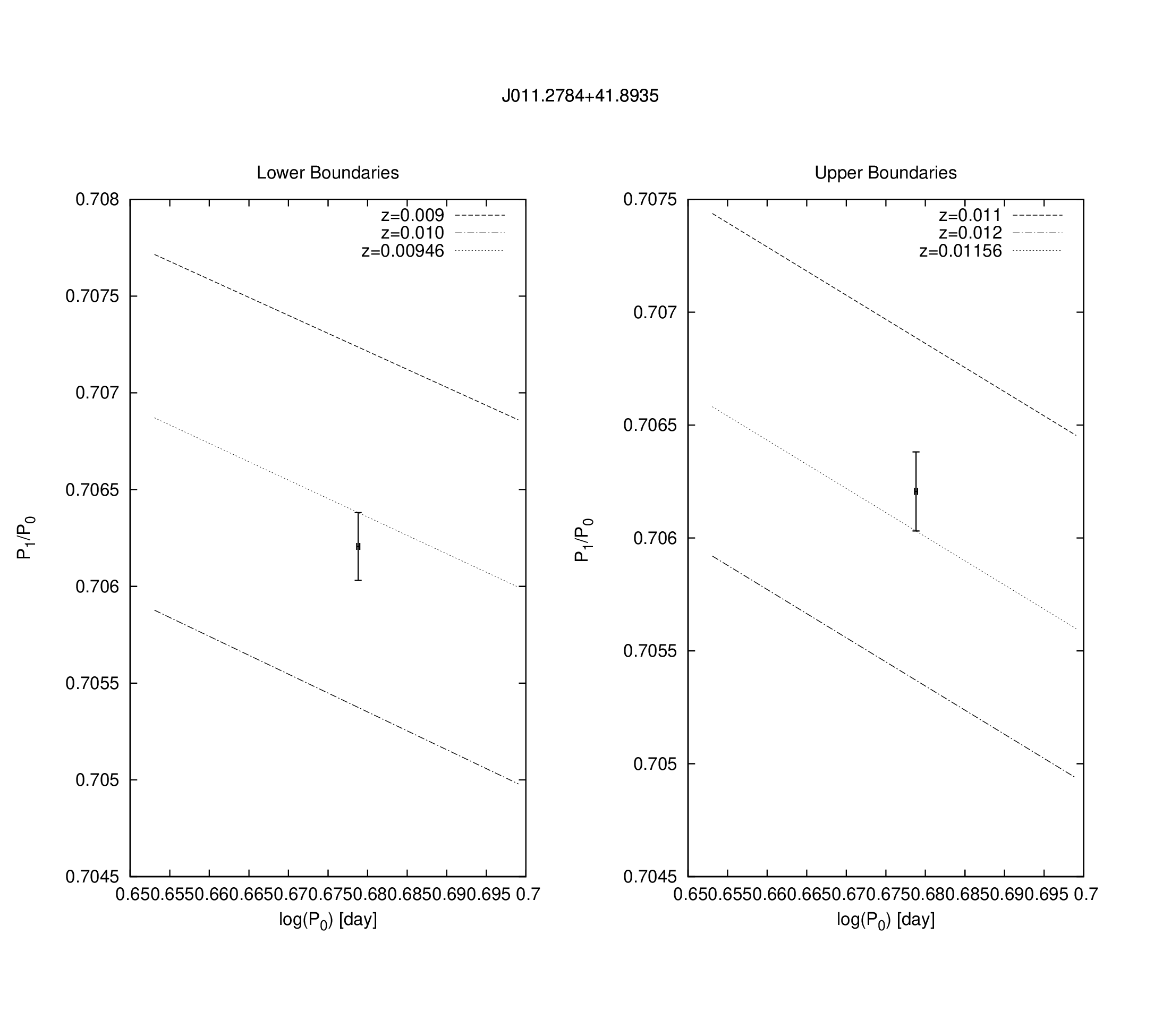}
  \includegraphics[scale=0.2]{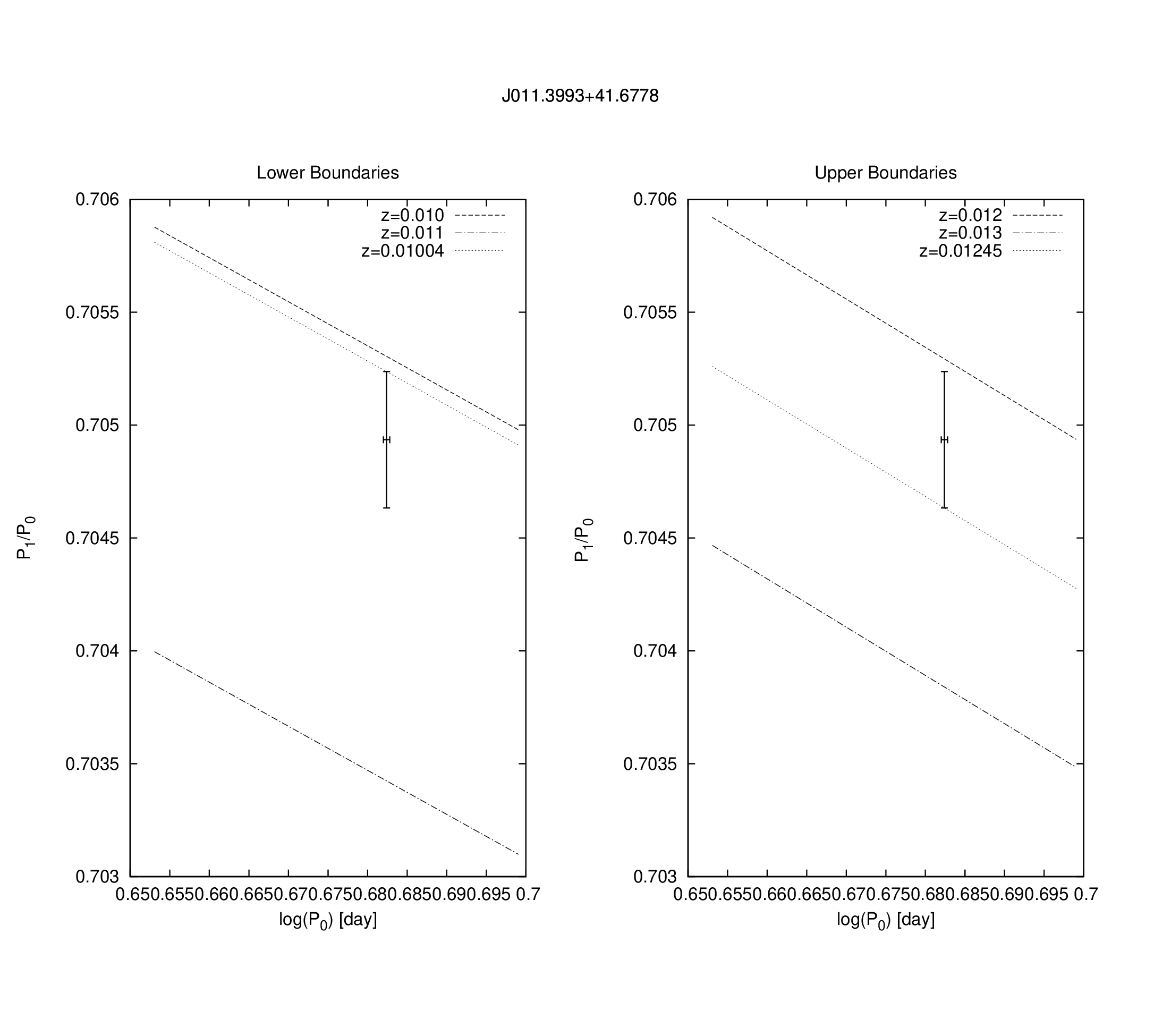}
  \includegraphics[scale=0.2]{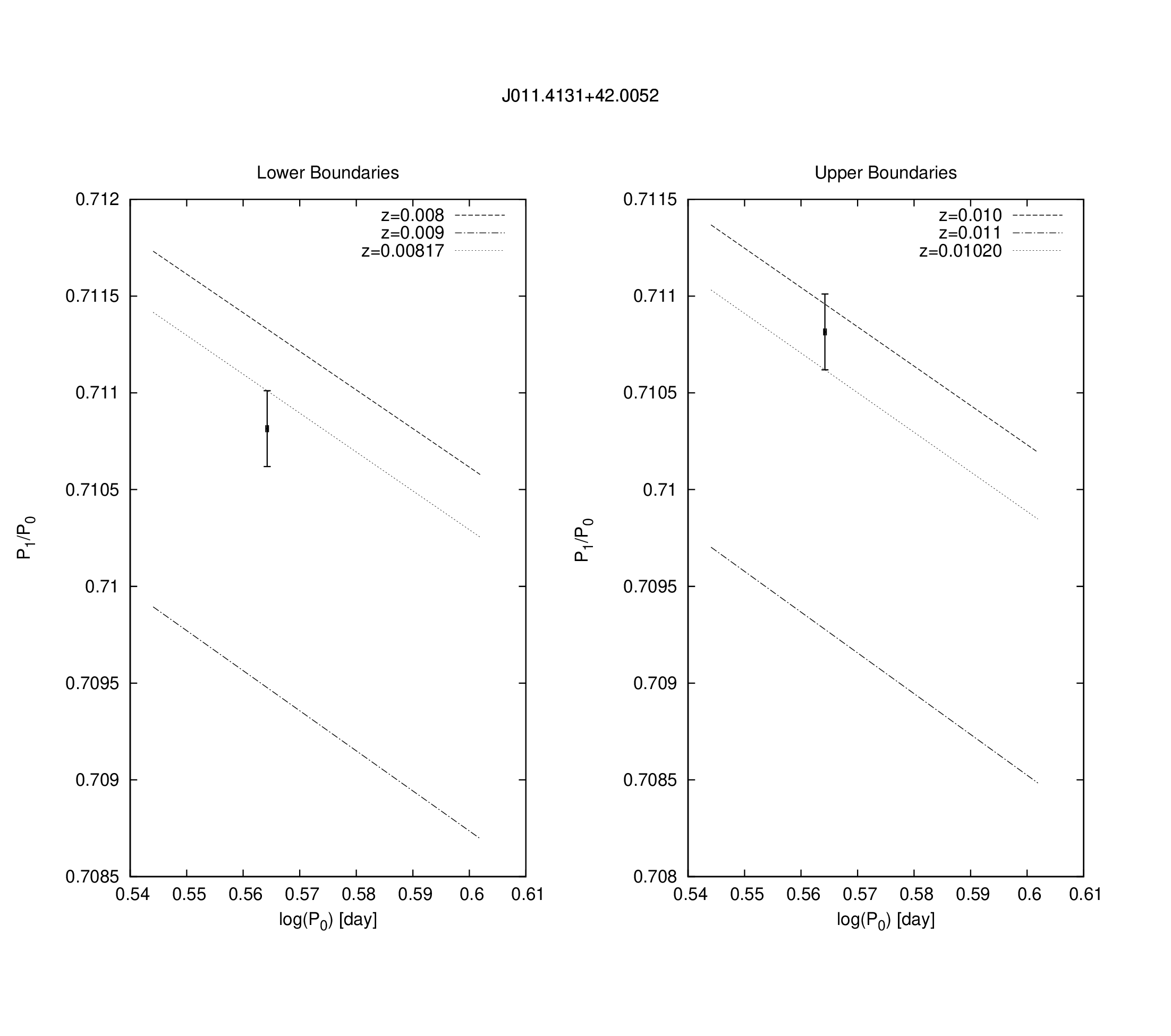}
  \includegraphics[scale=0.2]{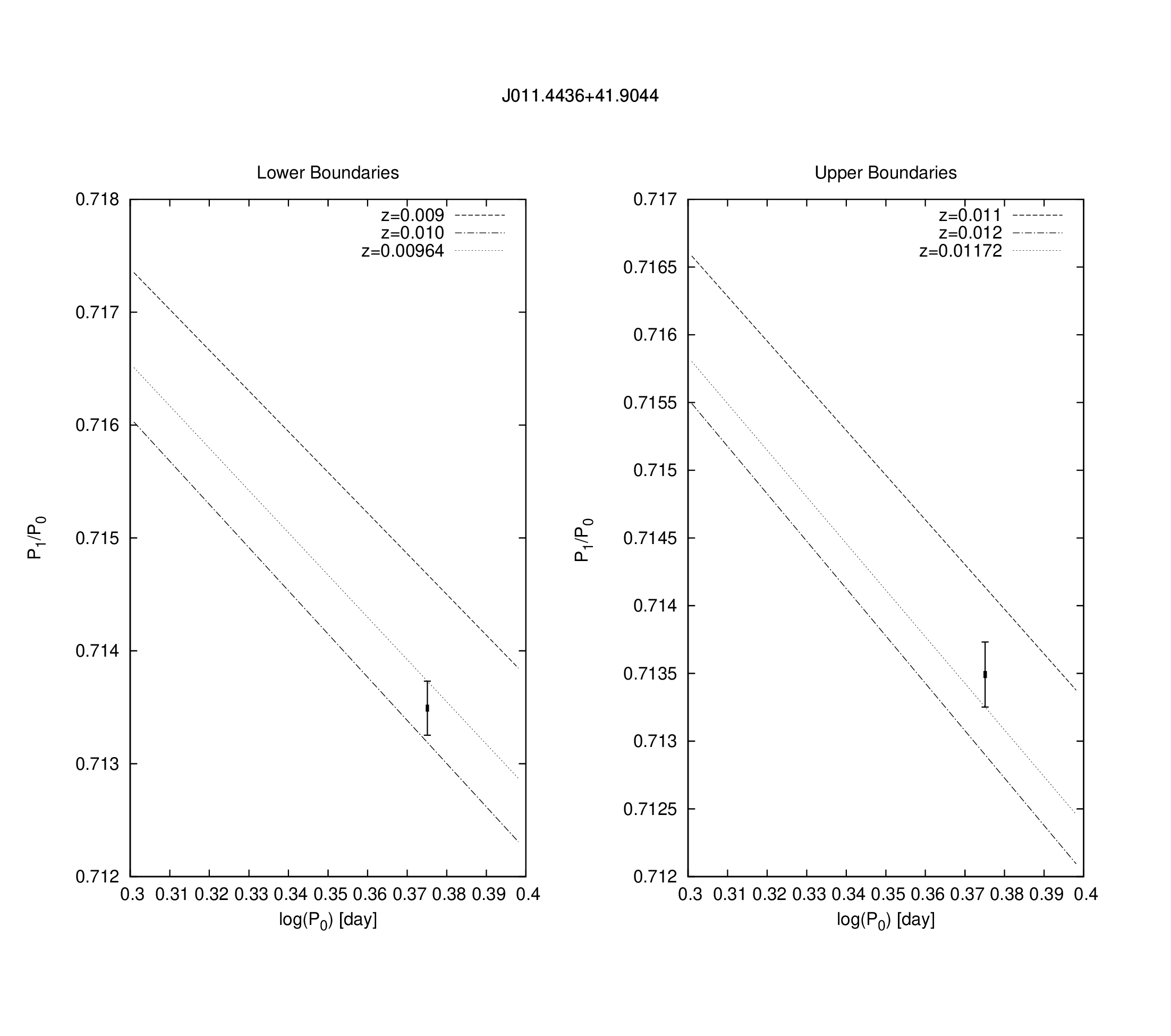}
  \includegraphics[scale=0.2]{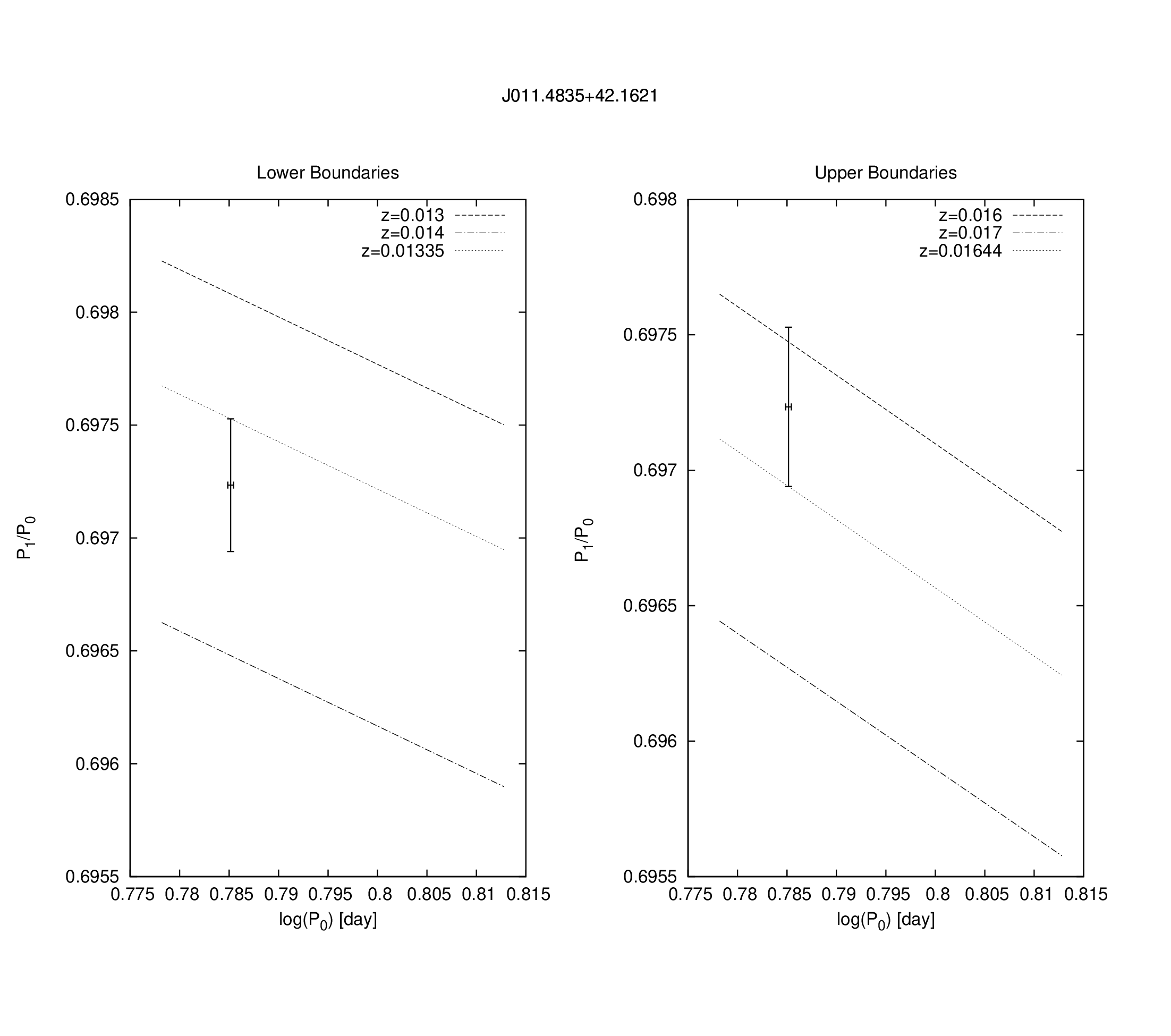} 
  \caption{Zoom-in of the Petersen diagram - \textit{continued}.}
  \label{fig.lc}
\end{figure*}

\end{document}